\documentclass[twocolumn,twocolappendix]{aastex63}      %

\usepackage{listings,amsmath,amssymb,upgreek,xspace}
\usepackage{newtxtext,newtxmath}
\usepackage[T1]{fontenc}
\usepackage{enumitem}

\DeclareRobustCommand{\VAN}[3]{#2}
\let\VANthebibliography\thebibliography
\def\thebibliography{\DeclareRobustCommand{\VAN}[3]{##3}\VANthebibliography}

\newcommand{\srevise}[1]{#1}
\newcommand{\threvise}[1]{#1}
\def\e{\mathrm{e}}                                 %
\def\mH{m_\mathrm{H}}                              %
\def\kB{k_\mathrm{B}}                              %
\def\MJ{M_{\mathrm{J}}}                            %
\def\RJ{R_{\mathrm{J}}}                            %
\def\ME{M_{\mathrm{E}}}                            %
\def\MSun{{M_\odot}}                               %
\def\LSun{{L_\odot}}                               %

\def\yr{\mathrm{yr}}

\def\Lya{\mbox{Ly\,$\alpha$}\xspace}                      %
\def\Baa{\mbox{Ba\,$\alpha$}\xspace}                      %
\def\Ha{\mbox{H\,$\alpha$}\xspace}                        %
\def\Hb{\mbox{H\,$\beta$}\xspace}                         %
\def\Paa{\mbox{Pa\,$\alpha$}\xspace}                      %
\def\Pab{\mbox{Pa\,$\beta$}\xspace}                       %
\def\Pag{\mbox{Pa\,$\gamma$}\xspace}                       %
\def\Bra{\mbox{Br\,$\alpha$}\xspace}                      %
\def\Brg{\mbox{Br\,$\gamma$}\xspace}                      %
\def\HII{\mbox{H\,\textsc{ii}}\xspace}                    %
\def\HeI{\mbox{He\,\textsc{i}}\xspace}                    %
\def\CaII{\mbox{Ca\,\textsc{ii}}\xspace}                  %
\def\NaI{\mbox{Na\,\textsc{i}}\xspace}                    %
\def\OI{\mbox{O\,\textsc{i}}\xspace}                      %
\def\LkCa{LkCa\,15\,A\xspace}                             %
\def\LkCab{LkCa\,15\,b\xspace}                            %
\def\PDS{\mbox{PDS\,70\,A}\xspace}                        %
\def\PDSb{\mbox{PDS\,70\,b}\xspace}                       %
\def\PDSc{\mbox{PDS\,70\,c}\xspace}                       %

\def\kms{\mathrm{km}\,\mathrm{s^{-1}}}             %
\def\MdotU{\ME\,\mathrm{yr}^{-1}}                  %
\def\MdotUJ{\MJ\,\mathrm{yr}^{-1}}                 %
\def\MdotUS{\MSun\,\mathrm{yr}^{-1}}               %
\def\MMdotUJ{\MJ^2\,\mathrm{yr}^{-1}}              %

\def\Mdot{\ensuremath{\dot{M}}\xspace}             %
\def\Mdotdisc{\ensuremath{\dot{M}_{\mathrm{disk}}}\xspace}      %
\def\lgMd{{\mathrm{lg}\Mdot_2}}                    %
\def\lgnz{{\mathrm{lg}n_{12}}}                    %
\def\MP{M_{\mathrm{p}}}                            %
\def\RP{R_{\mathrm{p}}}                            %
\def\ffill{\ensuremath{f_{\mathrm{fill}}}\xspace}  %
\def\ff{\ffill}                                    %
\def\Tacc{T_{\mathrm{acc}}}                        %
\def\Lacc{\ensuremath{L_{\mathrm{acc}}}\xspace}    %
\def\Facc{\ensuremath{F_{\mathrm{acc}}}\xspace}    %
\def\Lint{\ensuremath{L_{\mathrm{int}}}\xspace}    %
\def\Lphot{\ensuremath{L_{\mathrm{phot}}}\xspace}  %
\def\Lshock{\ensuremath{L_{\mathrm{shock}}}\xspace}  %
\def\LHa{\ensuremath{L_{\mathrm{H}\,\alpha}}\xspace} %
\def\FHa{\ensuremath{F_{\mathrm{H}\,\alpha}}\xspace} %
\def\LHb{\ensuremath{{L_{\textnormal{H}\,\beta}}}\xspace}              %
\def\LPab{\ensuremath{{L_{\textnormal{Pa}\,\beta}}}\xspace}            %
\def\LBra{\ensuremath{{L_{\textnormal{Br}\,\alpha}}}\xspace}           %
\def\LBrg{\ensuremath{{L_{\textnormal{Br}\,\gamma}}}\xspace}           %
\def\Lline{\ensuremath{L_{\mathrm{line}}}\xspace}                      %
\def\WHa{{\Delta \lambda_{\textnormal{H}\,\alpha}}} %

\newcommand{\Rtrunc}{{R_{\textnormal{trunc}}}}     %
\newcommand{\RAkk}{{R_{\textnormal{acc}}}}         %
\newcommand{\RHill}{{R_{\textnormal{Hill}}}}       %
\newcommand{\RBondi}{{R_{\textnormal{Bondi}}}}     %
\newcommand{\kLiss}{{k_{\textnormal{Liss}}}}       %
\newcommand{\Tint}{{T_{\textnormal{int}}}}         %
\newcommand{\Teff}{\ensuremath{{T_{\textnormal{eff}}}}\xspace}         %
\def\cs{c_{\mathrm{s}}}                            %
\def\Mach{\mathcal{M}}                             %
\def\Pram{P_{\textnormal{ram}}}                    %
\def\yt{y_\mathrm{t}}                              %
\def\AV{A_\mathrm{V}}                              %
\def\AR{A_\mathrm{R}}                              %
\def\AHa{A_{\Ha}}                                  %
\def\etaklassisch{ \eta^{\rm kin} }                %
\newcommand{\vff}{{v_{\textnormal{ff}}}}           %
\def\fdown{f_{\mathrm{down}}}                      %
\def\Fdown{F_{\mathrm{down}}}                      %
\def\Fshock{F_{\mathrm{shock}}}                    %
\def\tcool{\ensuremath{t_{\mathrm{cool}}}\xspace}  %
\def\tpop{\ensuremath{t_{\mathrm{pop}}}\xspace}    %
\def\Eint{\ensuremath{E_{\mathrm{int}}}\xspace}    %
\def\LambdaLya{\ensuremath{\Lambda_{\mathrm{Ly}\,\alpha}}\xspace}

\defcitealias{Storey+Hummer1995}{SH95}

\received{---}
\revised{---}
\accepted{---}
\submitjournal{ApJ}

\shorttitle{Planetary accretion shock spectra}
\shortauthors{Aoyama et al.}

\begin{document}

\title{Spectral appearance of the planetary-surface accretion shock:\\
Global spectra and hydrogen-line profiles and luminosities}

\author[0000-0003-0568-9225]{Yuhiko Aoyama} 
\affiliation{%
Department of Earth and Planetary Science,
University of Tokyo,
7-3-1~Hongo, Bunkyo-ku,
Tokyo~113-0033, Japan
}
\affiliation{Institute for Advanced Study, Tsinghua University, Beijing 100084, People's Republic of China}
\affiliation{Department of Astronomy, Tsinghua University, Beijing 100084, People's Republic of China}

\author[0000-0002-2919-7500]{Gabriel-Dominique Marleau}
\affiliation{%
Institut f\"ur Astronomie und Astrophysik,
Universit\"at T\"ubingen,
Auf der Morgenstelle 10,
72076 T\"ubingen, Germany
}
\affiliation{%
Physikalisches Institut,
Universit\"{a}t Bern,
Gesellschaftsstr.~6,
3012 Bern, Switzerland
}
\affiliation{%
Max-Planck-Institut f\"ur Astronomie,
K\"onigstuhl 17,
69117 Heidelberg, Germany
}

\author[0000-0002-1013-2811]{Christoph Mordasini}
\affiliation{%
Physikalisches Institut,
Universit\"{a}t Bern,
Gesellschaftsstr.~6,
3012 Bern, Switzerland
}

\author[0000-0002-5658-5971]{Masahiro Ikoma}
\affil{%
Department of Earth and Planetary Science,
University of Tokyo,
7-3-1~Hongo, Bunkyo-ku,
Tokyo~113-0033, Japan
}
\affiliation{Division of Science, National Astronomical Observatory of Japan, 2-21-1~Osawa, Mitaka, Tokyo~181-8588, Japan}

\begin{abstract}
Hydrogen-line emission
from an accretion shock  %
has recently been observed at planetary-mass objects.
Our previous work
predicted the shock spectrum and luminosity for  %
a shock on the circumplanetary disc.
We extend this to the
planet-surface shock.
We  %
calculate the global spectral energy distribution (SED) of accreting planets by combining our model emission spectra with photospheric SEDs, and %
predict the line-integrated flux for several hydrogen lines, especially H$\alpha$, but %
also H$\beta$, Pa$\alpha$, Pa$\beta$, Pa$\gamma$, Br$\alpha$, and Br$\gamma$.
We apply our non-equilibrium emission model
to the surface accretion shock for a wide range of
accretion rates $\dot{M}$ and masses $M_\mathrm{p}$. 
Fits to formation calculations provide radii and effective temperatures.
Extinction by the surrounding material 
is neglected,
which is arguably often relevant.
We find that the line luminosity increases monotonically with $\dot{M}$ and $M_\mathrm{p}$,
depending mostly on $\dot{M}$
and weakly on $M_\mathrm{p}$
for the relevant range of parameters.
The Lyman, Balmer, and Paschen continua can %
exceed
the photosphere.
The H$\beta$ line is fainter by 0--1~dex than H$\alpha$,
whereas other lines are weaker (by $\sim1$--3~dex). %
Shocks on the planet %
or the CPD surface are distinguishable at very high spectral resolution, but
the planet surface shock %
likely dominates %
if both are present.
Applied to recent non-detections of H$\alpha$, our models imply looser constraints on the $\Mdot$ of putative planets than \threvise{when extrapolating fits from the stellar regime}.
These
hydrogen-line luminosity predictions %
are useful
for interpreting
(non-)detections of accreting planets.
\end{abstract}


\section{Introduction}
\label{sec:intro}
Recent instrumental improvement have enabled the observation of forming planets \citep[e.g.][]{Kraus12,Quanz+2013a,Currie+2015, Wagner+2018}.
Because it is expected to yield information on how planets grow,
the detection of \Ha\ is particularly important
\citep{Sallum+2015,Wagner+2018,Haffert+2019,Cugno+2019,Zurlo+2020,Xie+2020}.

In the context of forming low-mass protostars (Classical T Tauri stars: CTTS), \Ha is known as an indicator of accretion and used to estimate mass accretion rate \citep[e.g.][]{Gullbring+1998}.
The \Ha\ from CTTS is brighter than the photospheric continuum by a few mag and has large width ($\gtrsim 200~\kms$). The magnetospheric accretion model \citep{Uchida&Shibata1984,Koenigl1991} can explain these characteristic features when the accretion funnel is hot enough to emit \Ha\ \citep[e.g.][]{Hartmann+1994,Muzerolle+2001}. Furthermore, the line-integrated luminosity (\LHa) or the spectral width ($\WHa$) of the \Ha line shows a correlation to the mass-accretion rate (or accretion luminosity, %
\Lacc) estimated with modeling of continuum emission \citep[e.g.][]{Valenti+1993,Calvet+Gullbring1998}. Therefore, \Ha is used to estimate the accretion rate for protostars that are too far for their continuum emission to be observable \citep[e.g.][]{Gullbring+1998,Herczeg+Hillenbrand2008,Fang+2009,Rigliaco+2012,Alcala+2014,alcal17,Natta+2004}. Similar statements hold for further hydrogen lines from the Balmer, Paschen, Brackett, or other series.

As for the stellar case, an \Ha\ excess was reported from \threvise{(candidate)} protoplanets, and the observed luminosity was used to estimate the accretion rate by applying the results of the CTTS observations  \citep{Sallum+2015,Wagner+2018,Haffert+2019}. 
However, there is no guarantee that relationships between \Lacc and \LHa or $\WHa$ given in CTTS are valid for protoplanets.
\citet{Thanathibodee+2019} applied the stellar \Ha\ emission model of \citet{Muzerolle+2001} %
to a planetary-mass object (\object{\PDSb}) and argued the \LHa--\Lacc relationship shows a different trend from that of protostars (\citealp{Ingleby+2013,Rigliaco+2012}; see also \citealp{szul20} and the discussion of their work in \citealt{AMIM20L}).

Planetary gas accretion is qualitatively different from the stellar one in some points.
An important characteristic feature is that protoplanets and their surrounding gaseous disk (circum-planetary disk, CPD) are embedded in the stellar surrounding disk (protoplanetary disk, PPD).
On the way of gas accreting towards the protoplanet, the gas preferentially enters the planetary gravitational sphere in high altitudes above the disk midplane \citep[e.g.][]{Tanigawa+2012}.
When the gas falling from the PPD to CPD vertically hits the CPD surface,
it yields a strong shock, which can be hot enough to emit \Ha\ \citep{szulmorda17}. 
\citet{Aoyama+2018} constructed a model of shock-heated gas with cooling, chemical reactions, and radiative transfer, estimated hydrogen line luminosity depending on the gas velocity and density, and estimated the \LHa depending on the shock properties.

On the other hand, %
the magnetospheric accretion may occur even in the planetary accretion, bringing about a strong shock %
also on the planetary surface. 
If protoplanets have dipole magnetic fields strong enough to control the gas dynamics, vertical accretion can occur directly onto the planetary surface \citep{batygin18}.
While the accretion shock on the CTTS surface is too strong and makes gas too hot to emit \Ha\ \citep[see e.g.,][]{Hartmann+2016}, the weak gravity of protoplanets leads to moderate free-fall velocity ($\sim 100~\kms$) and to emitting a significant amount of \Ha. 
In contrast, in the CPD surface shock model, only a small fraction ($\lesssim 1 \%$) can contribute to the \Ha\ emission, because most gas hits the CPD far from the planet \citep{Aoyama+2018}. Also, in the magnetospheric accretion-funnel model, the heating mechanism is still an open question \citep{Muzerolle+2001}. Therefore, the gas in the accretion funnel could be too cool to emit \Ha, perhaps especially for protoplanets not much more massive than Jupiter.

This motivates us, in this study, to model the hydrogen line emission coming from the planetary surface shock, considering a wide range of parameters. We focus on %
\Ha first and then explore other hydrogen line emission.
We combine these results with models of the photospheric emission and discuss when the shock lines are visible above the photosphere emission. 
Note that part of the planetary surface shock model presented here was used already in \citet{Aoyama+Ikoma2019} for the case of PDS~70~b and c. A more extensive investigation is done in this study.

The paper is organized as follows: In Section~\ref{sec:model} we discuss the properties of the planetary-surface shock and of the planets
and review our numerical shock model, which was introduced in \citet{Aoyama+2018}.
In Section~\ref{sec:manyspectra} we present emission spectra of accreting gas giants for a large grid of models, before applying this in Section~\ref{sec:obs} to a few objects, especially to their detection at \Ha. In Section~\ref{sec:furthobsabs} we explore other observational aspects, including line strengths for lines other than \Ha and the possibility of breaking some degeneracies. Finally, we present a critical discussion in Section~\ref{sec:Discussion} before summarizing in Section~\ref{sec:summconc}.
The appendices present further material:
a  %
discussion of our approach compared to \citet{Storey+Hummer1995} (Appendix~\ref{sec:cfSH95}),
the inverse relationship between the shock-microphysical and planet-formation parameters (Appendix~\ref{sec:inverse n0v0}),
a map of the \Ha luminosity %
for the cold-start radius fits (Appendix~\ref{sec:HaContourboth}),
and the calculation of the \Ha luminosity in \citet{Wagner+2018} (Appendix~\ref{sec:Wagner}).

\section{Description of the combined model}
 \label{sec:model}

We model the spectral energy distribution (SED) of an accreting gas giant with a surface accretion shock.
The radiation from the accreting gas giant is mainly composed of two components, namely the photospheric radiation \srevise{(Section~\ref{sec:photo emission})} and the shock excess \srevise{(Section~\ref{sec:shock emission})}.
We assume that the two components can be computed separately, i.e., that the layers heated by the shock do not affect significantly the rest of the emission.

The input parameters for our combined spectra of the accretion shock and the photosphere
are \srevise{the following five:} mass accretion rate $\Mdot$, planet mass $\MP$, planet radius $\RP$, filling factor $\ffill$ of the shock on the planet surface, and photospheric effective temperature $\Teff$.
However, taking them as independent would result in an impractically large parameter space and may lead to unlikely combinations (e.g., small radius and mass but large luminosity).
\srevise{Therefore, in this study, only %
($\Mdot, \MP$)
will be considered as free parameters via the modeling described in Section~\ref{sec:Fit}.}
We consider here the case that the planetary emission (photosphere and shock) is not extincted, and detail in Section~\ref{sec:neglectextinc} when this is relevant.

\subsection{Fitting of planetary properties}
\label{sec:Fit}
For convenience, the planet radius and effective temperature are derived from a specific detailed planet formation and structure model.
\srevise{The planet radius $\RP$ is fitted as a function of ($\Mdot,\,\MP$) by using the Bern model \citep{alibert05,morda12_I,morda12_II,morda15,mordasini17,Emsenhuber+2020a,Emsenhuber+2020b,Schlecker+2020}. 
We use a constant intrinsic temperature of $1000$~K and the accretion heating to predict the effective temperature $\Teff$.
For consistency, we assume, in the incoming mechanical energy at the shock, that the residual that is not radiated from the shock emission heats the photosphere.
More details are given in Section~\ref{sec:Fittin} (see Equation~\ref{eq:fluxconserv}).}

\subsection{Photospheric emission}
\label{sec:photo emission}
For the photospheric radiation model, we use the \texttt{CIFIST2011\_2015} BT-Settl models, which calculated spherical radiative transfer in atmospheres with solar metallicity\footnote{From \url{https://phoenix.ens-lyon.fr/Grids/BT-Settl/CIFIST2011_2015/}.}
\citep{Allard+2012,Baraffe+2015}.
\srevise{This requires the %
effective temperature, surface gravity, and the emitting area. They are derived from ($\Teff,\,\MP,\,\RP,\,\ffill$).}

\srevise{%
The BT-Settl model simulates the photospheric emission from isolated objects. The accretion heats the top of the atmosphere and, in general, will change the temperature structure. Since the detailed absorption feature could highly depend on the temperature structure in the upper layer, they are less reliable for accreting objects. However, this model can show how bright the shock emission is relative to the photospheric emission, i.e., the detectability of shock emission.
}

In this study, we focus on the shock-heated gas on the planetary surface. 
We treat only the emission from the photosphere and shock-heated gas but not from the CPD, whose temperature is lower than those of the photosphere and the shock.
Continuum emission from a (simplified) CPD model has been calculated in \citet{zhu15}, \citet{Eisner2015}, and \citet{Szulagyi+2019}, and the line emission from the shock on the CPD has been calculated in \citet{Aoyama+2018}.

\subsection{Shock emission}
\label{sec:shock emission}
The shock excess is calculated from the 1D radiation-hydrodynamic model developed by \citet{Aoyama+2018} and \citet{Aoyama+Ikoma2019}, which is outlined in Section~\ref{sec:shockmodel}.
\srevise{This model predicts the shock emission flux from the gas velocity $v_0$ and number density $n_0$ just before the shock. These shock parameters and the emitting area can be derived with ($\Mdot,\,\MP,\,\RP,\,\ff$), assuming $v_0$ to be the free-fall velocity. The details are described in Section~\ref{sec:shockmodel}. For simplicity, we use $\ff=1$ because $\ff$ hardly changes the line flux (see also Section~\ref{sec:geoeffect}). A discussion of how the accretion geometry sets \ff is given in \citet[][see their Figure~1]{maea21}.}

\srevise{In this study, we treat only the shock on the planetary surface rather than that on CPD surface, which \citet{Aoyama+2018} studied.
We  %
compare the emission from both
in Section~\ref{sec:distinguishsurfaceshockCPD}.
}

\srevise{Also, we contrast our model of the shock emission to \citet{Storey+Hummer1995} in Appendix~\ref{sec:cfSH95}.
In short, for the shock heated gas, collisional excitation from the ground state is a crucial element that we include, while \citetalias{Storey+Hummer1995} did not.}

\subsection{Neglecting extinction}
 \label{sec:neglectextinc}

In this work, extinction by material between the shock surface (the planet) and the observer is not considered. Several components contribute to it. In principle, the contribution from the interstellar medium (ISM) can be determined for a given source, from the stellar spectrum or by statistical tools such as \texttt{Stilism} of \citet{Lallement+2019}. Thus this component is relatively easily accounted for. To what extent the gas or dust surrounding a forming planet may weaken the shock signal is an important question, as the recent observational results and theoretical modeling in \citet{Hashimoto+2020}, \citet{Stolker+20b}, and \citet{Sanchis+2020} highlight. However, considering extinction adds an entire level of complexity and brings many uncertainties, in particular concerning the radiative transfer geometry and the dust opacity. Therefore, we deal with extinction by the gas and the dust in a dedicated paper \citep{maea21}.

Nevertheless, the extinction-free case is relevant in itself.
As we show in \citet{maea21}, an accretion flow free of extinction at \Ha is a plausible assumption for a wide range of accretion rates and masses.
We look at this in detail but,
heuristically, the transition disk gaps in which planets are found are usually dust-free \citep{Close2020}, and gas cooler than a few thousand kelvin can be optically thin for non-Lyman-series hydrogen lines.
Also, while \citet{Hashimoto+2020} derived for \PDSb an extinction of $>2.0$~mag at \Ha, we should recall that this is for a single object (\PDSb), and that this estimate depends on the wide spectral width of the observed \Ha line (see Section~\ref{sec:PDS70}), which can be overestimated due to the finite instrumental resolution \citep{Thanathibodee+2019}. More generally, it is conceivable that for some accretion and viewing geometries the \Ha produced at the shock could leave the system without passing through any absorbing material that could be present.
For these reasons, it seems sensible to treat the extincted case separately.

\section{Theoretical spectra of forming gas giants}
 \label{sec:manyspectra}

We now turn to results from the methods described above.
We look first at one representative example in detail (Section~\ref{sec:ein Beispiel}; \citealt{Aoyama+2018} showed three other cases for the CPD case)
and then survey a large part of the relevant $(\Mdot,\MP)$ parameter space (Section~\ref{sec:grid}).

\subsection{One example}
 \label{sec:ein Beispiel}

\subsubsection{Postshock structure}
 \label{sec:Postshock structure}
 
\begin{figure*}%
\plottwo{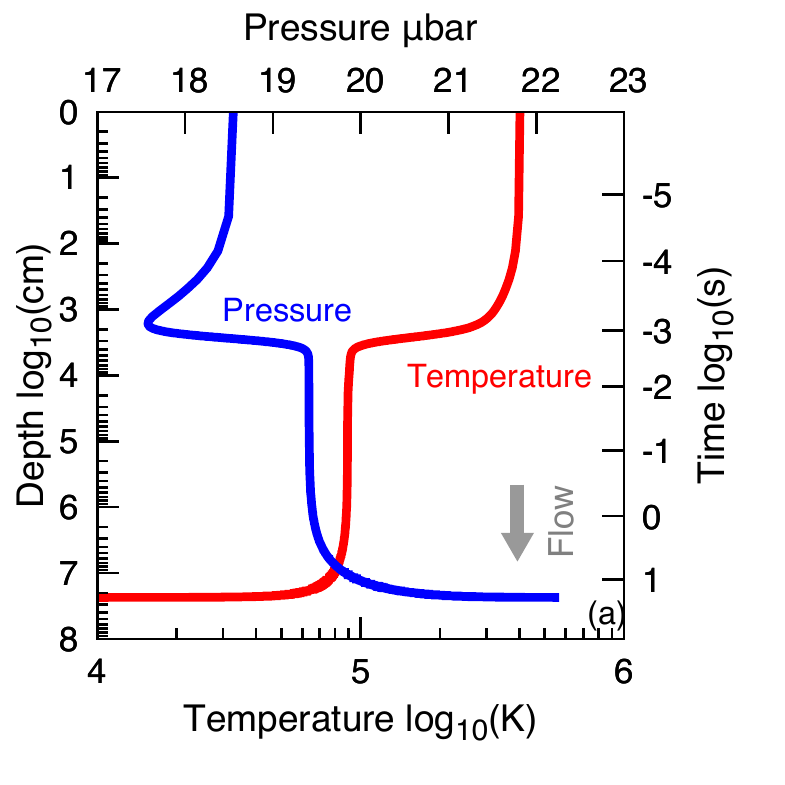}{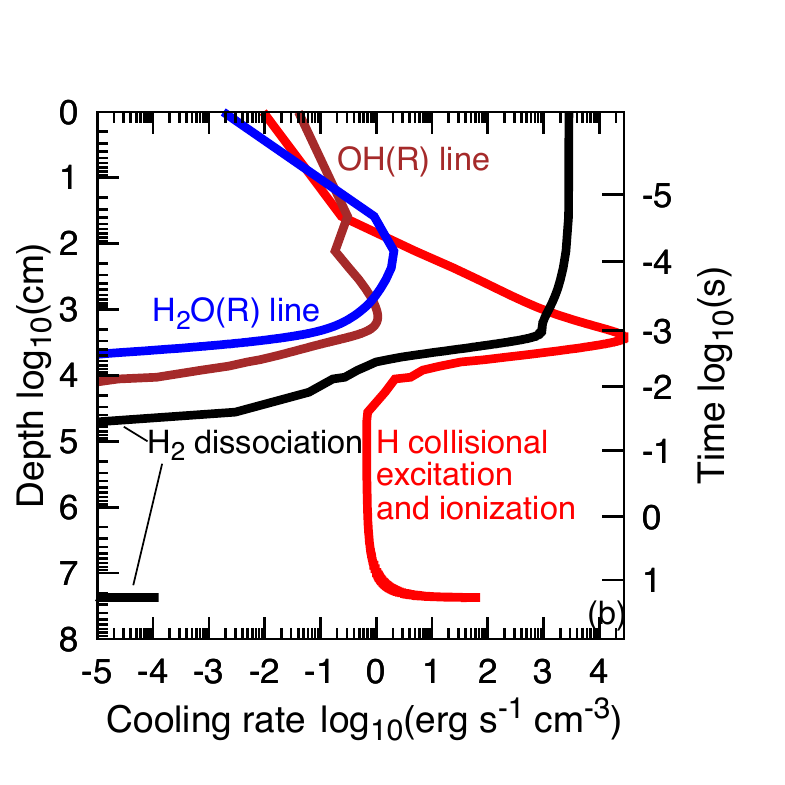}
\plottwo{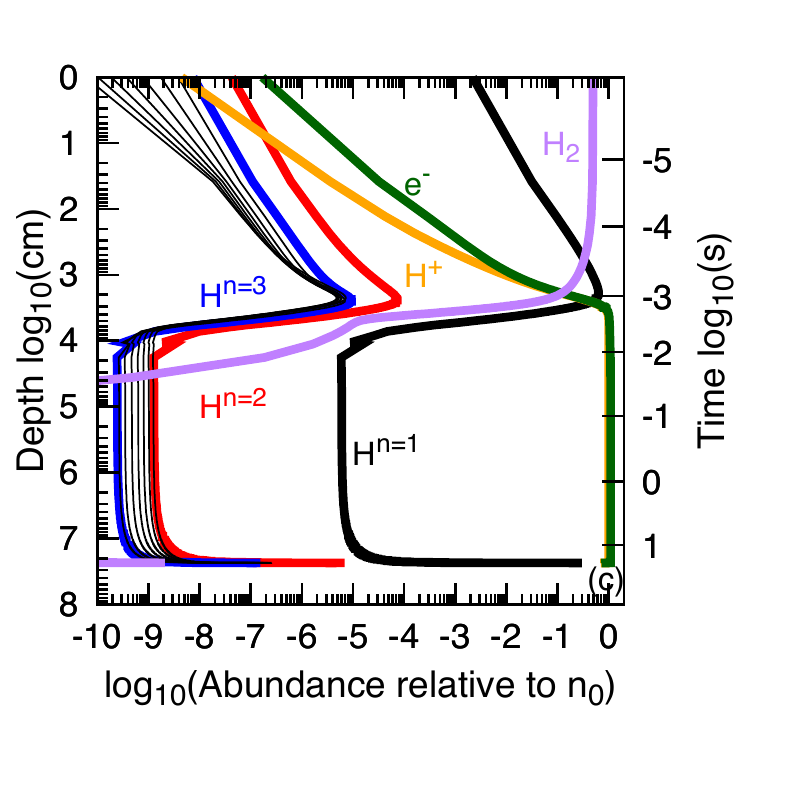}{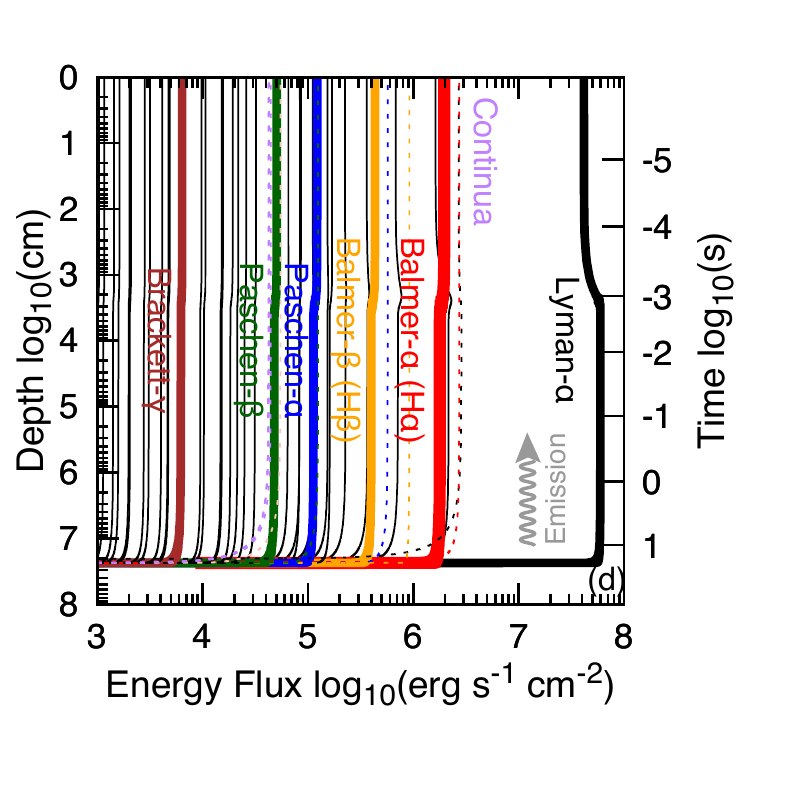}
\caption{Postshock flow structure, beginning immediately after the hydrodynamical jump, for $v_0=100~\kms$ and $n_0=10^{11}$~cm$^{-3}$.
This corresponds for example to $\Mdot=10^{-8}~\MdotUJ$, $\MP=5~\MJ$, and $\RP=1.7~\RJ$ with $\ffill=1$.
The left (right) axes are
the depth below (time elapsed after) the shock surface $\Delta z$ ($\Delta t$).
\textit{(a)}: %
Temperature $T$ (red line, bottom axis) and pressure $P$ (blue line, top linear axis).
The preshock $T$ (not shown) is $T_0\approx1190$~K.
The $P$ profile changes inversely to $T$ because the density increases faster than $T$ drops.
\textit{(b)}: %
Cooling rates of H collisional excitation and ionization (red line), dissociation of H$_2$ (black), and OH (brown) and $\mathrm{H_2O}$ (blue) rotational line emission.
Throughout the simulation, molecules hardly affect the cooling because they are minor relative to neutral hydrogen.
\textit{(c)}: %
Number density relative to $n_0$ (see Section~\ref{sec:parameters})
for $\mathrm{H}_2$ (purple), $\mathrm{H}^{n=1}$ (black), $\mathrm{H}^{n=2}$ (red), $\mathrm{H}^{n=3}$ (blue), $\mathrm{H}^{+}$ (orange), and $\mathrm{e}^-$ (green), with $n$ the principal quantum number.
Thin black lines show $n\geqslant4$ states. %
At intermediate depths, the H ionization fraction approaches unity.
\textit{(d)}: %
Upward energy flux of \Lya\ (black line), \Baa\ (\Ha; red), \Paa\ (blue), \Pab\ (green), \Brg\ (brown), %
recombination continua (thin dotted; Lyman (black), Balmer (red), Paschen (blue), Blackett (orange), Pfund (green), Humphreys (pink), and others (purple)), and other lines (thin black)).
At intermediate depths,
the line fluxes hardly change due to the optical thinness of the gas. %
The deep region %
in which the fluxes originate %
is well resolved by steps of 10\,\%\ in $T$,
but %
the details are not seen easily %
due to the logarithmic scale.
}
\label{fig:postshock}
\end{figure*}

The postshock structure and hydrogen line emission were detailed by \citet{Aoyama+2018}. 
Although the results shown here are basically the same as theirs, we review their findings in this subsection for the reader's convenience.
Also, the input parameters are chosen to be appropriate for the detected planet \PDSb. The $v_0$ and $n_0$ are higher than in \citet{Aoyama+2018}, where we focused on the CPD surface shock rather than the planetary surface.

In Figure~\ref{fig:postshock} we show one example of the postshock structure for $v_0=100~\kms$ and $n_0=10^{11}$~cm$^{-3}$.
This corresponds to, for example, $\Mdot=10^{-8}~\MdotUJ$, $\MP=5~\MJ$, and $\RP=1.7~\RJ$ with $\ffill=1$, assuming the radius fit to the warm population (see Figure~\ref{fig:n0v0} and Section~\ref{sec:Rfit}).  %
Following Equation~(\ref{eq:T0 historical...}), the preshock temperature is %
$T_0\approx 1200$~K, 
and with $\rho_0=2.3\times10^{-13}~$g\,cm$^{-3}$ this implies $\mu=2.3$ and $\Mach\approx45$.  %
The hydrodynamic shock heats the gas to $T_1\approx 4\times 10^5$~K (panel~(a)).
The gas density (not shown) increases as the temperature drops. 
Although the gas cools, the gas pressure increases slightly, by only~30\,\% at the end of this simulation, because of the density enhancement of compression. Notice that the pressure gets low once around the depth of $2\times10^{3}$\,cm ($10^{-3}$\,s) because H$_2$ dissociation results in expansion. %

Immediately %
after the shock, the dissociation of H$_2$ is the main process responsible for the cooling of the gas. However, it can bring down the temperature only by a small amount before the collisional excitation of the atomic hydrogen and its ionization take over some $\Delta t\sim 10^{-3}$~s after the shock (Figure~\ref{fig:postshock}b). The excitation and dissociation dominate until the end of the simulation when $T$ reaches $10
^4$~K.
Throughout the simulation, molecules hardly affect the cooling because they are minor relative to neutral hydrogen.

Figure~\ref{fig:postshock}c shows the abundance of each form of hydrogen relative to all hydrogen protons. At the preshock temperature $T_0\approx 1200$~K %
and thus at the beginning of the evolution in the postshock region, the molecular form H$_2$ (purple curve) is the most abundant, orders of magnitude more so than the ground state of atomic hydrogen H$^{n=1}$ (black) or ionized hydrogen H$^+$ (dark yellow).
This is why immediately after the shock hydrogen dissociation is the most important process. Near $\Delta t\approx10^{-3}$~s, neutral hydrogen dominates, but very quickly, by $\Delta t \approx 2\times10^{-3}$~s ($\Delta z\approx 5000$~cm), the ionization fraction has nearly reached unity. Excitation from the ground state to the first excited state $n=2$ nevertheless proceeds simultaneously.
The main processes increasing and decreasing the $n=3$ population is collisional excitation from $n\leqslant2$ to $n=3$ and radiative de-excitation from $n=3$ to $n<3$.
The $n=2$ level is special because \Lya\ ($n=2\rightarrow1$) is optically thick at this location in this example, which prevents radiative de-excitation. If the gas were even denser, also \Ha\ would become optically thick, which would prevent radiative de-excitation and lead to a higher $n=3$ population.
At depths $\Delta z\gtrsim1\times10^7$~cm, the dropping temperature and the longer timescale let the hydrogen recombine and the electrons fall back down to lower levels.

As Figure~\ref{fig:postshock}d reveals, the (potentially observable) hydrogen lines originate mainly in the deepest part, at $\Delta z\approx 2 \times 10^7$~cm (or $\Delta t\approx 30$~s), from the de-excitation of the electrons.
This region seems narrow on a logarithmic scale but we recall that the adaptive step size in time or space is set by a temperature criterion \srevise{that ensures} sufficient resolution \citep{Aoyama+2018}.

Above this region, i.e., for most of the postshock flow as visible in Figure~\ref{fig:postshock}, the line fluxes remain approximately constant, with only small modulations that can be related to the cooling processes shown in panel~(b).
\srevise{However, the shallower region plays an important role in the ``wing'' of the spectral profile because of its hot temperature. On the other hand, while the emission from the deeper region %
carries most of the energy, its Doppler width is only a few tens of~$\kms$ corresponding the temperature of a few $10^4$~K, with hardly a dependence on $v_0$.}

In general, to zeroth order, when the gas is optically thin and all the energy is in hydrogen lines, a half goes outward as \Lya\ and the other half goes inward also as \Lya. In this example, at the shock surface, the upward-travelling Lyman-$\alpha$ flux represents around 76\,\%\ of the incoming (mostly kinetic) energy, and \Ha\ carries only around 1\,\%. The other part of the energy influx travels downward, towards the photosphere (see also Section~\ref{sec:Tefffit} for more precise fractions).

In our models, we currently do not include cooling from He or metal lines but this ultimately does not matter.
In most regions, hydrogen lines are almost the only coolant, so that
when the abundance of neutral hydrogen becomes low enough, cooling by hydrogen becomes inefficient.
In Figure~\ref{fig:postshock}, this is between $\Delta z \approx 8\times10^3$ and $5\times10^7$~cm.
This leads to a plateau in the temperature,
which ends where the hydrogen recombines.
In that temperature region (at $T\sim10^5$~K), cooling by lines of ionic C, O, and He (specifically, the He~\textsc{i}~$\lambda16404$ line) or other metals lines would be more important \citep[see Figure~3 in][]{Gnat+Ferland2012} 
so that there would not be a temperature plateau.
Indeed, while the ionisation of C and O is included in the chemistry subroutine,
it is not included in the radiation transfer,
and neither is the cooling by lines of C and O in the energy equation.
For helium, the ionisation is included in the energy equation but, also here, the lines are not.
Also, note that for the case presented in Figure~2 of \citet{Aoyama+2018}, with $(v_0=40~\kms,n_0=10^{11}~\mathrm{cm}^{-3})$, in the early parts of the flow
the electron abundance is higher than the H$^+$ abundance.
These electrons are coming from ionized helium.

However, the gas in the region of the temperature plateau only contributes to the recombination continuum but not to the hydrogen lines.
Therefore, even including helium or metal lines (and thus changing the temperature structure of that region) would not modify the strength of the hydrogen lines.

\subsubsection{Radiative properties}

\begin{figure}
\epsscale{1.15}
\plotone{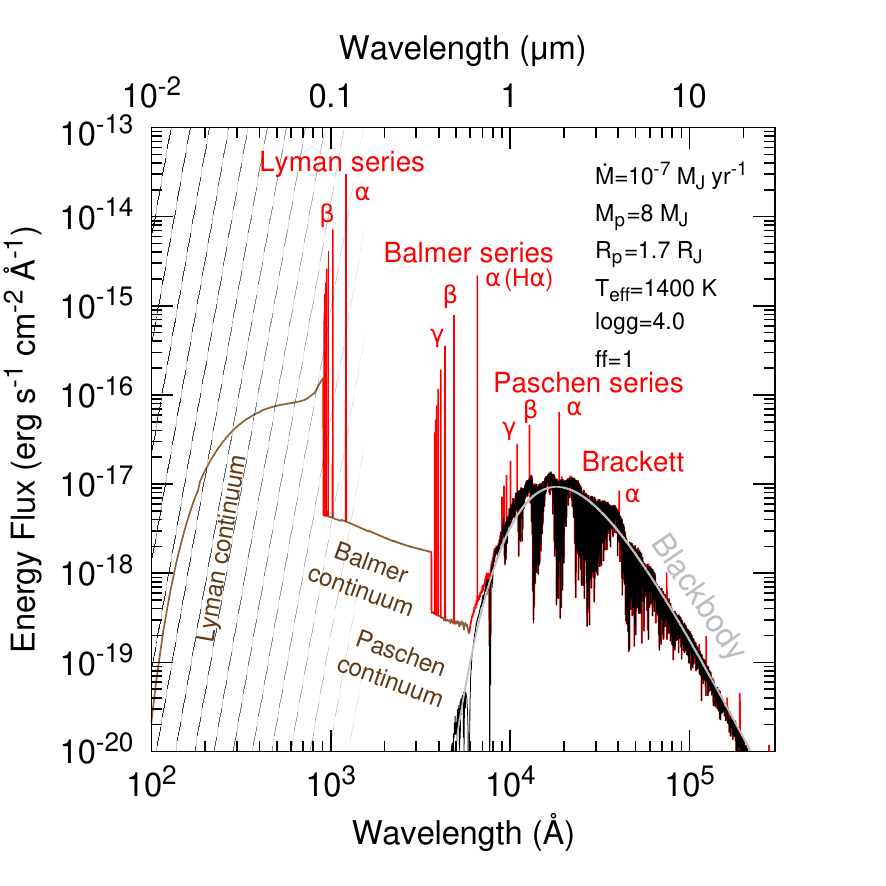}
\caption{%
Spectral energy distribution of an accreting gas giant,
with effective temperature $\Teff=1\threvise{4}00$~K and surface gravity $\log g=4.0$~(cm\,s$^{-2}$).
This corresponds for example to a $\MP=8~\MJ$ planet accreting at $\Mdot=10^{-7}~\MdotUJ$ with $\ffill=1$ %
in the hot-start population.
The flux is for a distance of 150~pc to the system.
The black line shows the pure photospheric radiation (BT-Settl; \citealp{Allard+2012})
and the red shows the photospheric radiation with the shock excess. %
The pale gray line is a blackbody at $\Teff=1\srevise{4}00$~K.
The continua are plotted darker to indicate that they are less certain. Only the Lyman, Balmer, and Paschen continua are brighter than the photosphere.
\srevise{Extinction via material between the planet surface and the observer %
is not %
taken into account. %
UV photons will be strongly affected, as the gray hatch indicates.
}
}
\label{fig:SED}
\end{figure}

Figure~\ref{fig:SED} shows the entire corresponding SED, including the contribution from the photosphere.
\srevise{The $\Teff$ is determined by the accretion heating with a constant intrinsic temperature of 1000~K (see Section~\ref{sec:Fittin}), \threvise{leading to $\Teff=1400$~K}. %
\threvise{Wherever the}
photospheric emission is brighter than the blackbody (gray \threvise{curve}),
\threvise{for instance} around $\lambda\approx 10^4$~\AA, \threvise{the photons are coming from layers higher than the photosphere. In those regions, the actual temperature structure could be different from the one in BT-Settl due to the shock heating, but this is difficult to estimate without detailed modeling.}}
For optical or longer wavelengths ($\lambda\gtrsim 4000$~\AA), the radiation is dominated by the photospheric contribution (black line) except for some hydrogen lines (red peaks).

On the other hand, at shorter wavelength, the dominant component is \Lya, stronger by tens of orders of magnitude than the thermal photospheric emission. The other Lyman and Balmer line series also clearly exceed the photospheric contribution. Notice also the clearly visible Lyman and Balmer recombination continua. These are the smooth parts of the SED between roughly 500 and 900~\AA\ (with 921~\AA\ the edge of our Lyman continuum at $n>10$ instead of 912~\AA\ for $n=\infty$) and 1000~and 4000~\AA, respectively. The continua for the other series are hidden by the photospheric contribution and are thus negligible. 
In any case, as we discuss in Section~\ref{sec:metal}, the strength and shape of the continua are approximate---within a factor of a few, but also possibly exceeding this---and are only meant to provide some guidance.
Also, note that the continua (hydrogen recombination and H$^{-}$) are thought to mainly come from the deeper and cooler region, \threvise{the} heated photosphere after (or \threvise{below}) the end of our calculation.

\subsection{Grid of models}
 \label{sec:grid}

We now present results from a large grid in accretion rate and mass,
\begin{align}
 \label{eq:parspac}
 10^{-8}~\MdotUJ \lesssim &~\Mdot \lesssim 10^{-4}~\MdotUJ\\
 0.5~\MJ \lesssim &~\MP \lesssim 20~\MJ.
\end{align}
This wide parameter space is chosen to cover present and future observations, which might reveal a population of closer-in accreting planets \citep{Close2020}.
The $\RP(\Mdot,\MP)$ and $\Teff(\Mdot,\MP)$ are given by the relations of Section~\ref{sec:Fittin},
where $\Teff$ is chosen
so that the total flux in the SED is equal to the sum of the internal and the incoming kinetic energy flux (Equation~(\ref{eq:fluxconserv})).
We take as the standard case the radius fit from the ``accretion hot-start'' population and set $\ffill=1$ for simplicity.
The upper accretion rate of $\Mdot\approx10^{-4}~\MdotUJ$ represents the highest values in the runaway phase (Phase~II) of classical in formation calculations,
with a dependence on the viscosity and the scale height of the PPD, as reviewed in \citet{helled14ppvi}. Included within this bound are thus the common maximum values near $10^{-2}~\MdotU\approx10^{-4.5}~\MdotUJ$ (\citealp{Bodenheimer+2000,Marley+2007,lissauer09,morda12_II,Tanigawa+Tanaka2016}; see also \citealt{schulik20}) so that our upper value represents a conservative choice. It also equals the typical accretion rate through the PPD $\Mdotdisc\sim10^{-8}~\MdotUS$ \citep{Hartmann+2016}; the planet would be intercepting the full typical \Mdotdisc or a smaller fraction of a higher global value.

The smallest $\Mdot=10^{-8}~\MdotUJ$ will turn out to be roughly the lower limit needed to explain the \PDSb and~c observations of \citet{Haffert+2019}.
Also, towards low $\Mdot$ we expect
the photospheric noise to dominate over the accretion lines, in addition to the total line luminosity becoming small, making it a less interesting case to study.

The lower mass of $\MP=0.5~\MJ$ corresponds to a free-fall velocity near %
$30~\kms$ (see Figure~\ref{fig:n0v0}), which is the lower limit on $v_0$
for hydrogen line emission when the preshock gas is in molecular form \citep{Aoyama+2018}.
As discussed in Section~\ref{sec:parameters}, it is not certain at these masses $\lesssim\MJ$ to what extent the accreting gas is indeed in free-fall, and the preshock velocity could be smaller. Thus this is an optimistic choice, especially since extinction by the upper layers of the PPD could be important for small masses, which may not carve out a deep gap.

Finally, we take as an upper limit $\MP=20~\MJ$ for a few reasons.
One is to focus on objects that are predominantly formed as planets, during the formation of which an accretion shock should occur, whereas this is less clear for brown dwarfs (see discussion in Section~4.2 of \citealt{bbmm16}). The planet and brown-dwarf mass functions overlap\footnote{Core accretion can form objects up to several tens of $\MJ$, with a low frequency \citep{morda12_II,Emsenhuber+2020b}.} and have a minimum near $\MP=25~\MJ$ \citep{reggiani16} before increasing towards low masses \citep{nielsen19}. Thus most objects with $\MP\lesssim20~\MJ$ are likely to have formed by core accretion \citep{schlaufman18,wagner19} and thus to have experienced an accretion shock, making them more observationally relevant. Secondly, we will see that towards higher masses, the line fluxes are relatively insensitive to the mass; thus stopping at 20 or 30~$\MJ$ makes little difference.

In any case, we emphasise that the range of $\Mdot$ and $\MP$ in Equations~(\ref{eq:parspac}) is not a prediction but rather meant as a conservative \textit{choice} for the input parameters, that is to say, a generous range of possibly relevant values. We are not making statistical predictions for the \Ha luminosities as in \citet{mordasini17}.

\begin{figure*}
\begin{center}
 \epsscale{1.15}
 \plotone{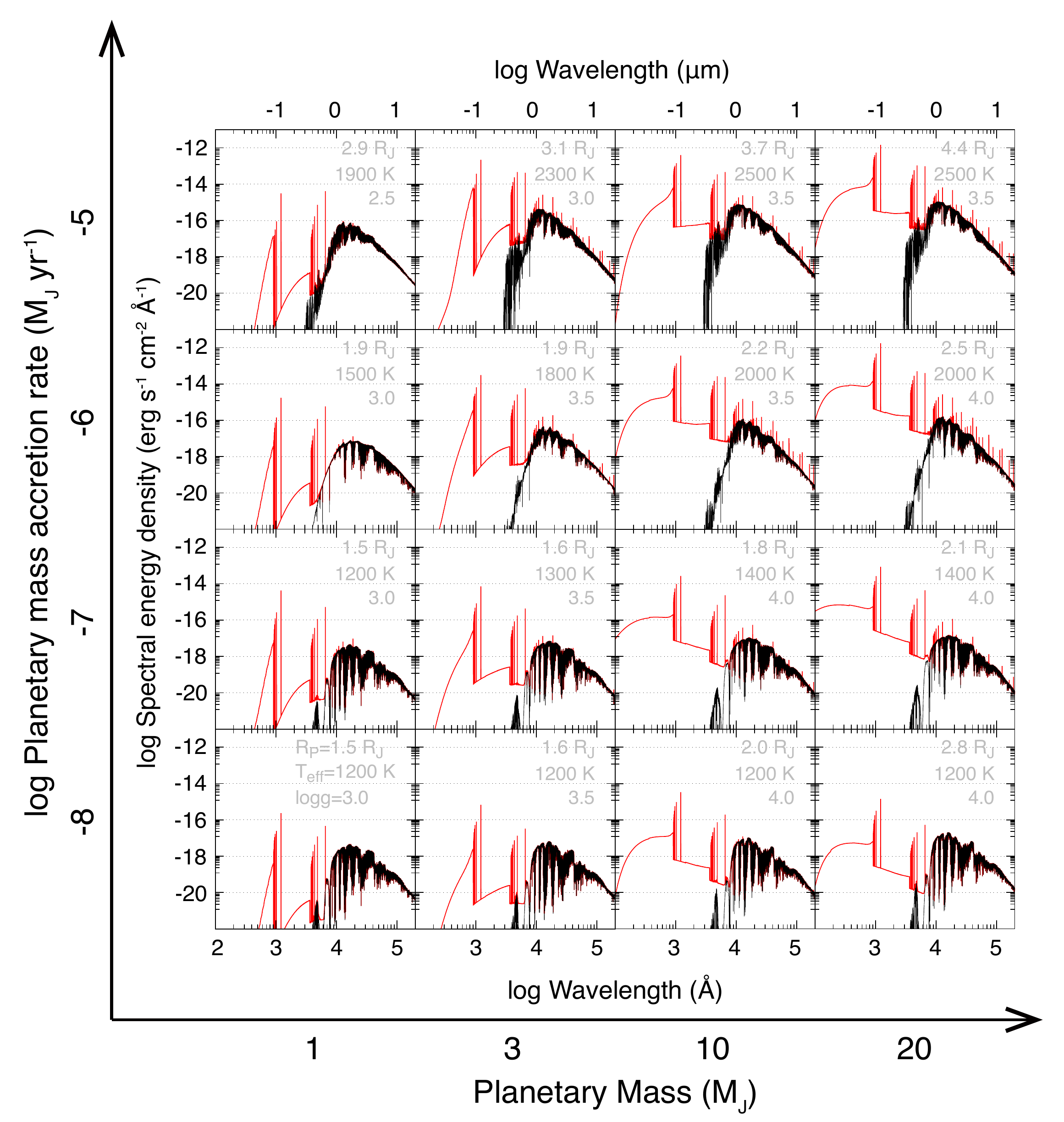}
\end{center}
\caption{
SEDs of accreting planets for a grid of accretion rates and masses: %
photospheric emission only (black lines)
or with %
the shock-heated gas emission (red).  %
We took $\ffill=1$, a distance of 150~pc, and the hot-start fits (Section~\ref{sec:Rfit}) for the planetary radius $\RP$ and effective temperature $\Teff$ (see subpanels), %
rounded to the nearest 100~K or up to $\Teff=1200$~K.
No contribution \srevise{from a CPD} \citep{zhu15} is shown.
\threvise{The ISM will absorb the Lyman lines and continuum but locally they will affect the disc chemistry \citep{Cleeves+2015,Rab+2019}.}
} 
\label{fig:multiSED}
\end{figure*}

Figure~\ref{fig:multiSED} shows a grid of SEDs for $\MP/\MJ=1$, $3$, $10$, $20$ and $\Mdot/(\MJ\,\mathrm{yr}^{-1})=10^{-8}$, $10^{-7}$, $10^{-6}$, $10^{-5}$.
The peak intensities of the \Ha\ and other hydrogen lines are significant relative to photospheric emission and increase with $v_0$.
In all panels except $(\Mdot=10^{-5}~\MdotUJ,\MP=1~\MJ)$, the \Lya\ line at $\lambda=1215$~\AA\ has the highest peak value.
The effective temperature $\Teff$ is almost a monotonic function of both $\Mdot$ and $\MP$ (cf.\ Section~\ref{sec:Tefffit}).
With increasing planet mass and decreasing accretion rate, the Lyman continuum blueward of $\lambda\approx912$~\AA\ becomes stronger relative to the hydrogen lines. As discussed in Section~\ref{sec:Postshock structure}, this means that a large fraction of the hydrogen is ionized and that, in reality, a large fraction of the energy should be converted into He and metal lines instead.
However, again, this does not affect the strength of the hydrogen lines.

In all cases shown, Lyman and Balmer lines have significant peaks above the photospheric emission, because the peak of photospheric emission is at a longer wavelength than these lines.

Figure~\ref{fig:multiSED} also shows that the ratio of the \Lya\ to the \Ha\ peaks increases with planet mass,
and that for $\MP\approx1~\MJ$, the ratio also increases with decreasing $\Mdot$.
This is because at high postshock temperatures $T_1$ (high $\MP$; see Equation~(\ref{eq:T1})),
hydrogen excitation occurs. This increases the abundance of the absorber of \Ha,
namely electrons in the $n=2$ state,
while depopulating the absorber of \Lya, electrons in $n=1$. This leads to a lower \Ha/\Lya\ ratio.
Towards high postshock densities (high $\Mdot$; see Equation~(\ref{eq:n0 b})),
both \Ha\ and \Lya\ are more strongly absorbed.
However, this absorption occurs in the upper regions (small $\Delta z$),
where the temperature is high but the excitation degree is low.
Normally, hotter gas emits and cooler gas absorbs, but since the hot gas has a low excitation degree, the hot gas can absorb. This is a non-equilibrium (NLTE) effect not captured by a time-independent approach.  %
Therefore, since the lower levels of hydrogen are more populated, \Lya\ absorption is stronger than \Ha\ absorption. %
This leads to the increase of \Ha/\Lya\ with $\Mdot$.

In Figure~\ref{fig:multiSED}, we also see that longer-wavelength series (e.g., Paschen or Brackett) are embedded in the photospheric signal but tend to emerge towards larger masses and accretion rates. 
This suggests that high-resolution spectroscopy of strongly accreting or massive planets might be able to detect lines from these other series (see also Sections~\ref{sec:L_others} and~\ref{sec:multiline}), unless infra-red excess from dust particles in CPD is significant enough.

Particularly with the hot-start radii, the Mach number is not monotonically proportional to $\Mdot$ because of the non-monotonic dependence of $\Mdot$ on the radius. Although increasing the mass flux of accreting gas $\Mdot$ increases the amount of shock-heated (and thus emitting) gas, it is associated with a larger planet radius at the same time. In turn, the larger planet radius leads to a slower free-fall velocity at the planet surface $v_0$.
Figure~\ref{fig:multiSED} reflects this, given that, as we verified,
the \Ha\ continuum is an at least roughly monotonic function of $\Teff$ at fixed planet mass.

\begin{figure*}
\begin{center}
  \includegraphics[width=0.7\textwidth]{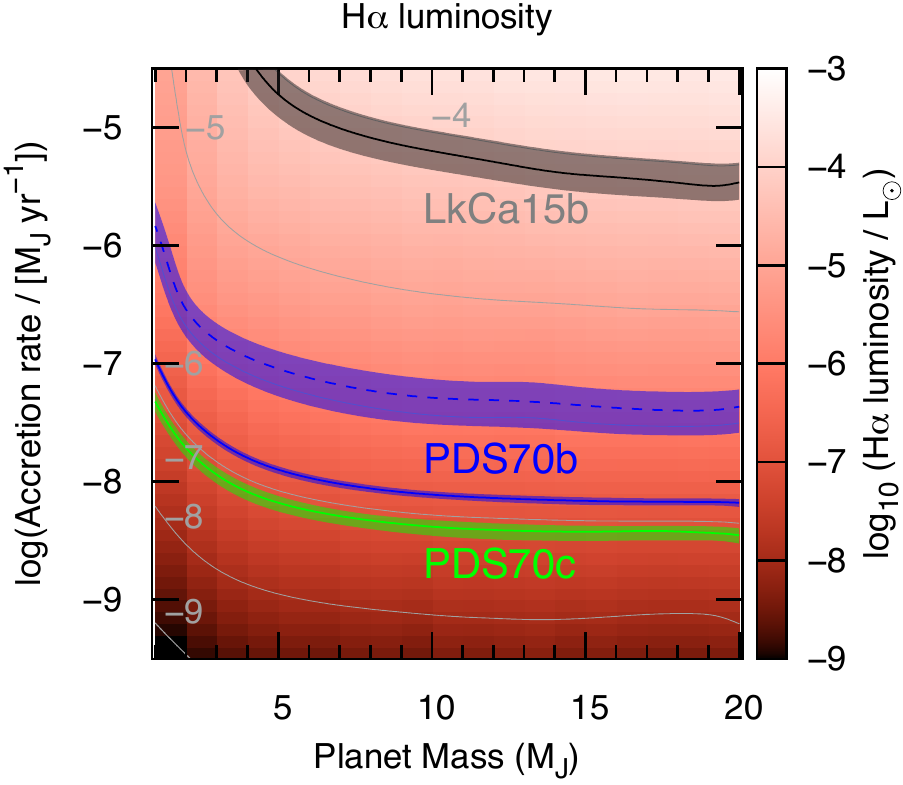}%
\end{center}
\caption{Non-extincted \Ha\ luminosity from the planet-surface shock as a function of accretion rate and planet mass (colorscale).
Thin gray lines highlight $\log \LHa/\LSun=-9$ to $-4$ in steps of 1~dex.
Shaded bands show %
non-dereddened
1-$\sigma$ contour regions for  %
\PDSb (dashed blue line: $10^{-5.9}~\LSun$; \citealp{Wagner+2018}; solid: $10^{-6.8}~\LSun$; \citealp{Haffert+2019})
and \PDSc (green: $10^{-7.1}~\LSun$; \citealp{Haffert+2019}).
The contour for the \citet{Hashimoto+2020} value of $\LHa=10^{-6.5}~\LSun$
(not shown) would lie between the two blue bands. %
The gray region is  %
for the less secure protoplanet candidate \LkCab (black: $10^{-4.1}~\LSun$; \citealp{Sallum+2015}, but see text).
}
\label{fig:HaContour}
\end{figure*}

As one of the most important results of this work,
Figure~\ref{fig:HaContour} shows the \Ha\ line luminosity as a function of $\Mdot$ and $\MP$.
We extrapolated the model results down to $\Mdot\sim10^{-9.5}~\MdotUJ$.
Note that Figure~3 of \citet{Aoyama+Ikoma2019} is similar but was made for a fixed $\RP$ = 2~$\RJ$, %
independently of $\Mdot$ and $\MP$.
The \Ha\ luminosity \LHa ranges from %
$\sim10^{-9}~\LSun$ to %
$\sim10^{-3}~\LSun$ over the grid,
overall increasing monotonically with both $\Mdot$ and $\MP$.
The contours show that for $\MP\gtrsim3$--$5~\MJ$,
the \Ha\ luminosity is independent of $\MP$ and is roughly linearly proportional to $\Mdot$.
The first part of the reason for this is that the \Ha\ luminosity turns out to be roughly linearly proportional to the incoming kinetic energy flux $\Lacc=G\MP\Mdot/\RP$ \citep{Aoyama+2018}, especially at a fixed mass (see Figure~\ref{fig:fdown} in Section~\ref{sec:Tefffit}).
The second part is a simple one: the mass coordinate is on a linear scale, with only a limited range ($\approx1.3$~dex) relevant to planetary detections, while the accretion rate axis is logarithmic and chosen to cover five orders of magnitude.

Our $\LHa(\Mdot,\MP)$ relation (Figure~\ref{fig:HaContour}) is robust to changes in the model choices. We compare in Appendix~\ref{sec:HaContourboth} the luminosity as obtained with the hot- and the cold-start $\RP$ and $\Teff$ relationships and find very little difference. Similarly, varying $\ffill$ from $\ffill=1$ to $\ffill=0.01$ (not shown) changes the \Ha\ fluxes by at most
a factor of two\footnote{
  At extremely low $\ffill\lesssim10^{-4}$, self-absorption becomes very important. For more moderate values $\ffill\gtrsim10^{-3}$ as inferred for young accreting stars \citep{Ingleby+2013}, self-absorption is not a significant effect.}.
This is because the incoming gas mass at the shock ($\Mdot$) is independent of $\ffill$, and \Ha\ emission is almost proportional to the mechanical energy of incoming gas \citep{Aoyama+2018}, as mentioned above.
See also the discussion in Section~\ref{sec:geoeffect}.

Figure~\ref{fig:HaContour} is meant as a tool for interpreting \Ha\ detections in terms of fundamental planet parameters.
Therefore, we also compare the luminosities with those of a few low-mass objects (labeled contours),
which we discuss in the next section.
Section~\ref{sec:L_others} presents
the \Brg, \Paa, \Pab, and \Hb luminosities in a similar fashion.

\section{Application to observations}
 \label{sec:obs}

In this section, we apply our results to a few (candidate) protoplanets\footnote{A recent addition to this list may be \object{2M0249~c} \citep{Chinchilla+2020}, with the caveat that chromospheric activity could be contributing to the \Ha.}.
Implications of the non-detections in dedicated surveys \citep{Cugno+2019,Zurlo+2020,Xie+2020} are discussed in Sections~\ref{sec:nondetect}.

\subsection[PDS 70 b and PDS 70 c]{\PDSb\ and \PDSc}
 \label{sec:PDS70}

\subsubsection{Brief partial review of observations}

\citet{Wagner+2018} reported the detection using MagAO of an \Ha\ signal from \PDSb,
a companion in the gap in the transitional disc around a young ($5.4\pm1.0$~Myr; \citealt{mueller18}) pre-main sequence 0.9--1.0-$\MSun$ star \citep{keppler19,wang21vlti} discovered by \citet{keppler18}.
Then, the \Ha detection at \PDSb was confirmed by \citet{Haffert+2019} using VLT/MUSE \citep{Bacon+2010}.
They also reported the discovery in \Ha of \object{\PDSc}, a companion at the edge of the gap.
From new VLT/SINFONI K-band data from \citet{Christiaens+2019a}, \citet{Christiaens+2019b} inferred the presence of a circumplanetary disc around \PDSb,
the first observational evidence for a disc around a planet in a circumstellar disc.
Using the near-infrared (NIR) SED and the models of \citet{Eisner2015},
they derived an accretion rate $\Mdot\sim10^{-7.5}~\MdotUJ$. 
Also, \citet{Wang+2020} observed this system with Keck/NIRC2 and estimated the mass accretion rate to be $\Mdot=(3$--$8) \times 10^{-7}~\MdotUJ$ by comparing to the luminosity-evolution model of \citet{ginzburg19b}. More recently,
\citet{Stolker+20b} added the first detection of \PDSb at 4--5~$\upmu$m and re-analyzed the other data from 1~to 4~$\upmu$m, confirming the finding by \citet{Wang+2020} that a blackbody fits well the SED. Given their modeling results, they concluded that \PDSb is likely surrounded by some dusty material, which nevertheless %
lets (some) \Ha pass through. Finally, thanks to $R\approx500$ $K$-band spectra and astrometry of \PDSb and~c from VLTI/GRAVITY, \citet{wang21vlti} found statistical support for a more complex (non-blackbody) SED and for a small mass for \PDSb ($\MP<10~\MJ$, but likely even lighter).

The \Ha signal of \PDSb has been detected with two different instruments, with different luminosity determinations.
From \citet{Wagner+2018}, the luminosity can be derived as $\LHa = (1.4 \pm0.6) \times 10^{-6}~\LSun$ following their data and description (see Appendix~\ref{sec:Wagner} for details); this agrees with $\LHa=(1.3\pm0.7)\times10^{-6}~\LSun$ derived by \citet{Thanathibodee+2019}.
As for \citet{Haffert+2019}, they obtain
$\LHa=(1.6\pm0.1)\times10^{-7}~\LSun$ for \PDSb\ and $(7.6\pm1.3)\times 10^{-8}~\LSun$ for \PDSc.
Thus the luminosity for \PDSb\ derived under the assumption of no extinction within the system, and ignoring the ISM contribution of $\AR\approx0.02$--0.12~mag (see Appendix~\ref{sec:Wagner}),
is about lower by one order of magnitude than in \citet{Wagner+2018}.

\citet{Hashimoto+2020} improved the data-correction method of the VLT/MUSE data and estimated higher values $\LHa=(3.3 \pm0.1) \times 10^{-7}~\LSun$ and $(1.3\pm0.1)\times 10^{-7}~\LSun$ for \PDSb\ and c, respectively. The value for \PDSb\ is still lower than in \citet{Wagner+2018} by a factor of four.  %
This could be due to intrinsic variability in the \Ha\ emission from \PDSb\ and/or
from the known variability of the star in the R band, combined with the way the contrast is measured. However, Haffert et al.\ (in prep.)  %
report that for a dozen \srevise{MUSE} measurements over a period of three months, there is no variability in the \Ha\ flux at the $\approx30$\% level. \srevise{More recently, \citet{Zhou+2021} detected \PDSb at \Ha with the \textit{Hubble Space Telescope} (HST) and also found no evidence for variability over an almost five-month baseline.}
Thus differences in the data reduction seem to be a likely explanation for the differences.

\subsubsection{No extinction at PDS 70 b and c?}

While we assumed no extinction in this paper, \citet{Hashimoto+2020} found the \Ha from \PDSb\ and~c to be likely strongly extincted ($\AHa > 2.0$~mag for \PDSb) based on the spectral width of the \Ha line and their upper limit on the \Hb/\Ha fraction. Repeating their analysis with a more realistic opacity law, \citet{maea21} derive even stronger constraints ($\AHa\gtrsim4$--8~mag for \PDSb).

However, if the observed spectral width is overestimated due to the instrumental resolution \citep{Thanathibodee+2019}, other solutions without extinction are allowed (see Figure~3 in \citealt{Hashimoto+2020}):
towards lower $n_0$ and $v_0$, both the flux ratio \Hb/\Ha and the line widths are smaller, and there are matching combinations with $\AHa=0$ and smaller $\ffill$.
Thus, our assumption can be consistent with the observational results. This solution without (or with weak) extinction is preferred by the mass estimate of \PDSb and c \citep{bae19,Stolker+20b,wang21vlti}.
To confirm whether the \Ha from \PDSb is strongly extincted, follow-up observations with a higher spectral resolution are needed.

\subsubsection{Derived accretion rate}

Given these \Ha\ luminosities,
our model yields $\MP$--$\Mdot$ relations, %
shown in Figure~\ref{fig:HaContour} as line contours: blue dashed \citep[\PDSb;][]{Wagner+2018}, blue solid \citep[\PDSb;][]{Haffert+2019}, and green \citep[\PDSc;][]{Haffert+2019}, respectively. If $\MP=5$--$9~\MJ$ for \PDSb as \citep{Wagner+2018} estimated and $\MP=4$--$12~\MJ$ for \PDSc \citep{Haffert+2019}, our model implies
$\Mdot=(8.0\pm4.8)\times10^{-8}$         for \PDSb's \LHa from \citet{Wagner+2018}, 
$\Mdot=(1.1\pm0.3)\times10^{-8}$         for \PDSb's \LHa from \citet{Haffert+2019}, and 
$\Mdot=(6.3\pm3.1)\times10^{-9}~\MdotUJ$ for \PDSc's \LHa from \citep{Haffert+2019}, respectively.
If instead $\MP\lesssim5~\MJ$, which is preferred for \PDSb\ \citep{wang21vlti} and~c \citep{bae19}
the constraints on \Mdot and $\MP$ are more accurately given in a joint form:
$\Mdot\MP\approx5\times10^{-7}~\MMdotUJ$ for the \LHa from \citet{Wagner+2018},
so that $\Mdot\approx1.7\times10^{-7}~\MdotUJ$ for $\MP=3~\MJ$,
and, for the \LHa from \citet{Haffert+2019},
$\Mdot\MP\approx6.3\times10^{-8}~\MMdotUJ$,
implying $\Mdot\approx2.1\times10^{-8}~\MdotUJ$ at $\MP=3~\MJ$. Towards low masses, these results depend somewhat more on the choice of the radius fit (see Figure~\ref{fig:HaContourboth}), but the main source of uncertainty is the one in the observed value of \LHa.

By extrapolating the empirical \Lacc--\LHa relationship for Young Stellar Objects (YSOs) from \citet{Rigliaco+2012}, \citet{Wagner+2018} estimated $\Mdot\approx10^{-9}~\MdotUJ$ for \PDSb. Also, with the $\Mdot$--linewidth relationship of \citet{Natta+2004}, \citet{Haffert+2019} derived $\Mdot\approx2\times10^{-8}~\MdotUJ$. 
Thus, applying stellar accretion models to planetary-mass observations yields a lower mass accretion rate than from our model by a few orders of magnitude. To estimate mass accretion rates, we suggest that our model constructed for planet accretion should be used rather than the extrapolation of empirical relationship from pre-main-sequence star studies. This is discussed briefly in Section~\ref{sec:preshock} but in more details in a companion publication, \citet{AMIM20L}.

Also, \citet{Thanathibodee+2019} constructed a model of \Ha\ emission focusing on \PDSb. They modeled the \textit{accreting gas} as the source of the \Ha\ rather than the postshock region that is the subject of this paper. The accretion rate they estimate, $\Mdot\approx~10^{-8.0\pm0.6}~ \MdotUJ$, is larger than the results of empirical \LHa--\Lacc relationships and in agreement with our results within the margin of error.
As discussed in Section~\ref{sec:preshock}, whether the \Ha emission originates from the accretion flow or the postshock region depends on whether the accreting gas is hot enough to emit \Ha. In fact, for \PDSb a contribution from both cannot be excluded \citep{AMIM20L}.

Finally, the upper limits on the \Bra \citep{Stolker+20b} and \Brg \citep{Christiaens+2019a,wang21vlti} emission are discussed in Section~\ref{sec:multiline}.

In summary, given an \LHa measurement, our model yields joint constraints on \Mdot and $\MP$ at low masses, which seem more likely for \PDSb and for \PDSc.
(For higher masses, \LHa becomes relatively independent of $\MP$.) The uncertainty in \Mdot is $\approx1$~dex.
Our estimated \Mdot values for \PDSb and~c are smaller than previously determined in the stellar literature and similar to the results of \citet{Thanathibodee+2019}. This is however a coincidence, since the two models have a very different physical basis and predict in general distinct \Mdot--\LHa relationships \citep{AMIM20L}.
The main limitations on determining \Mdot are the uncertainties on the line-integrated fluxes and the line widths, as well as the uncertainties about the true masses.

\subsection[LkCa 15 b]{\LkCab}
 \label{sec:LkCa15b}

Following the discovery of a companion to \object{\LkCa} by \citet{Kraus12}, \citet{Sallum+2015} reported the infrared detection of further sources
in the system using sparse-aperture masking (SAM).
Intriguingly, they also measured an \Ha\ signal which seemed to originate at the position of \LkCab. On the other hand, \citet{Thalmann+2016} analyzed scattered light from the disk and showed that the infrared detections of the planetary candidates around \LkCa\ could be false positives related to features of the disc in scattered light. 
In addition, %
observations by \citet{mendigut18} using spectro-astrometry suggest that the \Ha\ emission may not be coming from a point source but rather from  an extended region similar in size to the orbit of the claimed planet \LkCab.
Recently,
\citet{Currie+2019} conducted the first direct-imaging observations of the LkCa~15 system. They provided evidence that there is no point source at the location of the claimed planet (nor of the possible further companions) but that in fact the SAM signal originates from disc emission.

Despite the debate as to its origin, we will briefly analyze the \Ha\ signal at the position of a putative companion to \LkCa\ as originating from an accretion shock on the planet surface.
The reported \Ha\ luminosity is %
$\LHa=10^{-4.1\pm0.1}~\LSun$  %
from \citet{Sallum+2015} but using the updated Gaia distance determination of 158.8~pc \citep{GaiaBrown2018}.
From Figure~\ref{fig:HaContour} and assuming $\MP=10\,\MJ$, $\Mdot=4.0^{+2.5}_{-0.1}\times 10^{-6}\,\MJ\,\yr^{-1}$.
This accretion rate is
not implausible for a claimed forming gas giant, especially if it were undergoing an accretion outburst.

Using instead the \citet{Rigliaco+2012} approach as in \citet{Sallum+2015} and again with $\MP=10~\MJ$ as an example, yields $\Mdot = 3\times 10^{-7}~\MdotUJ$ for $\RP=1.6~\RJ$ as \citet{Sallum+2015} assumed. At this $(\Mdot,\MP)$, our fits (Section~\ref{sec:Rfit}) yield $\RP=1.9~\RJ$ ($\RP=1.4~\RJ$) for the hot (cold) population, so that $1.6~\RJ$ is a reasonable value, albeit perhaps on the small side. The upshot of the comparison is that the $\Mdot$ implied by the \citet{Rigliaco+2012} relationship is one order of magnitude smaller than derived with our approach; for \PDSb, it was a few orders of magnitude. As discussed in \citet{AMIM20L}, we suggest that
our models, which are tailored for the planetary case, should be used instead of extrapolations from the stellar regime.

\section{Further observational aspects} 
 \label{sec:furthobsabs}
 
We now discuss to what extent high-mass and high-$\Mdot$ planets can be distinguished (Section~\ref{sec:distinguish}) and the planet surface shock from the CPD shock (Section~\ref{sec:distinguishsurfaceshockCPD}), before presenting the line strengths and line ratios for accretion-generated hydrogen lines other than \Ha\ (Section~\ref{sec:L_others}). Finally, we discuss what information may be obtained from combining observations of several lines for the same object (Section~\ref{sec:multiline}).

\subsection{Distinguishing massive planets and strongly accreting planets from the line profile?}
 \label{sec:distinguish}
\begin{figure*}[ht]  %
    \centering
    \plottwo{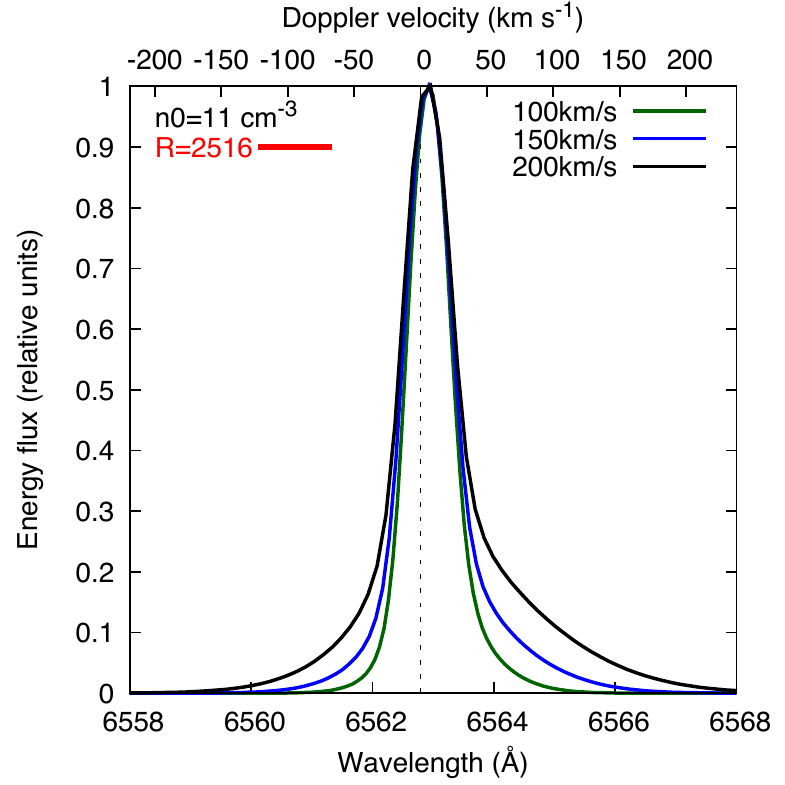}{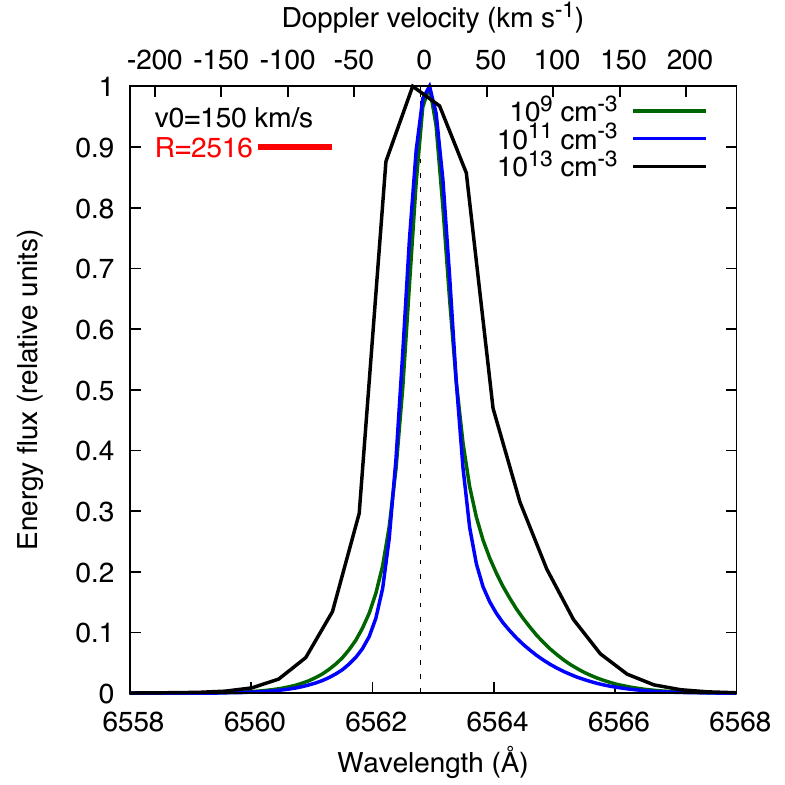}
    \caption{
    Spectral profile of \Ha\ for $v_0=100$, $150$, $200~\kms$ at $n_0=10^{11}$~cm$^{-3}$ (\textit{left panel}) and for $n_0=10^{9}$, $10^{11}$, $10^{13}~\mathrm{cm}^{-3}$ at $v_0=150~\kms$ (\textit{right panel}), typical for planetary masses (see Figure~\ref{fig:n0v0}).
    Each line is normalized by its peak value.
    The top axis shows the Doppler shift velocity, even though some features come from natural
    and not Doppler broadening. 
    The spectral resolution of MUSE at \Ha is indicated ($R=2516$; red bar).
    The \Ha line width depends on $n_0$ more than on $v_0$ because high $n_0$ leads to self-absorption near the peak.
    }
    \label{fig:SED_mass}
\end{figure*}

When characterizing gas giants from their \Ha\ luminosity, their mass and mass accretion rate are degenerate because the luminosity depends on their product. However, this degeneracy can be lifted by spectroscopic observations of \Ha, which was demonstrated by \citet{Aoyama+Ikoma2019} in the case of PDS~70~b and c.

Recall that the preshock velocity $v_0$ mainly depends on $\MP$, while the number density $n_0$ is mainly set by $\Mdot$ (see Figure~\ref{fig:n0v0 contours}).
Figure~\ref{fig:SED_mass} shows \Ha\ line shapes for several values of $v_0$ and $n_0$.
The preshock velocity $v_0$ sets the shock strength---and thus the temperature just after the shock (Equation~(\ref{eq:T1}))---but barely the line width, which is mainly set by Doppler broadening. This is because the gas that becomes ionized and then recombines is the main source of \Ha, and not the gas immediately after shock, even if there is some amount of excited hydrogen there (see down to $\Delta t\approx10^{-3}$~s in Figure~\ref{fig:postshock}).
Note that Figure~\ref{fig:SED_mass}a a small redshift (of several $\kms$) can be seen in the line profiles. This is due to the non-zero settling speed of the emitting gas.

The line profile can be divided into three parts: a narrower Gaussian, a broader Gaussian, and a Lorentzian profiles, with the latter visible further from the line center. The layers that emit the two Doppler profiles are separated by the highly ionized region (at $\Delta t\approx10^{-3}$--10~s in Figure~\ref{fig:postshock}). The $\Ha$ intensity coming from the deeper layers is much larger.
Consequently, even the width of the line where the energy density is 10\,\%\ of the maximum ($W_{10}$ in e.g.\ \citealt{Thanathibodee+2019}) reflects only the narrower Doppler component, coming from the hydrogen-ion recombination region at low temperatures, in the deep layers.
The gas temperature at which hydrogen recombination begins barely depends on $v_0$, because it corresponds to the hydrogen ionization energy of $13.6$~eV.
We can see the thermal broadening of the optically thinner gas just after the shock only far from the line center, at $\lambda\gtrsim6564$~\AA\ and $\lesssim 6562$~\AA\ (i.e., $|\Delta v|\gtrsim50~\kms$ away from the shock). Since the hot gas immediately below the shock has a high velocity and
is travelling away from the observer,
the red half of the line is more broadened by this mechanism than the blue half.
However, as seen in the right panel, the resolution of MUSE ($R=2516$ at \Ha; \citealp{Eriksson+2020}) is not sufficient to distinguish this.

As shown in the right panel of Figure~\ref{fig:SED_mass}, $n_0$ changes the width of the \textit{normalized} line dramatically. However, increasing $n_0$ hardly broadens the \Ha\ line because the pressure broadening is negligible relative to the natural broadening.
A high $n_0$ leads to \Ha\ self-absorption in the postshock gas (in the top part of the flow), which flattens the line peak. Since we normalized the line flux at the peak, the self-absorbed line looks broader (see \citealp{Aoyama+2018} for the non-normalized profile). However, the lines for higher $n_0$ are brighter than lower ones despite the absorption. This effect becomes significant for $n_0 > 10^{11}$~cm$^{-3}$ in the right panel.
Also, a lower $n_0$ ($10^{9}$~cm$^{-3}$) can lead to slightly broader profile. While the wider component that comes from the shallower region is independent of the density, the narrower component that comes from the deeper region gets weaker with decreasing $n_0$ due to a lower excitation degree. Thus, the wider component gets stronger relative to the narrower component.

As shown in Figure~\ref{fig:SED_mass}, the current spectral resolution of MUSE is not enough to distinguish the profiles clearly, while it barely resolves the spectral profile for higher density \citep{Eriksson+2020}.
However, it is not possible to determine in general what minimum spectral resolution is required for distinguishing high accretion rates from high masses because it depends on the relative uncertainty in the flux as well as on the planet properties through the dependence of the line profile on $(n_0,v_0)$.

\subsection{Distinguishing planetary-surface and CPD-surface shocks?}
 \label{sec:distinguishsurfaceshockCPD}
\begin{figure*}%
\epsscale{1.1}
\includegraphics[width=0.50\textwidth]{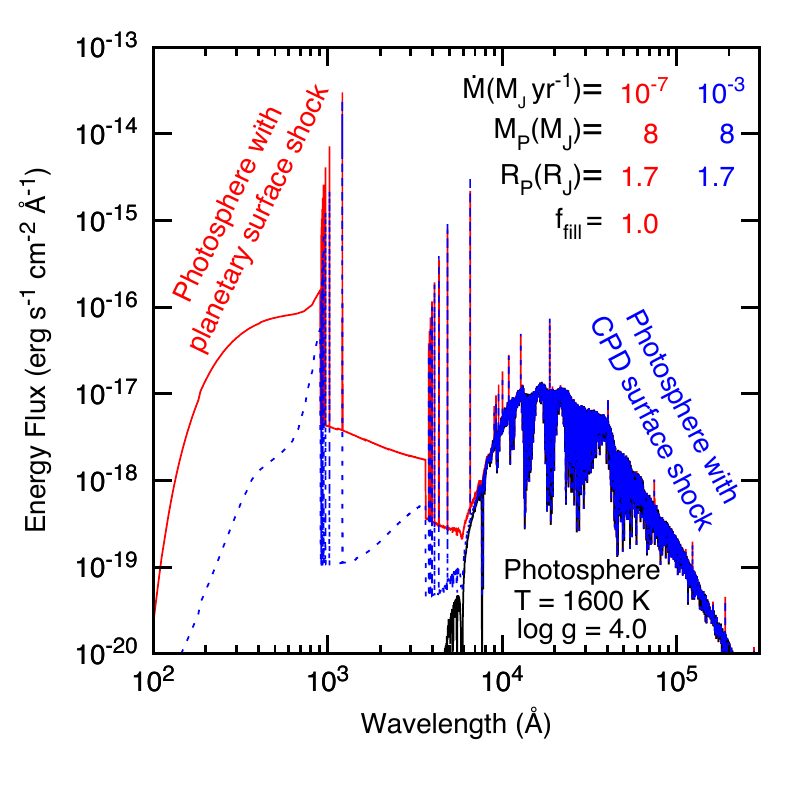}
\includegraphics[width=0.49\textwidth]{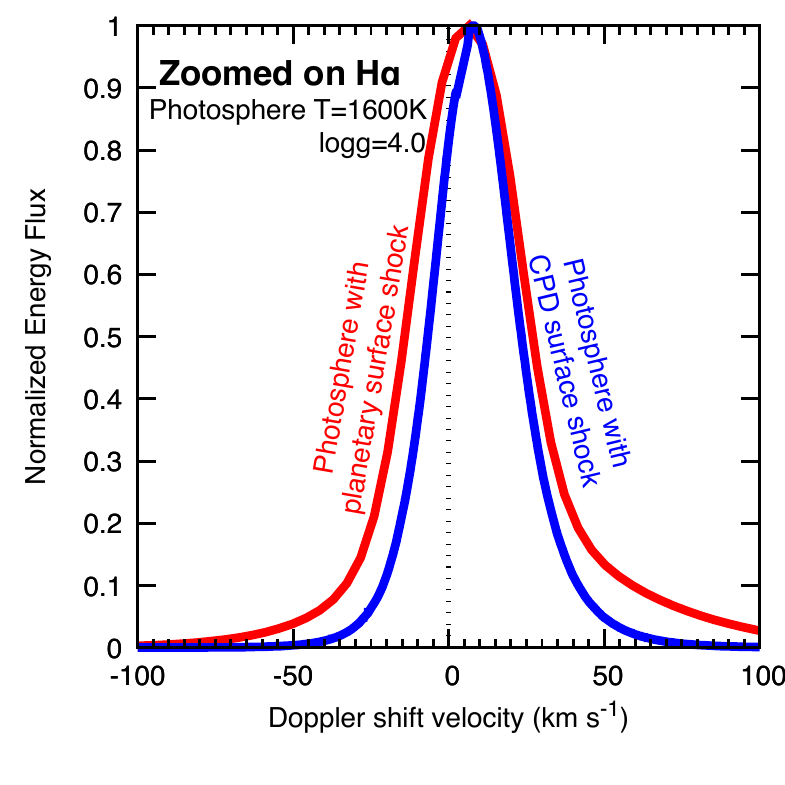}
\caption{SEDs for two different models of shock-heated gas emission: planetary photospheric emission plus planetary-surface shock (red) and CPD surface-shock (blue). The photosphere SED in the surface-shock case is also shown for reference (black).
The $\Mdot$, $\MP$, $\RP$ values (see figure and text) are chosen such that $\LHa\approx1.5\times10^{-6}~\LSun$ in both cases. The energy flux is calculated for the distance of 150~pc.
\textit{Left panel:} Global SED. \textit{Right panel}: \Ha\ line profile against Doppler shift velocity from the line center. The photospheric emission is not visible.
}
\label{fig:CPD_P}
\end{figure*}

Hydrodynamic simulations report that gas accreting toward proto-gas-giants goes through multiple shocks \citep[e.g.][]{Kley1999, Tanigawa+2012}. 
The gas that falls onto the CPD yields a shock, which can emit \Ha\ near the planet.
However, far from the planet, the shock is not strong enough to emit \Ha, and only the part of the shock close to the planet can contribute to the \Ha\ emission \citep{Aoyama+2018}. 
If the gas ultimately joining the planet passes firstly through a shock at the surface of the CPD and secondly through the planetary surface shock, the former is negligible for \Ha\ emission, because most of the gas hits the CPD at the far region. There, the free-fall velocity is too small for significant \Ha\ emission.
For example, when the CPD is truncated near the planet (and the gas is accreting by magnetospheric accretion or it is falling directly onto the planet from the PPD), the planetary surface shock dominates the emission.
However, when most gas passes through boundary-layer accretion (see, e.g., \citealp{Dong+2020} rather than a planetary surface shock, the CPD surface shock becomes significant.
Thus it would be desirable to distinguish the source of shock excess.

In Figure~\ref{fig:CPD_P}, we compare the SEDs from the planetary surface shock and the CPD surface shock. The shock excess from the latter highly depends on the gas accretion model. Here, guided by the results of an isothermal 3D hydrodynamic simulation \citep{Tanigawa+2012}, we assume the following:
\begin{enumerate}
\item All gas accretes vertically from the protoplanetary disk onto the CPD with a free-fall velocity starting from infinity set by the protoplanet's gravity, $\vff(r)=\sqrt{2G\MP/r}$, where $r$ is the radial distance from the protoplanet's center.
\item The mass accretion flux onto the CPD is spatially constant within 0.1~Hill radius, and zero outside of this. The inner disc radius does not matter
because the contribution of the outer region to the intensity dominates over that of the inner region \citep{Aoyama+2018}.
\end{enumerate}
Non-isothermal simulations might find a different flow pattern (especially concerning assumption~2.) but this should capture qualitatively the main differences between the planet-surface and CPD-surface shock cases.

As an example, we consider system parameters that could be appropriate for \PDSb.
The semi-major axis is $22.6$~au and the central-star mass is $0.82~\MSun$ \citep{Riaud+2006}.
The mass of \PDSb is very uncertain, with some indications for a low mass of a few~$\MJ$ \citep{Mesa+2019,Stolker+20b}. Here, we choose a somewhat high value within the range considered in the literature, namely
$\MP=8~\MJ$ \citep{Wagner+2018}. 
This places us in the flat part of the $\LHa(\Mdot,\MP)$ contours, making the choice of $\Mdot$ simple for
the surface-shock case, with $\Mdot=10^{-7}~\MdotUJ$. To have the same 
\Ha\ luminosity $\LHa\approx1.5\times10^{-6}~\LSun$ in the CPD-shock case \citep{Wagner+2018}, we take for the CPD-shock case $\Mdot=10^{-3}~\MdotUJ$, with the same mass.
In both cases, we set $\RP=1.7~\RJ$ and $\Teff=1600$~K (Section~\ref{sec:Fittin}).
For the surface-shock case, $\ff=1$ is also assumed.

What fraction of the accreting gas in the CPD-shock case can produce hydrogen lines? The chosen parameters yield $\RHill\approx7000~\RJ=3.3$~au. 
From Figure~\ref{fig:n0v0}, at this mass only the gas within $r\approx30~\RJ$ has $\vff\gtrsim30~\kms$ and thus contributes to \Ha. This region corresponds to
$0.005\RHill$, and thus
0.2~\%\ of the area over which the gas is assumed to accrete.
Thus only this small fraction of the accretion rate is available for producing \Ha. This partly explains the need for a high total $\Mdot$ compared to the surface-shock case to have the same \LHa. The other reason is that $v_0$ is smaller everywhere on the CPD than on the planet surface (see Figure~\ref{fig:n0v0}), so that $\Mdot$ must be higher to compensate since $\FHa$, the \Ha flux at the object's surface, roughly scales with the incoming kinetic energy flux $\Facc\propto n_0v_0^3$.

The left panel of Figure~\ref{fig:CPD_P} displays the global SEDs. The red and blue lines correspond to the SED assuming planetary surface shock and CPD surface shock, respectively. The black line corresponds to the SED without a shock excess for reference. %
In the Lyman and Balmer continua, namely for $\lambda\lesssim3000$~\AA, the two SEDs differ by more than a factor of ten. This comes from the difference in gas temperature after the two shocks.
For the CPD surface shock, the regions far from the planet dominate the shock excess emission because of their large emitting area. This is a relatively weak shock, associated with a low temperature.

This temperature difference can be also seen in the \Ha\ profile in Figure~\ref{fig:CPD_P}b: the profile is narrower in the CPD surface-shock case than in the planet surface-shock case when taking the different heights into account (i.e., looking at the full width at half-maximum (FWHM)).
The difference is small but might be detectable in the future with high-resolution observations. %
For example, for the case of Figure~\ref{fig:CPD_P}, the wavelength difference at the 10\% of the peak corresponds to $17~\kms$, which is distinguishable with the spectral resolution of $R\approx 18,000$ (which is much higher than the resolution of MUSE with $R\approx2500$). At lower flux levels relative to the peak, the difference is greater (e.g., $\approx25~\kms$ at 5\%) but it is not clear whether such levels can be robustly extracted because of the high contrast relative to the peak. The continuum emission from the CPD \citep{zhu15} or from the planet's photosphere are likely not important, but
the star's chromosphere could contribute \citep{manara13,Manara+2017,venuti19}. The stellar accretion-induced \Ha is possibly Doppler-shifted away from the planet's signal as in \citet{Haffert+2019}. However, the most important limitation is probably the maximal contrast allowed by the instrument.

The narrower \Ha\ and the weaker recombination continua in the CPD case means a weaker shock, which can also occur at the surface of a less massive planet. However, in such a case, the density should higher than the CPD case, and one can distinguish these two cases.

Our model does not include some continuum sources such as a heated photosphere \citep[e.g.][]{Koenigl1991,Calvet+Gullbring1998} or, if present, the boundary layer \citep[e.g.][]{Kenyon+Hartmann1987,Dong+2020}, which are well modeled in the stellar accretion context. Such continuum sources can change the spectral appearance, but it is unfortunately difficult to say how important this would be.

In summary, for a given \Ha\ luminosity, the resolution of MUSE is not sufficient to distinguish an accretion shock on the planetary surface from the one on the CPD, but high-resolution spectroscopy might be able to do so.

\subsection[Predictions for hydrogen lines other than H alpha]{Predictions for hydrogen lines other than \Ha}
\label{sec:L_others}

The \Ha\ line on which we have focused so far is only one of the 55~hydrogen lines we model. Recently, \citet{Eriksson+2020} reported an \Hb\ flux for the $\approx10$-$\MJ$ companion \object{Delorme~1~(AB)b} with the MUSE instrument on the VLT. 
Also, the upcoming, first-light HARMONI\footnote{See \url{https://harmoni-elt.physics.ox.ac.uk}.} integral field unit (IFU; \citealp{thatte16,rodrigues18}) on the ELT is expected
to provide $R\approx17,000$ spectroscopy between 0.8~and 2~$\mu$m (thus including for instance \Pab and \Brg);
the second-generation instrument HIRES for the ELT
\citep{Marconi+2016,Marconi+2018,tozzi18}
will cover 1--1.8~$\upmu$m and thus should observe %
Paschen lines with the high spectral resolution of $R\approx100,000$;
and the University of Tokyo Atacama Observatory (TAO) should be able to detect \Paa\ thanks to its location at 5,640~m \citep{TAO2010}.
Finally, the Keck Planet Imager and Characterizer (KPIC) \citep{Jovanovic+2019,Morris+2020} aims at obtaining $R=35,000$ spectroscopy in the $K$, $L$, and $M$ bands ($\approx2$--5~$\upmu$m).
Clearly, it is timely to extend the luminosity predictions to lines other than \Ha.

\begin{figure*}
    \centering
    \plotone{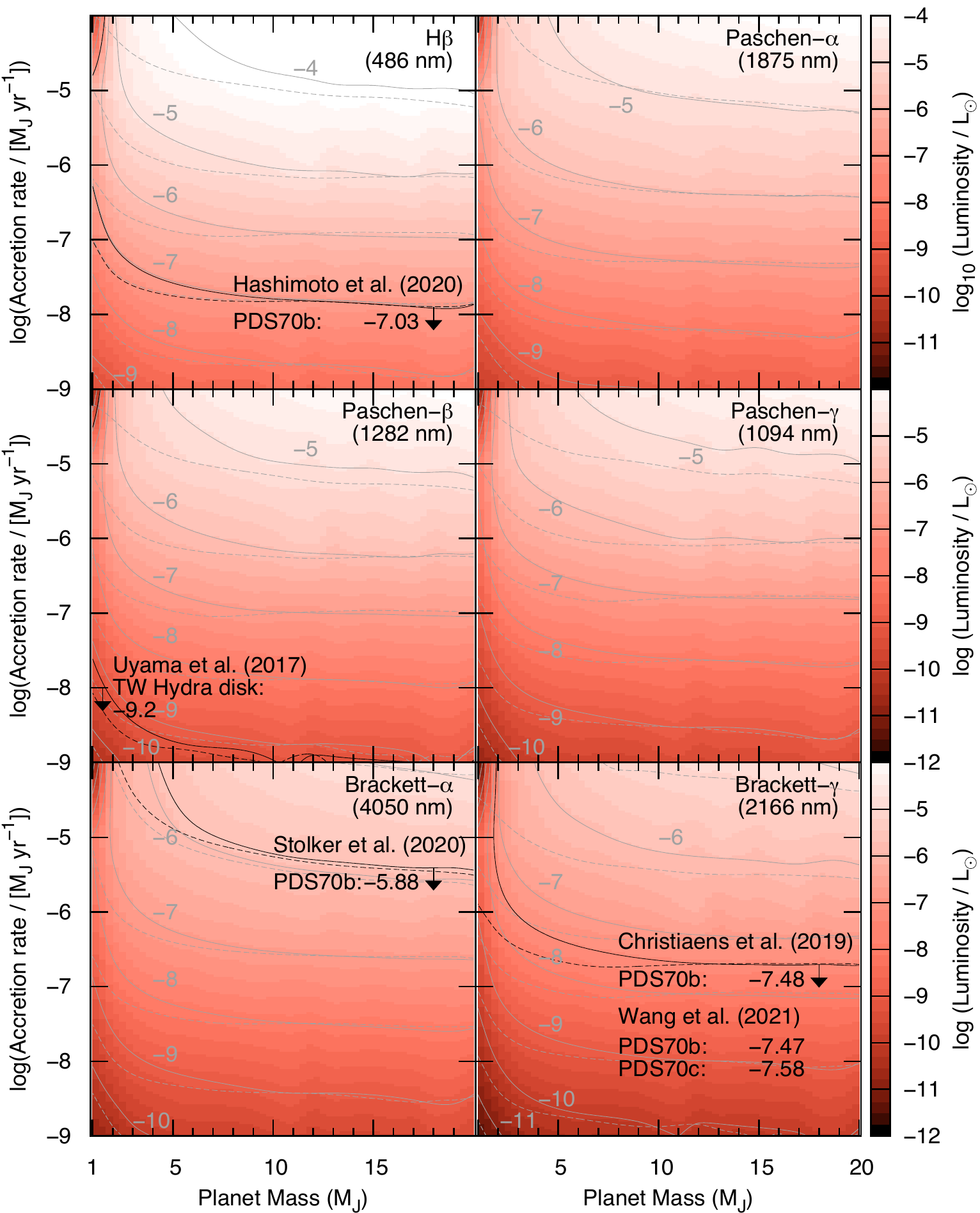}
    \caption{As in Figure~\ref{fig:HaContour} but for 
    \Hb, %
    \Paa, %
    \Pab, %
    \Pag, %
    \Bra, %
    and
    \Brg. %
    The color map and the solid contour lines show the results from the radius fit to the warm population, and the dashed contour lines use the cold population. %
    Black lines show 5-$\sigma$ upper limits of
    $\log(\LHb/\LSun) < -7.03$ \citep{Hashimoto+2020}
    and
    $\log(\LBrg/\LSun) < -7.48$ \citep{Christiaens+2019a}
    for \PDSb,
    and
    of $\log(\LPab/\LSun) < -9.20$ for a putative planet at 25~au in TW~Hya \citep{Uyama+2017}.
    \citet{wang21vlti} set \Brg upper limits for \PDSb and~c similar to the line of \citet{Christiaens+2019a}.
    In the \Bra\ panel, black contours are for the detected flux at NACO/NB4.05, $\log(\LBra/\LSun) =
    -5.88$
    \citep{Stolker+20b},
    which is an upper limit on the line flux if the continuum is much stronger, as is expected.
    }
    \label{fig:others}
\end{figure*}

In Figure~\ref{fig:others}, we show the line luminosity of \Hb, \Paa, \Pab, and \Brg\ for the same grid of $(\Mdot,\MP)$ as in Figure~\ref{fig:HaContour}. The luminosities range from $\Lline\sim10^{-12}$ to~$10^{-4}~\LSun$, increasing with $\Mdot$ and $\MP$ as for \Ha, and with the same qualitative shape of a very weak mass dependence for $\MP\gtrsim3$--$5~\MJ$.
The contours are very similar using the fit to the hot- or cold-start populations.
We show upper limits for a few objects but discuss them below in Section~\ref{sec:multiline}.

\begin{figure}
    \centering
    \epsscale{1.2}
    \plotone{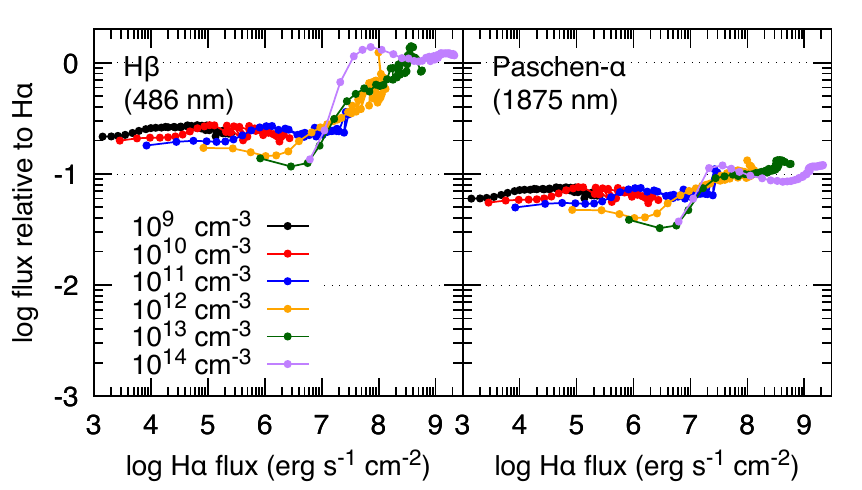}
    \plotone{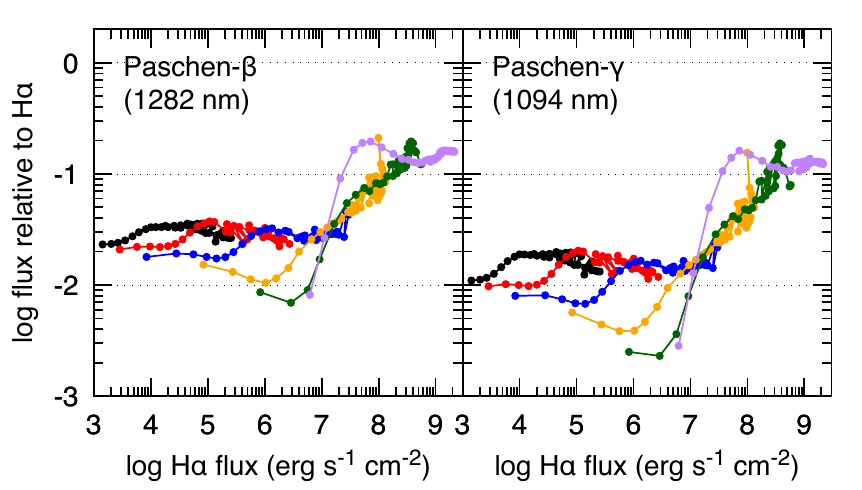}
    \plotone{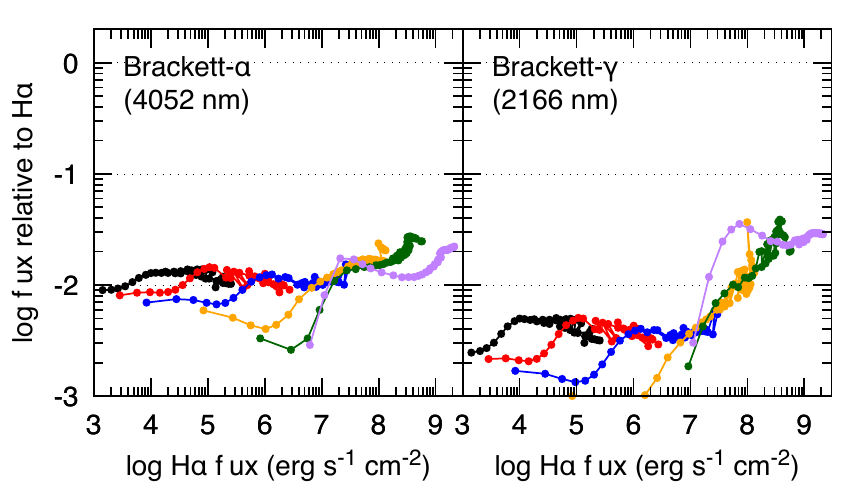}
    \caption{Flux of \Hb, \Paa, \Pab, \Pag, \Bra\ and \Brg\ relative to \Ha\ as a function of
    the \Ha\ flux at the surface of the object $\FHa$ and the preshock number density $n_0=10^9$--$10^{14}$~cm$^{-3}$ (colour). For each $n_0$, the preshock velocity $v_0$ is varied from 20~to 200~$\kms$.  %
    The \Ha\ luminosity is related to the $x$-axis by $\LHa = 4\times10^{-8} \FHa (\RP/1.5~\RJ)^2(\ff/1)~\LSun$.
    The apparent flux $\mathfrak{F}_{\mathrm{H}\,\alpha}$ of an 1.5-$\RJ$ object at 100~pc is given by
    $\log\mathfrak{F}_{\mathrm{H}\,\alpha}=\log\FHa-20.9$~dex.
}
    \label{fig:ratio2Ha}
\end{figure}

Next, Figure~\ref{fig:ratio2Ha} shows the intensity ratio of \Hb, \Paa, \Pab, and \Brg\ relative to \Ha\ as a function of the \Ha\ flux at the surface of the object. (Absolute fluxes for \Hb, \Paa, and \Pab\ as a function of $(n_0,v_0)$ can be found in \citet{Aoyama+2018}.) The ratios span \Hb/\Ha$\approx0.1$--2 to \Brg/\Ha$\approx0.001$--0.05, with typically $X/$\Ha$\approx$0.03--0.3 for $X=$\,\Hb, \Paa, \Pab.

All four ratios are more or less flat within 0.5~dex at low \Ha\ flux (within 0.7~dex for \Brg/\Ha) but start to increase
around an $\Ha$ flux $\FHa\sim 10^7\,\mathrm{erg\,s^{-1}\,cm^{-2}}$. For \Paa\ the rise is only moderate. This change of slope occurs because the \Ha\ saturates due to self-absorption in the shock-heated gas as $v_0$ increases, while the other lines do not saturate. Thus towards high $\FHa$,
$v_0$ must increase faster along the $x$ axis, which leads to stronger other lines (and thus ratios) since they  %
hardly display self-absorption.
Also, when the starting level of the transition is high (i.e., \Pab\ or \Brg: $n=5\rightarrow3$ or $n=7\rightarrow4$, respectively), some features at weaker \Ha\ ($\FHa\lesssim 10^6~\mathrm{erg\,s^{-1}\,cm^{-2}}$) become more clearly visible. %
To understand this, one should recall that for a given $n_0$, a low \Ha\ flux means a low gas temperature, and that for low temperatures, $\FHa$ increases faster than the other lines due to its lower excitation energy, and conversely at higher temperatures. Therefore, the ratio first decreases and then increases towards high temperature ($\FHa$).

Also, our model predicts that the $\Lya$ is much stronger than all lines and carries most of the incoming shock energy, which has been converted to radiation (see Figures~\ref{fig:postshock} and~\ref{fig:multiSED}). This is an important input for models of CPD chemistry \citep{Cleeves+2015,Rab+2019}. %
However, interstellar extinction is too strong for $\Lya$ to be detected, excepting a few young stellar systems such as TW~Hydra. However, other lines induced by planetary $\Lya$ might be detected as for some accreting stars, including NIR fluorescent molecular hydrogen lines \citep[e.g.,][]{Herczeg+2004}.

\srevise{%
Finally, \citet{Zhou+2021} detected for the first time
\PDSb in the Balmer continuum with the F336W ($U$-band) filter of the Wide-Field Camera 3 (WFC3) onboard HST. While our model does predict hydrogen recombination continua, it is constructed with a focus on the hydrogen \textit{lines}. The predicted continua are less reliable and likely overestimated especially for massive planets. This is because where the gas is almost ionized, the only effective coolant in our model is the recombination continua, while in reality metal lines 
should play an important role. Including them would decrease the fluxes currently predicted in the continua. To treat them more accurately, a model update is planned in the future.%
}

\subsection{Combining detections of multiple accretion lines}
 \label{sec:multiline}

We briefly discuss the few current application of these predictions in the planetary-mass regime.
Firstly, the 5-$\sigma$ \Brg upper limit for \PDSb from \citet{Christiaens+2019a},  %
$\log \LBrg/\LSun < -7.48$, is shown in Figure~\ref{fig:others}. From $R\approx500$ VLTI observations, \citet{wang21vlti} confirmed this upper limit to within 0.01~dex,
and also placed a similar 5-$\sigma$ limit of 
$\log \LBrg/\LSun < -7.58$ for \PDSc. Taken at face value, this implies for \PDSb that $\Mdot\leqslant2\times10^{-7}~\MdotUJ$ (and somewhat less for \PDSc) if $\MP\gtrsim8~\MJ$, but is much less constraining at lower $\MP$.

Also, \citet{Stolker+20b} detected \PDSb\ for the first time in the \Bra filter of NACO/VLT (NB4.05; the effective width is 0.0616~$\upmu$m and not 0.02~$\upmu$m\footnote{The too-narrow width came from \url{http://www.eso.org/sci/facilities/paranal/decommissioned/naco/inst/filters.html} (corrected in January 2021), which however provides the correct filter curve, and is often quoted in the literature \citep[e.g.][]{janson08,quanz10,meshkat14fomal,kervella14,Stolker+20a,Stolker+20b}. It however does not change the results in those studies. The correct value is from the SVO at \url{http://svo2.cab.inta-csic.es/theory/fps3/index.php?id=Paranal/NACO.NB405}.}). The \Bra line can be used as an accretion tracer \citep{Komarova+Fischer2020}, but in this case the flux is consistent with the   %
blackbody emission matching the global SED. Namely, given the observed \Ha\ flux, the \Bra\ is expected to be embedded in the continuum (see Figure~\ref{fig:ratio2Ha}), and the upper limit of the \Bra\ emission from the shock itself is the observed line-integrated luminosity of $L=1.3\times10^{-6}~\LSun$.
This implies that $\Mdot \leqslant 10^{-5}~\MdotUJ$ if $\MP\gtrsim5~\MJ$,
which is a mass range not favoured by the analysis of \citet{wang21vlti} but not completely unlikely. For $\MP$ of at most a few~$\MJ$, which seems more plausible \citep{Stolker+20b,wang21vlti}, the \LBra upper limit is effectively not constraining, with $\Mdot\leqslant10^{-4}~\MdotUJ$. Such high rates are barely expected from the theoretical side. Even the somewhat lower upper limits at high masses
are consistent with the results from Figure~\ref{fig:HaContour}, $\Mdot\approx8\times10^{-8}~\MdotUJ$ from the \citet{Wagner+2018} measurement or $\Mdot\approx1\times10^{-8}~\MdotUJ$ from \citet{Haffert+2019}.

For comparison, \citet{Hashimoto+2020} derived from their re-analysis of archival MUSE data an \Hb\ flux upper limit of $2.3\times10^{-16}$~erg\,s$^{-1}$\,cm$^{-2}$ for \PDSb, corresponding to $\LHb< 9.2 \times 10^{-8}~\LSun$ (3$\sigma$). 
This \Hb\ is inconsistent with the \Ha\ result (see Figure~\ref{fig:HaContour}). The same is true for \PDSc.
From this, \citet{Hashimoto+2020} concluded that there must be differential extinction. As justified in Section~\ref{sec:neglectextinc}, we emphasise that our results are \textit{without} taking extinction into account, which we do separately in \citet{maea21}.

In the same class as \object{CT~Cha~B/b} or \object{DH~Tau~B/b}, some of the only low-mass putative accretors for which other accretion lines have been detected \citep{Schmidt+2008,bonnefoy14lib,Zhou+2014,bowler14,wu15}, is Delorme~1~(AB)b, for which \citet{Eriksson+2020} measured \Hb\ (and also \HeI\ lines and upper limits on the infrared \CaII\ triplet). They infer a mass $\MP\approx12~\MJ$ and radius $\RP=1.6~\RJ$ from (hot-start) evolutionary models by combining their results to the photometry of \citet{Delorme+2013}. They report $\log(\LHa/\LSun)=-7.05\pm0.06$, from which they infer $\Mdot\approx(0.8$--$3.0)\times10^{-8}~\MdotUJ$ by combining their \LHa measurement with different models. Combined with ours (\citealp{Aoyama+2018}; \citealp{Aoyama+Ikoma2019}; this work), they derive $\Mdot=1\times10^{-8}~\MdotUJ$ and $\MP=11~\MJ$, using the radius $\RP=1.6~\RJ$ suggested by the photometry.
From Figure~\ref{fig:others}, the predicted \Hb\ luminosity is $\LHb\approx4\times10^{-8}~\LSun$. However, \citet{Eriksson+2020} measured $\LHb=10^{-8}~\LSun$, which is a factor four lower than the model prediction. This suggests that as for \PDSb, extinction is affecting the measurement of accretion tracers. The interesting difference is that there is up to now no evidence for an accretion disc around Delorme~1~(AB)b, but the constraints on a disc mass or surface density are not clear \citep{Eriksson+2020}. Thus a detection of (or upper limit on) a disc around Delorme~1~(AB)b, as well as more observational information on other hydrogen lines would be useful. On the theoretical side, predictions for the other currently available lines (\HeI\ and the infrared \CaII\ triplet) would be welcome.

Finally, in Figure~\ref{fig:others}, we show the 5-$\sigma$ upper limit on \Pab\ emission for the \object{TW~Hya} disc in its gap at 25~au, $\log(\LPab/\LSun) < -9.20$, derived by \citet{Uyama+2017} using Keck. At 95~au, where there is an other gap, the upper limit is 
$\log(\LPab/\LSun) < -9.79$ (not shown).
This is consistent with the mass constraints of $\MP\lesssim0.5~\MJ$ from \citet{vanBoekel+2017} for TW~Hya.
Looking at Figure~\ref{fig:others}, the interesting implication is that Keck at \Pab\ is sensitive to planets with a relatively low mass or accretion rate, at least for the nearest protoplanetary discs.

\section{Discussion}
 \label{sec:Discussion}

We now take a critical look at different aspects of the model and results presented in this work (Sections~\ref{sec:preshock}--\ref{sec:coolsize}), and discuss the non-detections of recent \Ha surveys (Section~\ref{sec:nondetect}).
Some caveats about the model were already discussed in Section~4.2 of \citet{Aoyama+2018}
and we do not repeat them here.
We compare with other recent models of \Ha\ emission from accreting planets \citep{Thanathibodee+2019,szul20} and discuss the validity of their approach in a different work \citep{AMIM20L}.
Appendix~\ref{sec:cfSH95} already details why the physical assumptions behind \citet{Storey+Hummer1995} do not apply to the planetary surface shock.

\subsection{Emission by the gas accreting onto the planet}
\label{sec:preshock}

Our model treats only the shock-heated gas. Several previous studies focusing on stellar-mass objects considered the gas flowing onto the accretor as the source of observed \Ha\ excess, assuming an unknown source of heating there \citep[e.g.][]{Hartmann+1994,Edwards+2013}, as opposed to heating provided by the shock.
Since the shock on accreting stars makes the gas too hot ($>10^6$~K) to emit hydrogen lines, the shock-heated gas is negligible for hydrogen-line emission. On the other hand, the planet-surface shock emits significant hydrogen lines, while the accreting gas should be cooler and emit weaker or no hydrogen lines compared to the stellar cases (see also the discussion in \citealt{AMIM20L}). This is the reason why we neglect hydrogen line emission from the region.

Even if the gas in the accretion flow is too cool to emit \Ha, the warm gas plays a significant role in excess emission other than from hydrogen lines. In the T~Tauri-star context, \citet{Calvet+Gullbring1998} found that a Balmer recombination continuum is emitted by the gas accreting onto the object rather than the postshock gas. The hydrogen recombination continua are well modeled and compared with observational results in the stellar accretion context.
As seen in Figures~\ref{fig:SED} and~\ref{fig:multiSED}, the planetary continua are predicted to be several (up to tens of) orders of magnitude stronger than the contribution from the planetary photosphere. Therefore, detecting in the planetary context a hydrogen continuum much stronger than the photospheric emission would lend support to our emission models.

\subsection[Effect of the accretion geometry: spherical versus magnetospheric]{Effect of the accretion geometry:\\spherical versus magnetospheric}
 \label{sec:geoeffect}

The predicted \Ha\ luminosity (see Figure~\ref{fig:HaContour}) was derived explicitly in the context of spherical accretion onto the protoplanet's surface. However, it also represents the signal expected for any accreting planet, regardless of the accretion geometry, i.e., spherical or magnetospherical.
In the case of magnetospheric accretion with a filling factor $\ffill<1$, relative to spherical accretion,
the kinetic-energy flux locally is higher by $1/\ffill$ but the accreting area is smaller by $\ffill$.
These effects cancel each other out to a large extent over most of parameter space.
However, more precisely, spherical accretion yields an upper limit to the \Ha\ intensity because of \Ha\ self-absorption in the shock-heated gas.
To emit more intense \Ha\ by avoiding self-absorption, the gas needs to be less dense, which, at a given $\Mdot$, will be the case for higher $\ff$.
Note that in a realistic situation, the infalling gas and dust could absorb a part of the flux emitted at the planet surface. This is explored systematically in \citet{maea21}.

Also, in the scenario of accretion onto a circumplanetary disk,
the \Ha\ emission is at most roughly  1\,\%\ of the \Ha\ coming from the planet surface for a similar planet mass,
at least for the (simple) disc model and scaling assumed in \citet{Aoyama+2018}.
(In Section~\ref{sec:distinguishsurfaceshockCPD} we had compared the line shapes at \textit{fixed} total luminosity.)
Therefore, when a strong shock occurs on the planetary surface,
regardless of the geometry, the CPD surface shock is negligible.

\subsection{Helium and metal lines}
\label{sec:metal}
For accreting stars, helium and metal lines (\HeI, \CaII, \NaI, \OI, etc.) are also detected and used as indicators of stellar accretion as for hydrogen lines \citep[e.g.][]{Kastner+2002}.
At the upper edge of the planetary-mass range, $\MP\approx20~\MJ$ leads to $\vff\approx200$~km/s for $\RP=2\RJ$ (see Figure~\ref{fig:n0v0 contours}), so that the postshock gas temperature can exceed $T_1\approx10^6$~K (Equation~(\ref{eq:T1})). In that case, metal lines instead of hydrogen lines are responsible for the dominant emission processes at ultraviolet (UV) wavelengths.

Our estimate of the hydrogen recombination continua would change when including metal lines.
When hydrogen ionization proceeds and neutral hydrogen is minor (e.g.\ at $T\approx10^5$~K in Figure~\ref{fig:postshock}), the gas should cool through hydrogen recombination continua and/or metal lines.
Presently, because we do not include metal lines, almost all the thermal energy is converted into the continua, so that they are overestimated.
The hydrogen recombination continua are mostly used as accretion indicator of protostars \citep[e.g.][]{Calvet+Gullbring1998}. For planetary accretion, they are currently expected to highly exceed photospheric emission at UV wavelengths, as in the example in Figure~\ref{fig:SED}, but this should be re-assessed once more complete models are available.

\subsection{On the physical size of the cooling region}
 \label{sec:coolsize}

Towards high $v_0$, the line-forming region can be at a depth of order of the planet size, which is clearly unrealistic. By contrast, in Figure~\ref{fig:postshock} it is at $\Delta z\approx3\times10^7$~cm, which represents $0.002\RP$ and is much smaller than the planet size and thus reasonable.
The reason for the large extent of the cooling zone in some of the other cases is that %
there is no cooling by helium nor atomic metals.
In any case, we remind that we always resolve the line-forming region because of adaptive time-stepping, which keeps the relative change in temperature within 10\,\%\ per step.

Despite the unrealistically large extent of the cooling region in some cases, the hydrogen-line fluxes should be relatively accurate within the other model assumptions.
This is because the hydrogen-line emission occurs in a spatially thin region (as in the example in Figure~\ref{fig:postshock}). Where exactly this region is located (i.e., possibly at too large a depth) is inconsequential for its emission properties. 
The presence of the heated photosphere (as in \citealt{Calvet+Gullbring1998}) is more likely to affect the thermal structure of the postshock region as well as the continuum emission but exploring this is beyond the scope of this work.

\subsection{Explaining the non-detections}
 \label{sec:nondetect}

Recent searches \citep{Cugno+2019,Zurlo+2020,Xie+2020} for accreting planets in \Ha\ have returned non-detections, with 5-$\sigma$ sensitivities down to line-integrated luminosities $\LHa\sim10^{-7}$--$10^{-6}~\LSun$ beyond $\sim15$~au ($\gtrsim100$~mas).
Using the \citet{Rigliaco+2012} correlation between \LHa and \Lacc derived for stars and fixing $\MP=5~\MJ$ and $\RP=1.5~\RJ$, \citet{Zurlo+2020} converted the \LHa non-detections to accretion rate upper limits $\Mdot\lesssim10^{-10}$--$10^{-9}~\MdotUJ$ for most systems.
For those luminosity upper limits, our model for $\MP=5~\MJ$ implies instead $\Mdot\lesssim10^{-8}$--$10^{-7}~\MdotUJ$ (see Figure~\ref{fig:HaContour}).
This is thus less constraining by a factor of $\sim100$, which is due to the lower conversion efficiency from accretion energy to \Ha in the planetary case compared to YSOs as commented by \citet[][see below and \citealt{AMIM20L}]{Zurlo+2020}.

\srevise{
The new upper limits of $\Mdot\lesssim10^{-8}$--$10^{-7}~\MdotUJ$ are still much lower than the \textit{average} mass accretion rate of $\Mdot\sim10^{-4.5}~\MdotUJ$ that accumulates a Jupiter mass within a reasonable disk lifetime (\citealp{haisch01}; see brief review in \citealt{Silverberg+2020}).
A time-dependent accretion rate can solve this apparent tension.
Namely, giant planets could almost reach their final mass %
with a much short timescale than the disk lifetime, which means their observational probability is much lower than that of the disk detection. This is a feature of
runaway gas accretion \citep[e.g.,][]{pollack96,mordasini17}, and holds also for  %
gravitational instability \citep[e.g.,][]{greaves10,kueffmeier17,nixon18,manara18,concha20,alves20,segura-cox20, schib21}.
Later on,
$\Mdot$ could remain low \citep{Tanigawa+07ApJ,Tanigawa+Tanaka2016}.
Also, alternating periods of high and low $\Mdot$ could occur during the main phase \citep[episodic accretion;][]{lubow12,brittain20,martin21}.%
}

\srevise{Another possibility is that}
the accretion rate \Mdot is high but only a small fraction of at most 1--10\,\%\  %
(the inverse of the 1--2~dex difference in \LHa quoted above) undergoes an accretion shock with sufficiently high preshock velocity $v_0\gtrsim30~\kms$ to generate \Ha (Section~\ref{sec:shockmodel}). This could occur if, as in Section~\ref{sec:distinguishsurfaceshockCPD}, most gas hits the (thick) CPD relatively far from the planet, perhaps due to angular momentum effects, and there is no surface shock. This was studied in \citet{Aoyama+2018}.
\srevise{Also, extinction can make the \Ha\ flux at the detector lower than the emitted one. This leads to an underestimate of $\Mdot$ in some geometry when the PPD, CPD, or accreting material interrupts the line of sight \citep[e.g.,][]{maea21}.
}

\srevise{Finally, }
one obvious \srevise{explanation} is that there are in fact no companions outside of the inner working angle of detectors up to now ($\gtrsim100$--200~mas; \citealp{Close2020,Zurlo+2020}) as mentioned by \citet{Zurlo+2020}.
From direct-imaging surveys, several-$\MJ$ objects at tens of~au or more are known to be intrinsically rare \citep[e.g.,][]{bowler16,nielsen19,Vigan+2020}.
The absence of distant gas giants is also predicted by classical core accretion theories \citep[e.g.,][]{Ikoma+00ApJ,Thommes+08Sci}.
This would leave the \Mdot of forming planets unconstrained by the \Ha survey results.

\section{Summary and conclusions}
 \label{sec:summconc}

Motivated by recent detections of accretion signatures at young planets or very-low-mass objects (\citealp{keppler18,Wagner+2018,Haffert+2019,Eriksson+2020}),
we have extended the NLTE shock emission model of \citet{Aoyama+2018} to the case that only the planet surface, as opposed to the circumplanetary disc, is the origin of the hydrogen lines.
By combining the shock spectrum with models for the photospheric emission, we predict global SEDs (UV to IR) of accreting planets.
A possible contribution from the CPD, relevant at far IR wavelengths \citep[e.g.,][]{zhu15}, is not included.
Extinction by the accreting material or the CPD or PPD is neglected in this work, which we argued (Section~\ref{sec:neglectextinc}) is a relevant case on its own and allows us to deal with the complex issue of extinction separately in \citet{maea21}.

The formation-relevant input parameters of our model are
the accretion rate $\Mdot$, the planet mass $\MP$, the planet radius $\RP$, and the filling factor of the accreting region $\ffill$.
To provide guidance,   %
we have fit the radius of forming planets as a simple but non-monotonic function of $(\Mdot,\MP)$ from the results of detailed planet structure calculations (Section~\ref{sec:Rfit}).
While the structure calculations are not definitive, the fit is an improvement over using a constant radius as is often done. Other $\RP(\Mdot,\MP)$ relations could be used.

The photospheric effective temperature $\Teff$ was derived approximately self-consistently\footnote{In the ``Suite of Tools to Model Observations of accRetIng planeTZ'' (\texttt{St-Moritz}) at \url{https://github.com/gabrielastro/St-Moritz}, the $\RP$ and $\Teff$ functions are implemented.}
from the energy transport from the shock model (Section~\ref{sec:Tefffit}).
Because the shock heats the planet, $\Teff$ can easily reach 3000--5000~K for the chosen ranges of $\Mdot$ and $\MP$ values.

Fixing $\ffill=1$, we have scanned the large parameter space
and shown
global SEDs of forming planets (Figure~\ref{fig:multiSED}).
We have %
also displayed line luminosities as a function of $\Mdot$ and $\MP$ for individual hydrogen lines, focusing on \Ha as well as \Hb, \Paa, \Pab, \Pag, \Bra, and \Brg (Figures~\ref{fig:HaContour} and~\ref{fig:others}) and discussing their ratios (Figure~\ref{fig:ratio2Ha}). The data for these and other hydrogen lines are available upon request.

Our main findings  %
are the following:
\begin{enumerate}

    \item Despite the high $\Teff$ of the planet heated by the shock, the shock contribution to the narrow and broad \Ha filters of SPHERE and MagAO dominates over the photospheric contribution (Figure~\ref{fig:SED}).

    \item At the surface of the planet, the Lyman and Balmer series clearly emerge above the photosphere over all parameter space, and the Paschen continuum is visible at low accretion rates.     Details depend on the $\Teff$ fit but many lines in the Paschen, Brackett, or other series are visible above the hot photosphere (Figure~\ref{fig:multiSED}).
    
    \item The \Ha\ line luminosity as a function of $\Mdot$ and $\MP$ is a monotonic function of both and makes it possible to constrain mostly $\Mdot$ given an observed value (Figure~\ref{fig:HaContour}). This is one of the key results. Applying this tool to current detections yields reasonable constraints (Section~\ref{sec:obs}).
    For example, the mass accretion rate of \PDSb\ is estimated as $\Mdot=(1.1\pm0.3)\times10^{-8}~\MdotUJ$ from $\LHa = (3.3\pm 0.1)\times 10^{-7}\LSun$ \citep{Haffert+2019}. The degeneracy between \Mdot and $\MP$ can be lifted at sufficiently high resolution, higher than afforded by MUSE \citep{Aoyama+Ikoma2019,Thanathibodee+2019}.
    
    \item If there is an accretion shock both on the planet surface and on a circumplanetary disk, the signal is likely to be dominated by the surface-shock contribution.
    The two shocks are expected to be spectrally distinguishable, with the CPD shock narrower. For \PDSb, the resolution of MUSE is not sufficient, but the minimum requirement depends on how sensitive future instruments are to the wings of the \Ha line
    (Figure~\ref{fig:CPD_P}).

    \item The line luminosity of other transitions such as \Hb, \Paa, \Paa, \Pab, \Pag, \Bra, or \Brg\ is a monotonic function of $\Mdot$ and $\MP$ (Figure~\ref{fig:others}).
    We compare this to upper limits for \PDSb\ and the TW~Hydra disk.
    The intensity ratios of these lines to \Ha\ range between $10^{-3}$ and~$\approx1$ (Figure~\ref{fig:ratio2Ha}), and their measurement can yield some constraints on the amount of extinction \citep{Hashimoto+2020}.

    \item Recent \Ha surveys have resulted in non-detections of planets outside of $\approx15$~au ($\gtrsim100$~mas).
    Applying the extrapolated YSO \Lacc--\LHa relationship of \citet{Rigliaco+2012} yields upper limits on the instantaneous accretion rate of $\Mdot\lesssim10^{-10}$--$10^{-9}~\MdotUJ$ \citep{Zurlo+2020}, but using our models made explicitly for planets implies instead $\Mdot\lesssim10^{-8}$--$10^{-7}~\MdotUJ$, which is a less strict constraint, assuming that planets indeed are present are the surveyed stars (Section~\ref{sec:nondetect}).
    
\end{enumerate}
We point out that a determination of \Mdot from accretion tracers yields a lower limit on the total accretion rate; the rest of the accreting mass could be joining the planet without a shock at all (e.g., by boundary-layer accretion) or with a shock with too-low velocity $v_0\gtrsim30~\kms$ (Sections~\ref{sec:parameters} and~\ref{sec:distinguishsurfaceshockCPD}).

In companion papers, we discuss the use of spectrally-resolved line profiles for inferring the physical parameters of planets \citep{Aoyama+Ikoma2019}, study the correlation between \LHa and the accretion luminosity \Lacc in the planetary case \citep{AMIM20L}, and assess the absorption of the \Ha\ flux by the infalling gas and dust \citep{maea21}. These models are applied to \PDSb also in \citet{Hashimoto+2020} and to Delorme~1~(AB)~b in \citet{Eriksson+2020}.

In \citet{AMIM20L},
we compare our work to other recent models for the \Ha emission associated with accreting planets \citep{Thanathibodee+2019,szul20}.
We emphasise that the postshock electron populations cannot reach equilibrium as the gas cools
\citep{Aoyama+2018}. This and other radiative properties invalidate the key assumptions behind \citet{Storey+Hummer1995}, which was developed for a physically very different context (e.g., planetary nebulae or \HII regions). Thus \citet{Storey+Hummer1995} is not appropriate for hydrogen-line luminosity predictions for the planetary accretion shock (Appendix~\ref{sec:cfSH95}). For the stellar context too, \citet{Storey+Hummer1995} has been suggested to not be appropriate \citep[e.g.,][]{Edwards+2013,Rigliaco+2015,antoniucci17}, despite its use in earlier analyses.

Continued searches with existing
and upcoming or proposed instruments 
are expected not only to reveal more sources, but will hopefully also increase the number of detected lines.
The first group includes the ZIMPOL subsystem of VLT/SPHERE \citep{Beuzit+2008,Schmid+2018}, as well as VLT/MUSE \citep{Bacon+2010}, LBT/MagAO \citep{Close+2014,Close+2014b}, SCExAO/VAMPIRES \citep{Uyama+2020},
while to the planned instruments belong
MagAO-X \citep{Males+2018,Close+2018,Close2020}, KPIC \citep{Jovanovic+2019,Morris+2020}, VIS-X
(for \Ha\ with $R=15,000$; PI: S.~Haffert, priv.\ comm.), RISTRETTO\footnote{See \url{https://zenodo.org/record/3356296}.} on the VLT and later possibly on the ELT (with $R>130,000$, possibly up to $R=150,000$, covering \Ha; PI: Ch.\ Lovis; \citealp{chazelas20}), 
NIRSpec\footnote{See \url{https://jwst-docs.stsci.edu/near-infrared-spectrograph}.} on the \textit{James Webb Space Telescope} (JWST; but note the moderate resolution $R\approx1000$--2700 and likely high demand for time), HARMONI\footnote{See \url{https://harmoni-elt.physics.ox.ac.uk}.} on the ELT (a first-light IFU that will cover \Ha at $R=3000$ and $\lambda=0.8$--2~$\upmu$m at $R=7000$ or $R=17,000$), and
HIRES/ELT (a second-generation spectrometer planned to cover 1--1.8~$\upmu$m at $R=100,000$; \citealp{Marconi+2018,tozzi18}, E.~Oliva 2020, priv.\ comm.).
 Combined with simulations of forming planets, these rich data sets are poised to help constrain observationally the complex accretion geometry and ultimately the origin of gas giants.

\acknowledgments

{\small
We wish to pay tribute to France Allard, who passed away unexpectedly in October 2020. Her world-leading atmospheric models are used widely in the observational and theoretical communities, and are an important input in this work too. Her kind nature and expertise will be missed by many.
We thank
R.~van~Boekel, M.~Keppler, A.~M\"uller, J.~Bouwman, B.~Husemann, M.~Samland, D.~Homeier,
H.~M.~Schmid, J.~Milli, J.~Girard,  %
S.~Quanz, G.~Cugno, T.~Stolker,
S.~Edwards, L.~Venuti, B.~Stelzberg,
Ch.~Rab,       %
C.~Manara,     %
N.~Turner,
W.~B\'ethune,
M.~Bonnefoy,   %
S.~Kraus,      %
and
O.~Ernesto     %
for useful discussions and explanations, insightful questions, and helpful sharing of data.
YA was supported by the Leading Graduate Course for Frontiers of Mathematical Sciences and Physics.
G-DM acknowledges the support of the German Science Foundation (DFG) priority program SPP~1992 ``Exploring the Diversity of Extrasolar Planets'' (KU~2849/7-1 and MA~9185/1-1).
G-DM and CM acknowledge support from the Swiss National Science Foundation under grant BSSGI0\_155816 ``PlanetsInTime''.
Parts of this work have been carried out within the framework of 
JSPS Core-to-Core Program ``International Network of Planetary Science (Planet$^2$)'' 
and
the NCCR PlanetS supported by the Swiss National Science Foundation.
This research has made use of the SVO Filter Profile Service (\url{http://svo2.cab.inta-csic.es/theory/fps/}) supported by the Spanish MINECO through grant AYA2017-84089 \citep{rodrigo12,rodrigo20}.
This research was supported in part by the Japanese--German visitor program of the NCCR PlanetS and by the visitor program of the German Science Foundation (DFG) priority program SPP~1992.
}

\appendix

\section{Detailed model description}
\subsection{Fitting of planetary properties}
\label{sec:Fittin}

\subsubsection{Radius fit}
\label{sec:Rfit}

In principle, the main parameter space for our calculations is %
$(\Mdot,\MP,\RP,\ffill)$ along with the choice of ``cold-start'' or ``hot-start'' accretion (high or low radiation efficiency of the accretion energy;
\citealp{Marleau+2017,Marleau+2019b}).
Here, 
we take for the sake of definiteness the $\RP(\Mdot,\MP)$ relations in
the cold- and hot-start populations of the Bern model \citep{alibert05,morda12_I,morda12_II,morda15,mordasini17}, 
using data from all time snapshots\footnote{
  The data can be visualized at, and downloaded from, the ``Evolution'' section of the Data Analysis Centre for Exoplanets (DACE)
  platform under \url{https://dace.unige.ch}.}.
We use the populations \texttt{CD752} (hot) and \texttt{CD753} (cold), described and analysed in \citet{morda12_I,morda12_II,mordasini17} (Generation~Ib). Recently, the first results from the Generation~III population syntheses of the Bern model
were released \citep{Emsenhuber+2020a,Emsenhuber+2020b,Schlecker+2020}, which all assume warm accretion. We verified that the distribution of points in $(\Mdot,\MP,\RP)$ space is very similar between the 1- and the 100-embryo-per-disk simulations \texttt{NG73} and \texttt{NG76}, respectively, on the one hand, and \texttt{CD752} on the other. Two small differences are that the accretion rates reached are not quite as high as in Generation~Ib (note however that there are fewer synthetic planets in the region of interest), and that in \texttt{NG76} the radii can be higher at a given $\Mdot$ and $\MP$, likely due to interactions between the embryos. Since the radii are overall similar, we will keep using the Generation~Ib populations in order to cover also high accretion rates, as could be relevant for instance to accretion outbursts \citep[e.g.,][]{lubow12,brittain20,martin21}.

These planet structure models were calculated assuming that the planet is at all times convective.
Recent work suggests that forming planets may be in fact in part radiative \citep{berardo17,berardocumming17,cumming18}
and thus have a different radius. Nevertheless, the relations $\RP(\Mdot,\MP)$
from the population syntheses provide a reasonable bracket and
reduce the dimensionality of the large parameter space $(\Mdot,\MP,\RP,\eta,\ffill,\ldots)$.
However, it is clear that this is not meant as a final answer and one could repeat this study with for example the $\RP(\Mdot,\MP)$ relations of \citet{ginzburg19b}, who find (much) larger radii at a given mass.

After some experimentation, we arrived at the following relatively simple form for the fitting function\footnote{For convenience, it is provided in different languages along with the fit coefficients in the ``Suite of Tools to Model Observations of accRetIng planeTZ'' (\texttt{St-Moritz}) at \url{https://github.com/gabrielastro/St-Moritz}.}:
\begin{align}
 \label{eq:R fit}
 R_1(\Mdot,\MP) = &~a_0 + b_0 \lgMd + c_0\e^{d_0\lgMd} \notag \\
                   & + \left(a_1+b_1 \lgMd + c_1\e^{d_1\lgMd}\right)\left(M_1-1\right)\notag \\
                   & + \left(a_2 + b_2 \lgMd  + c_2\e^{d_2\lgMd}\right)\left(M_1-1\right)^2,
\end{align}
where
$R_1\equiv\RP/\RJ$, $M_1\equiv\MP/\MJ$,
and %
$\lgMd\equiv\log_{10}(\Mdot/10^{-2}~\MdotU)$.
The fits were performed through \texttt{gnuplot}'s built-in \texttt{fit} routine. %
Only planets with $\Mdot>10^{-5}~\MdotU$, $1~\MJ< \MP<20~\MJ$, and $\RP<4~\RJ$ were used to obtain the fits.
We used for each planet a statistical weight inversely proportional to its radius
to have a more accurate fit at lower radii,
for which $v_0$ is higher and thus the accretion signatures a priori stronger.
The coefficients for the cold-nominal population are
\begin{lstlisting}
  a0 = 1.53;   b0 = 0.111;
  c0 = 1.06;   d0 = 0.906;
  a1 = -0.195; b1 = -0.0307;
  c1 = 0.0977; d1 = 0.000695;
  a2 = -0.250; b2 = 0.000276;
  c2 = 0.254;  d2 = 0.000214
\end{lstlisting}
and 
for the warm population the coefficients are
\begin{lstlisting}
  a0 = 0.411;  b0 = -0.244;
  c0 = 3.45;   d0 = 0.762;
  a1 = -0.489; b1 = -0.0961;
  c1 = 0.652;  d1 = 0.353;
  a2 = -0.228; b2 = -0.00106;
  c2 = 0.226;  d2 = 0.000220.
\end{lstlisting}
We verified that excluding from the fitting procedure the planets for which the accretion radius $\RAkk<10\RP$ (this concerns only a small fraction of the planets) changed neither the relationships nor the quality of the fit significantly. The accretion radius $\RAkk$ is a spherically-averaged estimate
of the typical distance from which the gas is effectively falling onto the planet.
It is defined through
\begin{equation}
 \label{eq:Racc}
 \frac{1}{\RAkk}=\frac{1}{\RBondi} + \frac{1}{\kLiss\RHill},
\end{equation}
where $\RBondi$ and $\RHill$ are the Bondi and Hill radii
and $\kLiss=1/4$ (and not, as written in \citealp{morda12_I}, $\kLiss=1/3$). 

The resulting relations $\RP(\Mdot,\MP)$ for the cold-nominal and the warm populations
are shown in Figure~\ref{fig:RP(Mdot,MP)}.
At high accretion rates, $\RP$ reaches $\approx5~\RJ$ ($\approx3~\RJ$) in the ``warm'' (``cold-nominal'') population.
As a function of mass, the radius monotonically increasing with mass for the ``warm population'' and has a minimum near 10--15~$\MJ$ for the ``cold-nominal'' population.

The fit is overall excellent, with a match to roughly 10\,\%.
For the cold-nominal population, which displays the largest deviations:
Only for masses $\MP\gtrsim15~\MJ$ near $\Mdot=10^{-4}$--$10^{-3}~\MdotU$
is the function too small, by at most only $\approx30\,\%$,
and at $\Mdot\gtrsim3\times10^{-2}~\MdotU$ for $\MP$ between 1 and 10~$\MJ$
larger or smaller by at most $\approx30\,\%$.
This reflects in part the intrinsic scatter in the population synthesis results.
For the warm population, the fitted function yields radii also at most 30\,\%\ too small
but only towards high masses and low accretion rates.
At lower accretion rates and, in the cold-nominal population, for lower masses than shown, the fit re-increases but this matches rather well overall the data (not shown).

\begin{figure*}
\includegraphics[width=0.49\textwidth]{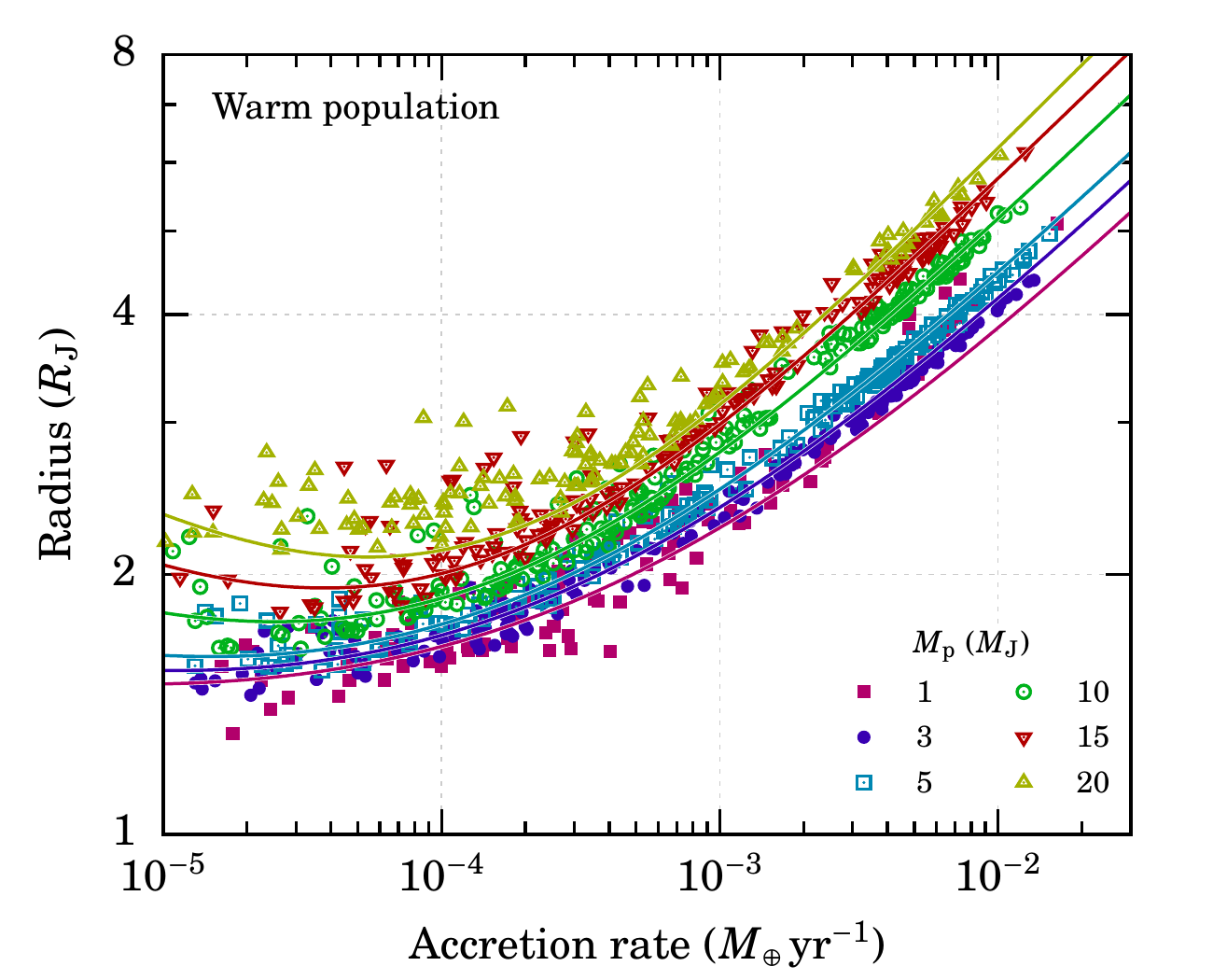}
\includegraphics[width=0.49\textwidth]{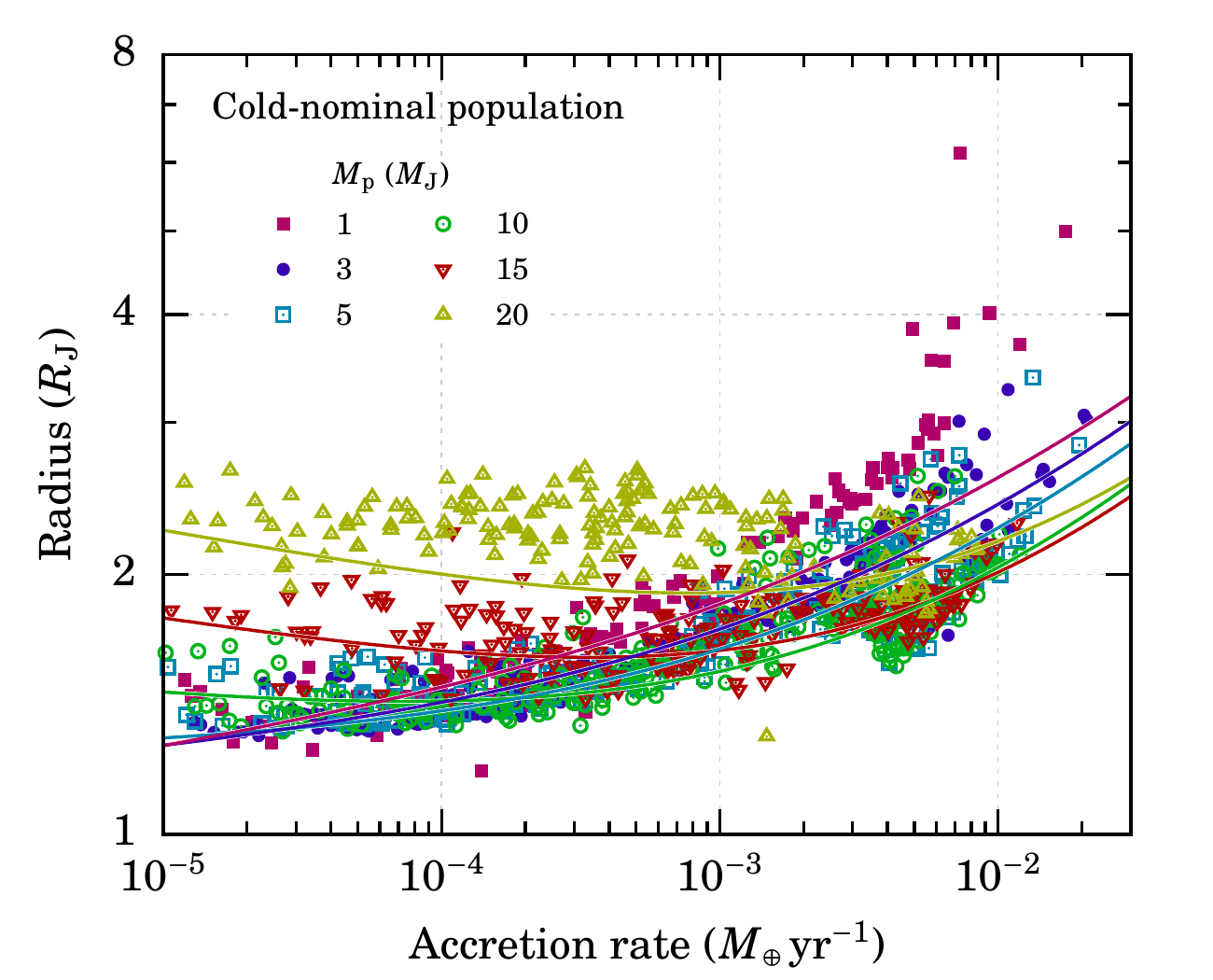}\\
\includegraphics[width=0.49\textwidth]{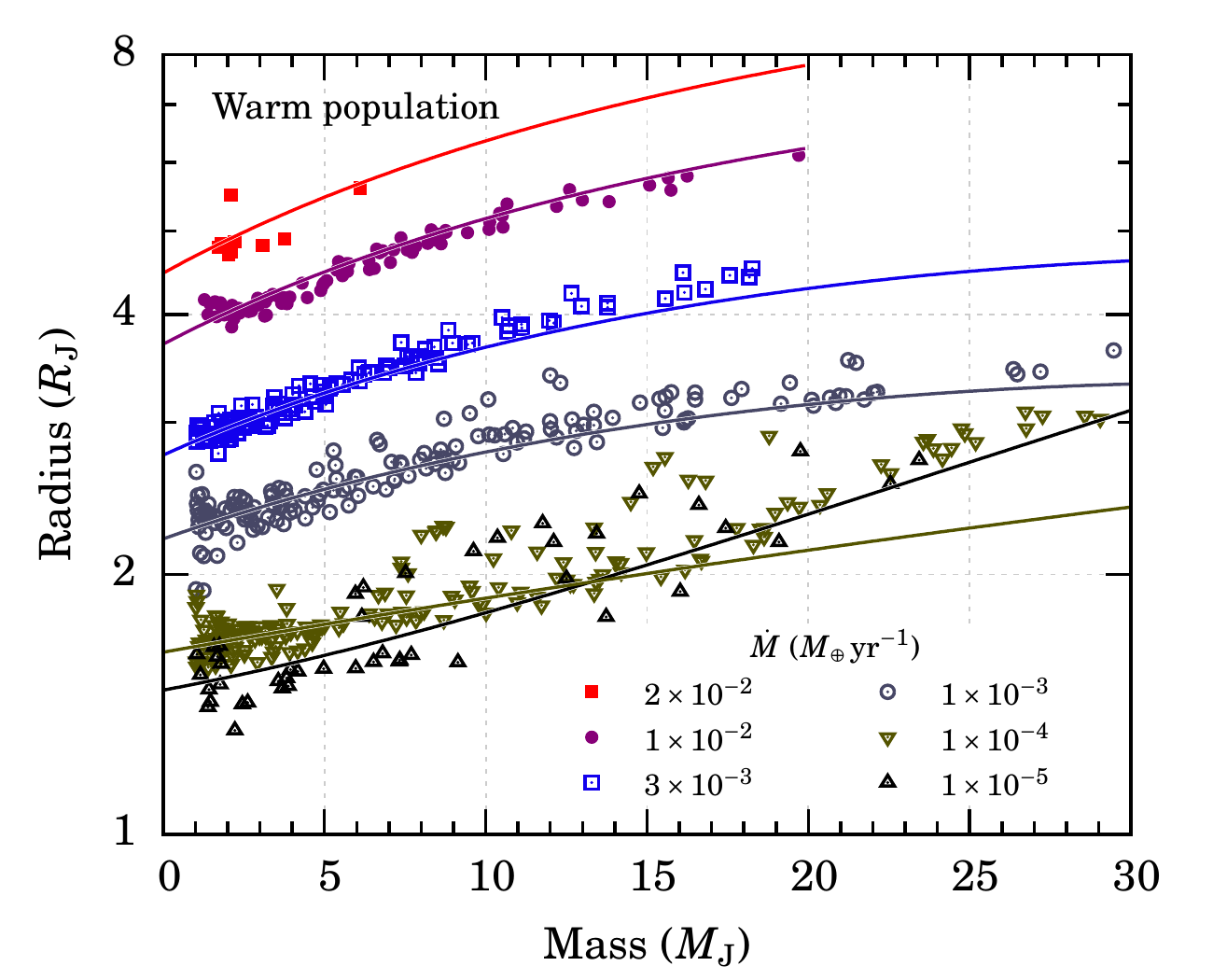}
\includegraphics[width=0.49\textwidth]{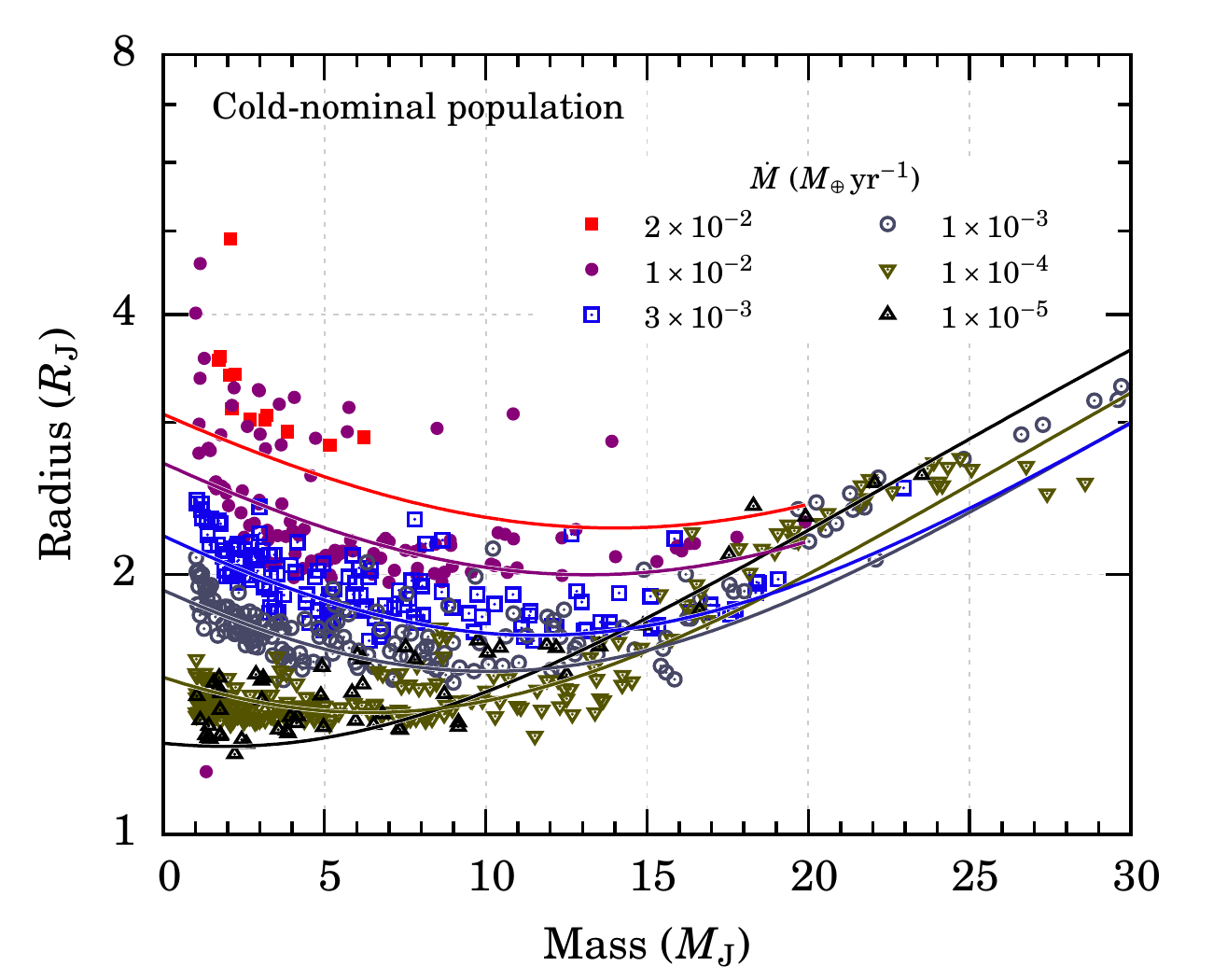}\\
\caption{
Dependence of the planet radius $\RP$ on accretion rate $\Mdot$
and planet mass $\MP$ for planets forming by core accretion,
for the warm- (cold-)population on the left (right).
The population syntheses of \citet{morda12_I} are used
but only planets with $\RAkk>20\RP$ are shown. %
In the top (bottom) row, six masses (accretion rates) are considered
(see respective legends).
The solid lines display the simple but accurate function of Equation~(\ref{eq:R fit})
with the fitted coefficient as given in the text.
Note the logarithmic vertical scale.
}
\label{fig:RP(Mdot,MP)}
\end{figure*}

\subsubsection{Effective temperature}
  \label{sec:Tefffit}
  
For the photospheric temperature $\Teff$, we adopt a semi-analytical prescription that ensures that, approximately, the total outgoing flux (shock and photosphere) is equal to the sum of the internal and incoming energy flux.
At the shock, a portion of the incoming energy is converted into hydrogen-line and recombination-continua emission.
The remaining portion travels downward into the atmosphere, where it is expected to be thermalised because most of the energy is in \Lya, which can easily be thermalised.
What matters for the structure of the planet is whether this radiation
goes deep into the planet, thereby heating it up,
or whether only the top layers are heated up and re-emit the radiation,
on a timescale that is short compared to the cooling time of the planet.
The former outcome corresponds to the ``Hot start (accreting)'' case of \citet[][their Equation~(13)]{morda12_I}, and the latter to their ``Cold start'' case.
In both cases, the total luminosity just outside the planet\footnote{This
	does not consider the energy recycling in the accretion flow
	discussed by \citet{Marleau+2017,Marleau+2019b},
	and ignores any  %
	absorption in the layers closest to the planet; we take
	$\etaklassisch=1$ for this discussion.}
is $L\approx\Lint+\Lacc$,
where $\Lint$ is the energy coming from the deep interior.

However, the spectrum of the accreting region (covering a fraction \ffill of the surface of the planet) differs between the two extreme cases, as noted by \citet{morda12_I}.
In the ``Hot start'' extreme case the spectrum is entirely thermalised,
as given by an atmospheric model with $\Teff^4=\Tint^4+\Tacc^4$,
where the ``accretion temperature''
\begin{equation}
 \label{eq:Tacc}
\Tacc=\left(\frac{\Lacc}{4\pi\RP^2\ffill\sigma}\right)^{1/4},  %
\end{equation}
with $\sigma$ the Stefan--Boltzmann constant.
Then there would be no emission-line spectrum as we have been computing in this work.
At the other extreme, in the ``Cold start'' case the spectrum from the accreting region is given by the sum of an atmosphere at $\Teff=\Tint$ (i.e., not heated up at all by the shock) and the shock emission. Clearly both are limiting cases. In either case the remaining fraction $(1-\ffill)$ of the planet surface has $\Teff=\Tint$, and the global spectral appearance will be a mixture of the spectrum from the accreting and the non-accreting components, with the proportion set by the (rotational-phase-dependent) viewing geometry. We set $\ffill=1$ in this work for simplicity.

Which scenario is likely more accurate? The single-stream, frequency-averaged simulations of \citet{Marleau+2017,Marleau+2019b} suggest that on the net,
only a small fraction of the incoming \Lacc will go in deeper, but that this small portion is likely (much) higher than the internal luminosity in the extreme cold starts by \citet{Marley+2007}. However, for the two-stream, frequency-dependent calculations presented here, %
\citet{Aoyama+2018} mentioned that roughly one half of the radiation goes down and one half goes up, which holds in the limit that the emitting region is optically thin.
Because the radiation transport is more detailed in the models of \citet{Aoyama+2018}, we will follow their results. Therefore, we will  %
quantify this fraction
more precisely and use it to derive $\Teff$.

\begin{figure}
    \centering
    \includegraphics[width=0.49\textwidth]{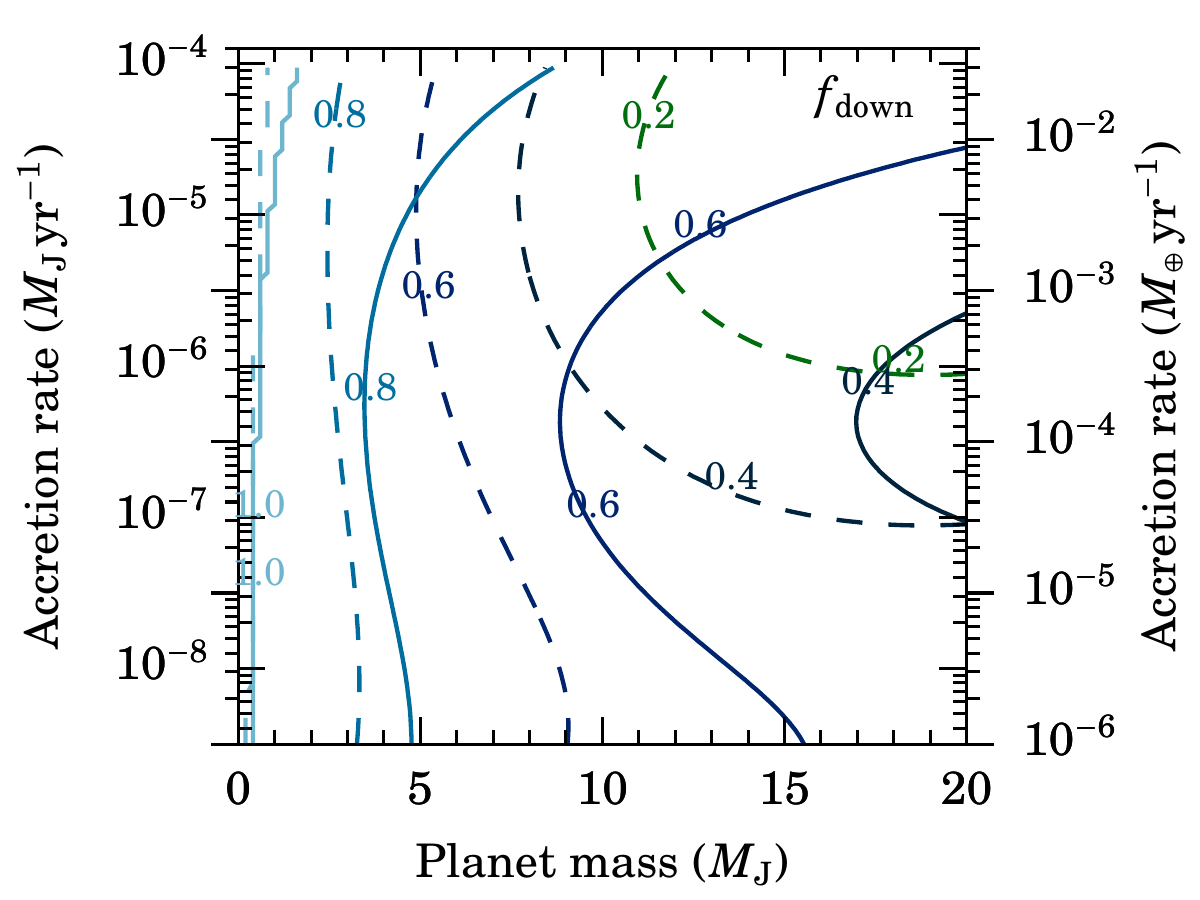}
    \caption{Fraction $\fdown$ of the incoming kinetic-energy flux that is in the \textit{downward}-travelling radiation field at the bottom of the atmosphere. This includes all hydrogen lines and continua. The solid (dashed) lines use the hot (cold) population radius fits, and for $\fdown(n_0,v_0)$ we use the fit from Equation~(\ref{eq:fdfit}).}
    \label{fig:fdown}
\end{figure}

Figure~\ref{fig:fdown} shows the fraction $\fdown$ of the incoming kinetic energy flux that is present in the downward-moving radiation field at the bottom of the computational domain, where $T$ reaches $10^4$~K. For this, we have first written
\begin{subequations}
\label{eq:fdfit}
\begin{align}
  \fdown' =&~a_0 + b_0 \lgnz + c_0 \lgnz^2  \notag\\
           &+ \left(a_1+b_1\lgnz+c_1\lgnz^2\right)\left(v_0-100~\kms\right)\notag\\
           &+  a_2\left(v_0-100~\kms\right)^2,\label{eq:fdfita}\\
  \fdown =&~\min\left( \max\left(\fdown',0\right), 1 \right),\label{eq:fdfitb}
\end{align}
\end{subequations}
where $\lgnz\equiv \log_{10} \left(n_0/10^{12}~\mathrm{cm}^{-3}\right)$.
\srevise{Equation~(\ref{eq:fdfita}) is a simple biquadratic function in the natural parameters $v_0$ and $\log(n_0)$ without the $b_2$ and $c_2$ terms (see below).}
Equation~(\ref{eq:fdfitb}) ensures that $\fdown$ remains between~0 and~1 and is needed only for a small part of the parameter space, for some models on the grid edge.
Using \texttt{gnuplot}'s built-in \texttt{fit} command yielded\footnote{This too is provided in different languages in the ``Suite of Tools to Model Observations of accRetIng planeTZ'' (\texttt{St-Moritz}) at \url{https://github.com/gabrielastro/St-Moritz}.}
\begin{lstlisting}
a0 = 0.703752;   a1 = -0.00527886;
b0 = -0.0967987; b1 = -0.00146833;
c0 = -0.0254579; c1 = -0.000321504;
a2 = -9.91492e-06,
\end{lstlisting}
which matches very well the model data (not shown). %
We had included at first
$\lgnz$ terms in the $v_0^2$ term but their coefficients 
\srevise{($b_2$ and $c_2$)} were consistent with zero.
We therefore repeated the fit with the form of Equation~(\ref{eq:fdfita}), yielding the reported coefficients.
The result of this fit depends only on our grid of models in $(n_0,v_0)$ space and is thus independent of the population.
Next, we used Equations~(\ref{eq:v0}) and~(\ref{eq:n0 a}), which relate $(n_0,v_0)$ and the macrophysical parameters $(\Mdot,\MP,\RP,\ffill)$ and are presented below, along with
the radius fits (Equation~(\ref{eq:R fit})) to obtain $\fdown(\Mdot,\MP)$.

We find that the fraction $\fdown$ does vary at low masses but that it covers mainly $\fdown\approx0.3$--0.8 between the border of the brown-dwarf region and a few $\MJ$. The fraction depends only relatively weakly on $\Mdot$ and much more on $\MP$ (through the $v_0$ dependence), reaching $\fdown\approx1$ at $\MP\approx\MJ$.
Note that $\fdown$ is affected to some extent by the non-inclusion of low-temperature ($T\lesssim10^4$~K) cooling processes at the bottom of the computation domain, for instance from molecules. Currently they are not included, so that in a future model iteration $\fdown$ could be different, once the inclusion of helium and metals will accelerate the cooling and thus make it computationally feasible to let the simulations cool down to lower temperatures than $T=10^4$~K. Nevertheless, Figure~\ref{fig:fdown} already provides some guidance.

Having computed $\fdown$, we use it to write the photospheric temperature of the accreting region as
\begin{subequations}
\label{eq:fluxconserv}
\begin{align}
  \Teff^4 &= \Tint^4 + \fdown \Tacc^4, \label{eq:Teff}\\
  \Rightarrow \Lphot + \Lshock &= \Lint + \Lacc,\label{eq:Lsum}
\end{align}
\end{subequations}
where \Lphot is the luminosity of the photosphere at \Teff and \Lshock is the total shock luminosity. In other words,
we assume that the component $\Fdown=\fdown \sigma\Tacc^4$ from the shock (lines and recombination continua) is thermalised and re-emitted. Equation~(\ref{eq:Teff}) ensures that the total upward-travelling radiative flux from our combined models is $F= \sigma\Teff^4 + (1-\fdown)\sigma \Tacc^4 = \sigma \Tint^4 + \sigma \Tacc^4$, with the shock flux $\Fshock = (1-\fdown)\sigma \Tacc^4$ contained in the models that are the main subject of this work (see Equation~(\ref{eq:Lsum})).
In Equation~(\ref{eq:fdfit}), to avoid $\Teff=0$~K when $\fdown=0$, which occurs at large $(\Mdot,\MP)$ in the cold-start case, we set somewhat arbitrarily $\Tint=1000$~K. We verified that this barely affects the function $\Teff(\Mdot,\MP)$. There is a slight effect at $\Mdot\lesssim10^{-8}~\MdotUJ$ but this region of parameter space is of lesser interest.

\begin{figure*}
\includegraphics[width=0.49\textwidth]{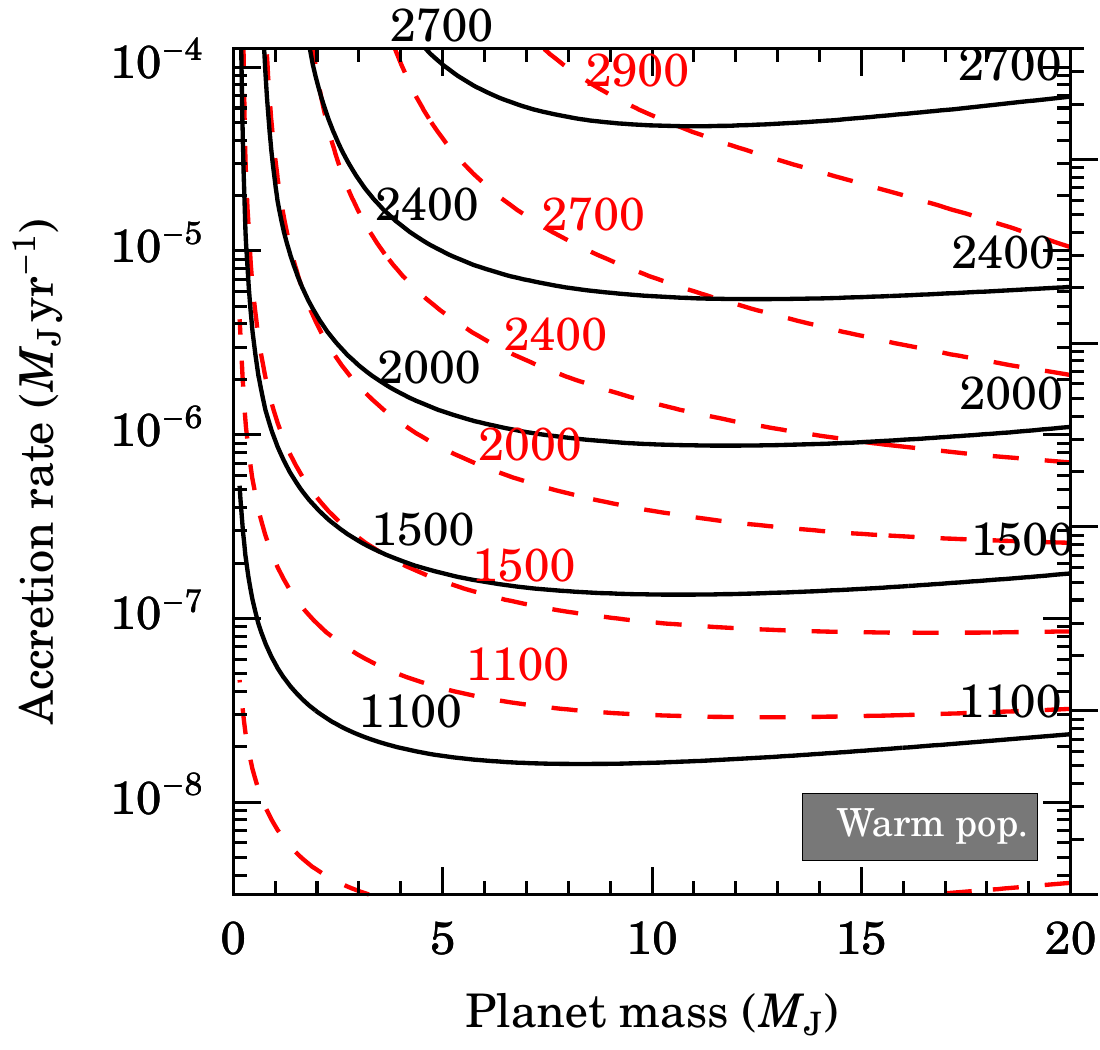}
\includegraphics[width=0.49\textwidth]{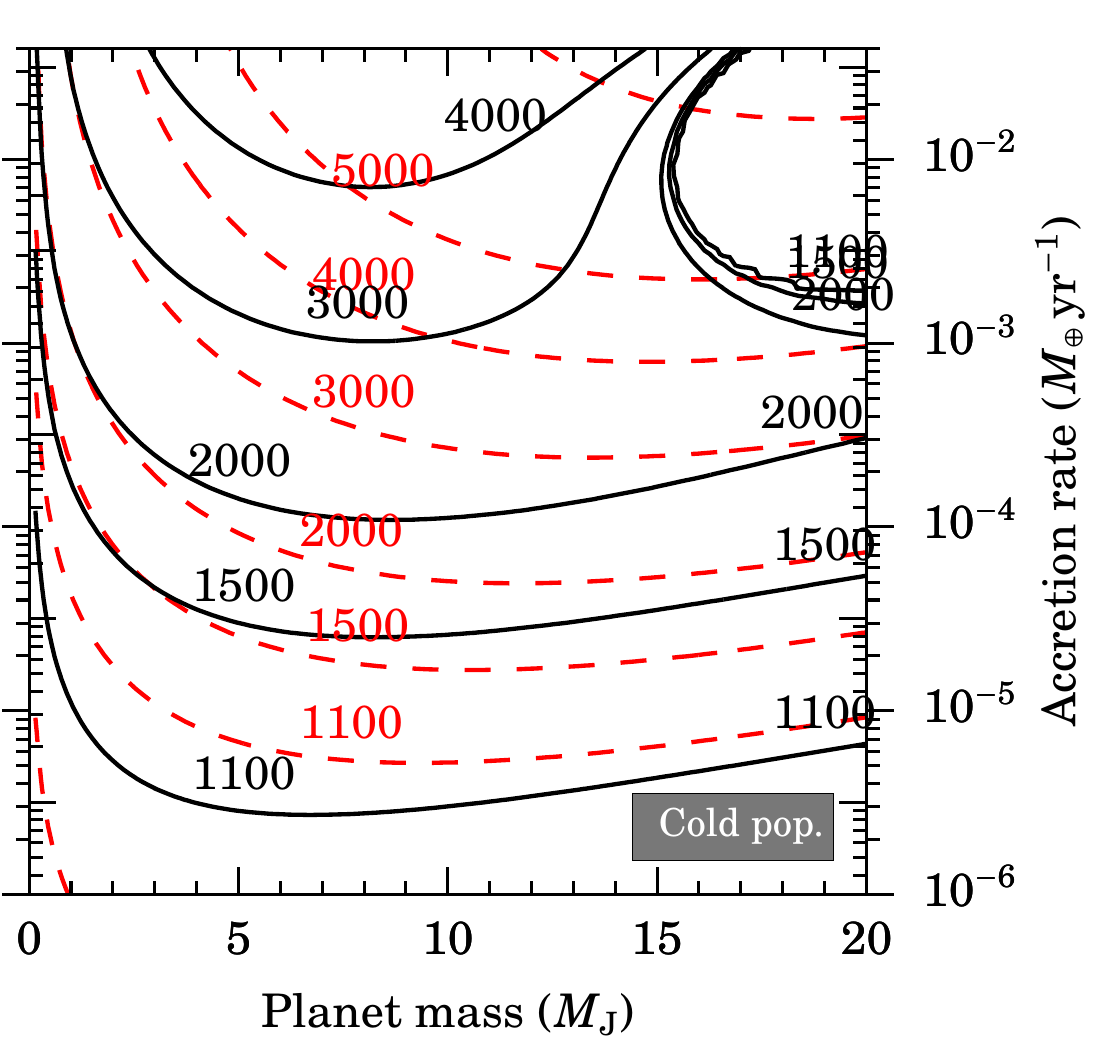}\\
\caption{
Approximate effective temperature of accreting planets (Equations~(\ref{eq:fdfit}) and~(\ref{eq:Teff}), using $\Tint=1000$~K; \textit{black contours}). This is compared to contours of constant $\Tacc$ (\textit{red dashed contours}).  %
We fix $\ffill=1$.
}
\label{fig:Teff}
\end{figure*}

Figure~\ref{fig:Teff} shows the resulting $\Teff$ from Equations~(\ref{eq:fdfit}) and~(\ref{eq:Teff}) for both populations for $\ffill=1$. %
For the warm-population fit, due to the large radii, $\Teff$ ranges only up to $\approx2500$~K, whereas in the cold-population $\Teff$ can reach up to 4000~K. Since a logarithmic scale is used for $\Mdot$ but a linear one for $\MP$, the dependence of $\Teff$ on $\Mdot$ appears stronger on $\Mdot$ than on $\MP$. The low $\Teff\approx1000$--2000~K for the cold-population fit at large mass $\MP\approx20~\MJ$ and radius is a consequence of $\fdown$ going to zero there, with automatically a large sensitivity on the choice of the fit. Thus it should not be taken very seriously, and is not of major concern anyway since this part of parameter is of lesser interest to the present work.

We comment briefly on an implicit assumption we make.
When we use standard atmospheric models with Equation~(\ref{eq:Teff}), we are assuming that the emission from the heated photosphere can be described by that of an isolated object, i.e., that if the shock changes the pressure--temperature structure of the atmosphere (as it is likely), the resulting spectrum is not entirely different. This may be a strong simplification, and it would be interesting to explore with dedicated radiative transfer calculations how this modifies the spectral shape.

Spectra for the BT-Settl models we use (\texttt{CIFIST2011\_2015}) exist only for $\Teff\geqslant1200$~K. Therefore, for the case $\Teff<1200$~K, we use $1200$~K.

\subsection{Shock model}
 \label{sec:shockmodel}
\subsubsection{Shock parameter space}
\label{sec:parameters}

Locally, a strong shock converts most of the mechanical energy into thermal energy, and the gas temperature increases by orders of magnitude compared to the preshock value. However, since the temperature much exceeds the radiative equilibrium temperature at that point, the shock-heated gas in the postshock region cools rapidly (compared to the postshock flow time) by emitting radiation. 

Given the thinness of the radiation-emitting shock layer compared to the planetary radius, we assume the shock on the planetary surface to be one-dimensional as in \citet{Aoyama+2018}.
Then, the shock structure is mainly determined by two input parameters,
namely the hydrogen proton number density $n_0$ and the velocity $v_0$ before the shock.
The preshock temperature of the gas $T_0$ hardly affects the shock properties, as we discuss below.

As in \citet{Aoyama+2018} and e.g.\ \citet{Shapiro+Kang1987} or \citet{Kwan+Fischer2011},
$n_0$ is defined as the immediate preshock number density of \textit{hydrogen nucleons (protons)}
(i.e., contained in H$_2$, H~\textsc{i}, and H$^+$ taken together).
Thus, contrary to the definition, common in the stellar-structure literature (e.g.\ \citealp{Hansen+2004}), of $n$ as the number density of all particles,
our $n_0$ is independent of the dissociation and ionization degrees of hydrogen.
This definition implies that $n_0$ is related to the (total) preshock gas mass density $\rho_0$ by
\begin{equation}
\label{eq:rho0 n0}
 X \rho_0 = n_0 \mH,
\end{equation}
where $X$ is the hydrogen mass fraction and $\mH$ is the mass of a hydrogen atom.
We use number ratios given by
$\textrm{H}:\textrm{He}:\textrm{C}:\textrm{O}=1:10^{-1.07}:10^{-3.48}:10^{-3.18}$ \citep{Allen2000}.
As in \citet{Aoyama+2018} we do not consider the other elements, which are neglible.
Thus the hydrogen, helium, and metal mass fractions are respectively $X=0.738$, $Y=0.251$,
and $Z=1-X-Y=0.011$.

In the chemistry module, the abundances of the species are defined by $y_i$ (see also \citealp{Iida+2001}).
The quantity $y_i$ is the relative abundance (in number) of species $i$ (of any particle)
with respect to the number of hydrogen protons.
For example, pure H$_2$ has $y_{\mathrm{H}_2}=0.5$.
Defining the total $\yt=\sum{y_i}$,
the usual total number density of all particles is given by $n=\yt n_0$;
the number density of particles of species $i$ is $n_i = y_i n_0$.
With the mean weight per particle given by
$\mu_0=\sum y_im_i/\yt$ for particle masses $m_i\mH$,
we have that $\rho_0 = \yt \mu_0 \mH n_0$, implying $X=1/(\yt\mu_0)$.

The shock input parameters $(n_0,v_0)$ are related to the macrophysical, planet formation parameters $(\Mdot,\MP,\RP,\ff)$
by
\begin{align}
 v_0 &= \sqrt{\frac{2G\MP}{\RP}} \label{eq:v0}\\
 n_0 &= \frac{X\Mdot}{4\pi\RP^2 \ff \mH v_0 } \label{eq:n0 a}\\
     &=\frac{X\Mdot}{\sqrt{32G}\pi\mH\ff\sqrt{\MP\RP^3}},\label{eq:n0 b}
\end{align}
where $G$ is the gravitational constant,
$\MP$ is the planet mass, $\RP$ is the planet radius, $\Mdot$ is the accretion rate,
and
$\ff$ is the filling factor of the shock on the planet surface. %
Equations~(\ref{eq:v0})--(\ref{eq:n0 b}) are valid in the limit that the accretion radius $\RAkk\gg\RP$,
with the gas free-falling from $\RAkk\sim \RHill$ (Equation~(\ref{eq:Racc}); \citealp{Bodenheimer+2000}).
Especially at low masses $\MP\lesssim1~\MJ$, the accreting gas could be falling in at less than the free-fall velocity, depending on the thermodynamics and the angular momentum conservation \citep[][but note that the finite smoothing length of the latter two works imply that converged results have not quite been reached yet]{b19,schulik19,schulik20}.
However, the difference should be small, and for the classical 1D models that we use as an approximation, the limit $\RAkk\gg\RP$ usually holds for $\MP\gtrsim1~\MJ$, especially for planets forming at large distances.
Similarly, our model is also applicable when the gas falls from the inner edge of the CPD, but, in such a case, the $v_0$ and the estimated \LHa is smaller by a factor of a few than the results in this paper.
Inserting typical values for stars and planets, the typical preshock number density is larger in the planetary than in the stellar case by about a factor of~100 %
\citep[see Eq.~(15) in]{zhu15}
but one should keep in mind that the parameter space is large. %

The assumption here is that all of the accreting gas is available for a shock, whether this turns out to produce \Ha\ or not. In reality, some fraction of the accreting gas could be added to the planet through boundary-layer accretion (BLA; e.g., \citealp{Kenyon+Hartmann1987,Kley1989a,Dong+2020}), which does not feature supersonic radial velocities. In this scenario, the temperature in the boundary layer would not be high enough for \Ha\ to be emitted. Thus converting an observed \Ha\ luminosity to planetary parameters such as accretion rate and mass needs to assume something about the fraction of the incoming gas that is able to produce \Ha. Put differently, a measured \Ha\ luminosity yields an estimate of the \Ha-emitting accretion rate, while the total accretion rate could be higher. However, it is not clear how likely BLA is in the planetary case; \citet{Owen+Menou2016} argue for BLA but this is based, through the \citet{christensen09} scaling, on magnetic field strengths appropriate of old and faint planets ($B\approx0.03$--0.06~kG), not forming nor young, high-luminosity \citep{mordasini17} objects, which could have $B\sim1$~kG (\citealp{katarzy16}). Thus, assuming that all the accreting gas can undergo a shock seems reasonable, but high-resolution studies are required to help settle the question.

\begin{figure*}
\begin{center}
\includegraphics[width=0.49\textwidth]{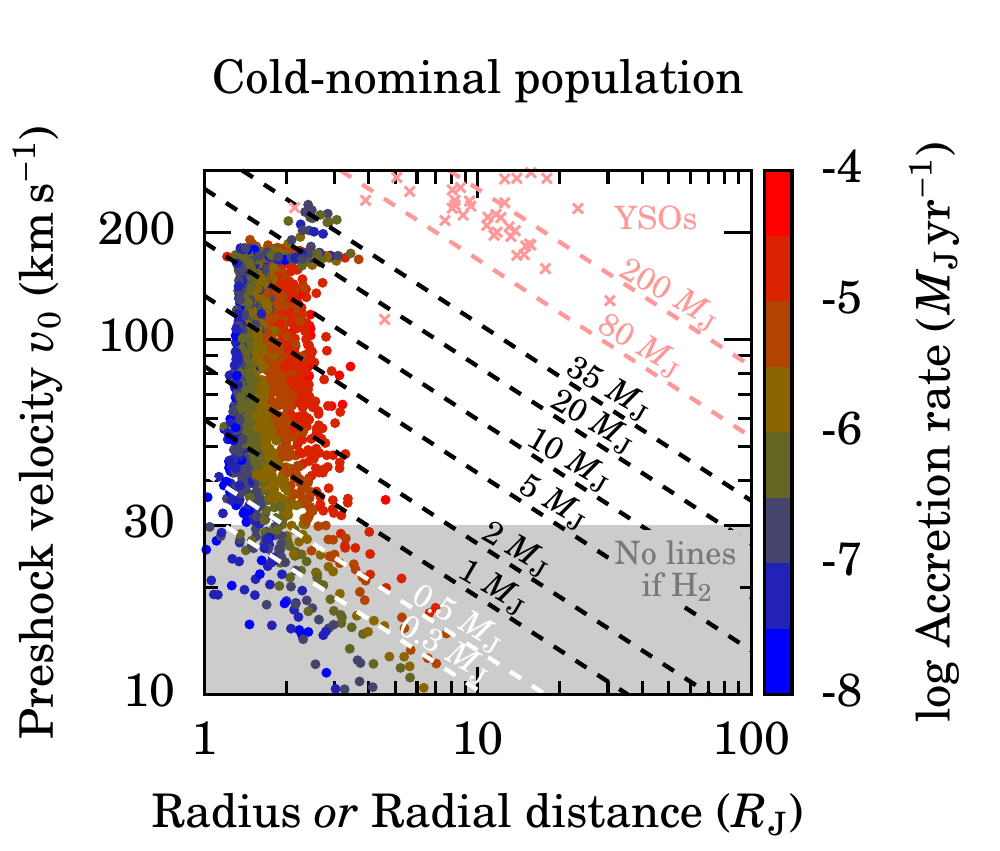}
\includegraphics[width=0.49\textwidth]{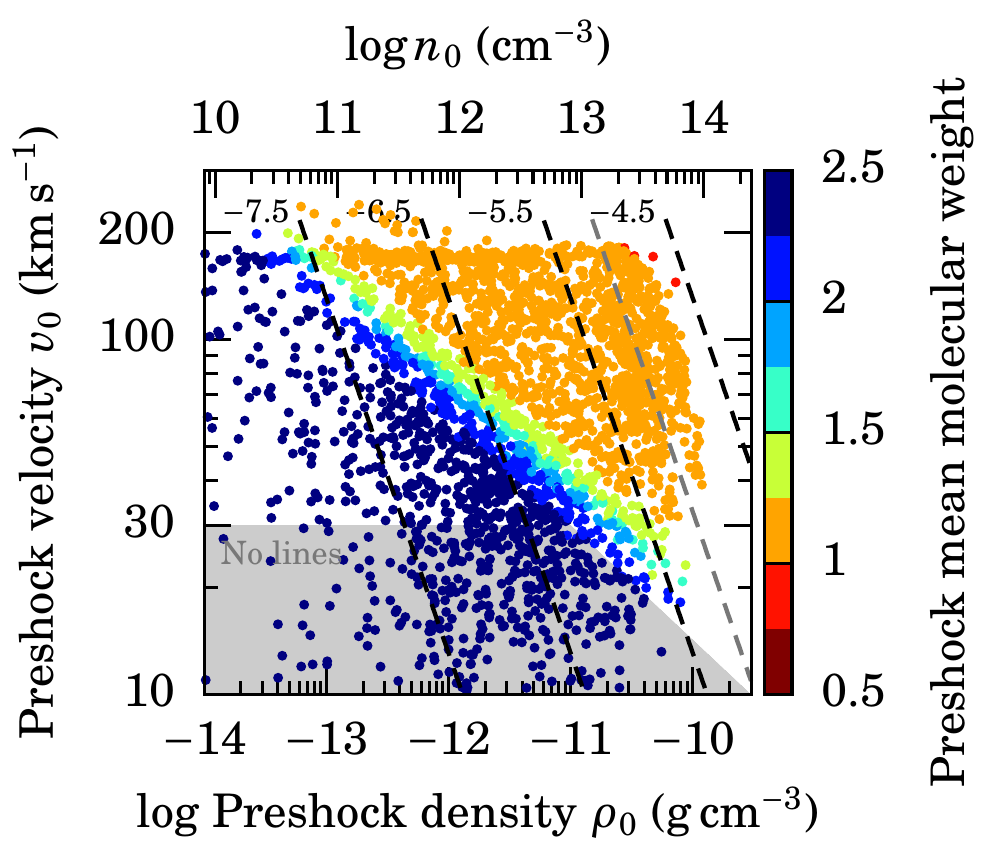}
\end{center}
\caption{
Preshock velocities of forming planets.
\textit{Left panel}:
Preshock velocity %
against planet radius %
or radial distance (for a shock on the CPD).
Bern cold-nominal population synthesis planets with $\RAkk\geqslant10\RP$
are shown, coloured by \Mdot,
and the free-fall velocity (Equation~(\ref{eq:v0})) is drawn %
for different planet masses
(dashed lines).
Pink crosses are from the \protect\citet{alcal17} YSO sample. Below $v_0\approx30~\kms$, there is no line emission (\Ha, \Hb, etc.) if the gas comes in molecular (gray region).
\textit{Right panel}:
Velocity against mass density %
or hydrogen proton number density (\textit{lower, upper axes}; Equation~(\ref{eq:rho0 n0})).
Population synthesis points are shown, %
colored by the preshock $\mu_0$ (for $\mu_0\approx1.3$, only half the points are shown).
Dashed lines: constant $\log\Mdot=-7.5$ to~$-4.5$ ($\MJ\,\mathrm{yr}^{-1}$)
and $\RP=1.5$ (black) or $3~\RJ$ (gray).
We fix $\ffill=1$.
}
\label{fig:n0v0}
\end{figure*}

In Figure~\ref{fig:n0v0} we plot the preshock velocity as a function of \Mdot (colour), mass (to be read off from the dashed lines), and radius planet ($x$ axis) for the cold-nominal Bern population synthesis \citep{morda12_II,mordasini17}.
The hot population objects (not shown) have larger radii but an otherwise similar distribution.
All snapshots are included but we restrict $\Mdot\geqslant10^{-8}~\MdotUJ$ and $\MP\geqslant0.1~\MJ$.
Planets with $\RP<10\RAkk$ are excluded, but this affects only a small fraction of the points; for example, planets with $\MP\approx0.1~MJ$ and with a small Hill sphere might be in the attached phase even at high \Mdot. Essentially all points above 0.5~$\MJ$ have $\RP>10\RAkk$.
Thus, $v_0$ is indeed given by Equation~(\ref{eq:v0}) to 5\,\%\ or better.

Figure~\ref{fig:n0v0}a shows that
typical preshock velocities are $v_0\approx50$--200~km\,s$^{-1}$,  %
and that there is a maximal $v_0\approx170~\kms$
for masses between $\MP\approx13$~and $\approx35$~$\MJ$.
(For the hot population, it is also $v_0\approx170~\kms$ but with a thicker spread of the horizontal portion there, down to $v_0\approx150~\kms$.)
The points along this locus of approximately constant $v_0$ (which implies that $\MP\propto\RP$ approximately)
are objects which have not yet burned their deuterium and are all younger than 10~Myr.
These points span a range of ages and masses. Thus their distribution is different from the D-burning ``shoulder'' in plots of the radius of an object of a given mass as a function of time, in which the D~burning stalls the cooling \citep[e.g.,][and references therein]{moll12}.
Also, these points with $\MP\propto\RP$ are not in conflict with the classical result $M\propto R^{-1/3}$ in that mass regime \citep{zs69} since that holds for ``zero-temperature'' (i.e., degenerate) objects, whereas the objects here are still forming and thus hotter, also in the sense of not being completely degenerate.
Older, post-deuterium-burning objects have a smaller radius and thus a higher $v_0$ but they are not plotted because of the selection on $\Mdot$.
This group of constant-$v_0$ points could show up as a pile-up in a histogram of $v_0$ for forming planets.

Figure~\ref{fig:n0v0}a compares preshock velocities and object radii to young stellar objects (YSOs). As an example, the properties of some of the targets in the \citet{alcal17} sample are shown (crosses), which covers %
$M\approx10$--180~$\MJ$ and %
$R\approx2$--30~$\RJ$, with a preshock velocity $\vff\approx150$--$500~\kms$.
There is no significant overlap between the region occupied by planets and the \citet{alcal17} sample.

Next, Figure~\ref{fig:n0v0}b situates the planet accretion shock in the input parameter space of preshock velocity $v_0$ and hydrogen proton density $n_0$.
The latter is in the range
$n_0=10^{10}$--$10^{14}~\mathrm{cm}^{-3}$,
which
corresponds to $\rho_0 \sim10^{-14}$--$10^{-10}$~g\,cm$^{-3}$.

We also indicate
the preshock state (molecular, atomic, or ionic) of the hydrogen %
in Figure~\ref{fig:n0v0}b.
We calculate
the mean molecular weight $\mu_0$ from $\rho_0$ and $T_0$
($x$ axis and Equation~(\ref{eq:T0 historical...}), respectively)
and the Saha equation.
For $\ffill=1$, the incoming hydrogen is molecular at low densities or velocities and atomic above this. In a few high-velocity cases, it arrives at the shock significantly ionized (bright red points).
Overall, according to the $(\Mdot,\MP,\RP)$ combinations found in the Bern population synthesis, for large filling factors $\ffill\approx1$ the gas reaches the planet in an atomic form in the majority of cases.
Below $v_0\approx30~\kms$, the hydrogen is usually molecular, so that there will be no line emission (\Ha, \Hb, etc.) because the shock energy is used up to dissociate the molecules \citep{Aoyama+2018}. %
However, for the cases where the preshock gas is atomic, there is no $v_0$ emission threshold and the \Ha flux continuously gets weaker with decreasing $v_0$. This is relevant only for a few points.
The limit of $v_0=30~\kms$ is indicated by the gray areas in Figure~\ref{fig:n0v0}a and~b,
with the slanted right edge drawn approximately in panel~b.
\srevise{Note that the shock radiation can change the preshock hydrogen state compared to what the Saha equation predicts from the local $(\rho,T)$. Especially, \Lya\ absorption by the preshock atomic hydrogen decreases the cooling efficiency via \Lya\ radiation. It could enhance the non-Lyman hydrogen lines such as \Ha\ when $v_0$ is close to the critical velocity,
but this is not taken into account.}

Finally, we note in passing that the minimum velocity for \Ha of $v_0\approx30~\kms$ puts a constraint on the CPD size needed to have \Ha in the CPD-shock case studied in \citet{Aoyama+2018}.
Figure~\ref{fig:n0v0} shows that if the inner edge of the CPD is further out than 3--5 (7--15) planetary radii for 2-~(5)-$\MJ$ planets, there will not be any \Ha emission. For his fiducial values, \citet{batygin18} found that the magnetospheric truncation radius $\Rtrunc\approx2\RP$. Thus the requirement $v_0>30~\kms$ should be easy to meet already at a few $\MJ$ and all the more for higher masses. This implies that the CPD could be an emitter of \Ha. We discuss how to distinguish the emission from a CPD from the one from the planet surface in Section~\ref{sec:distinguishsurfaceshockCPD}.

\begin{figure*}
\begin{center}
\includegraphics[width=0.49\textwidth]{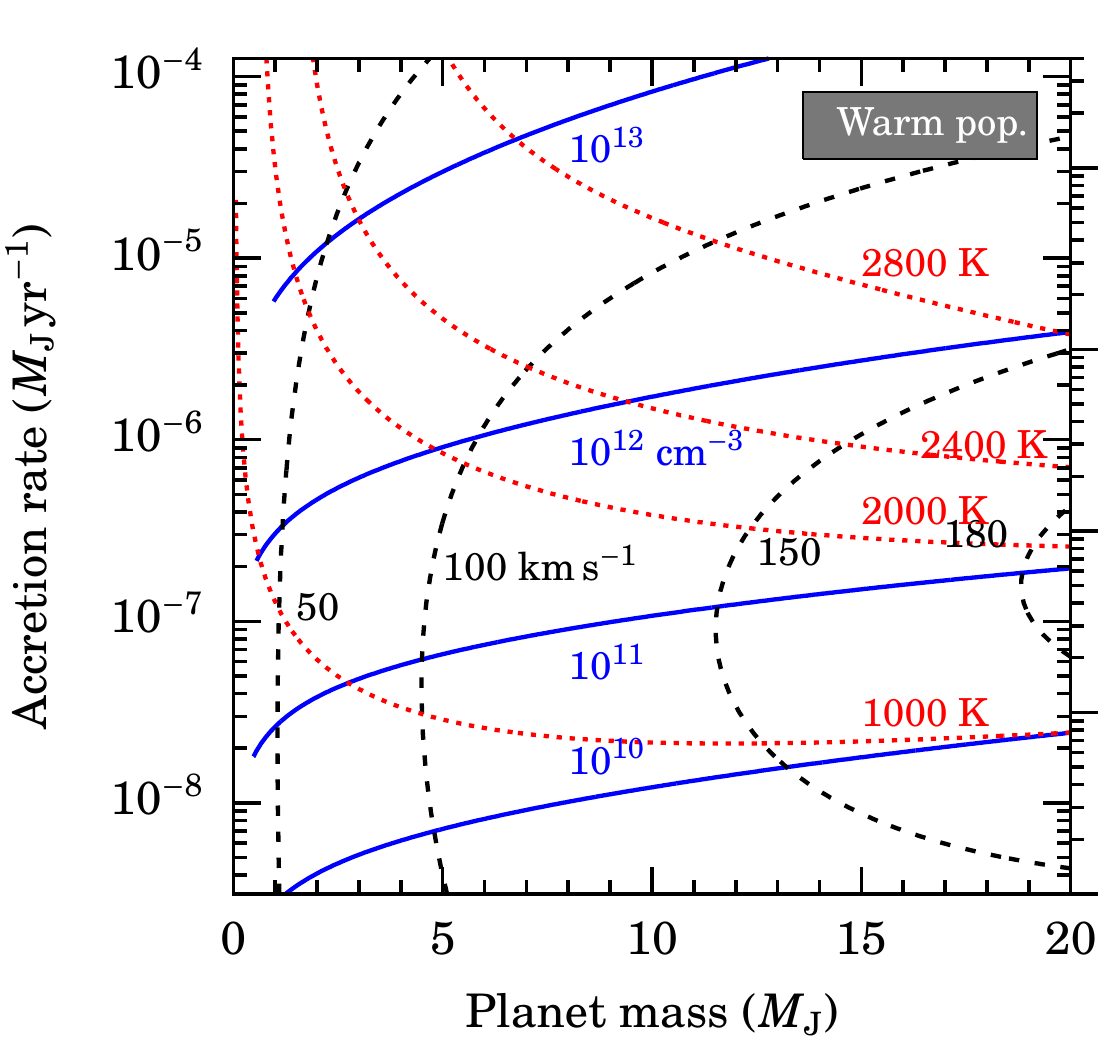}
\includegraphics[width=0.49\textwidth]{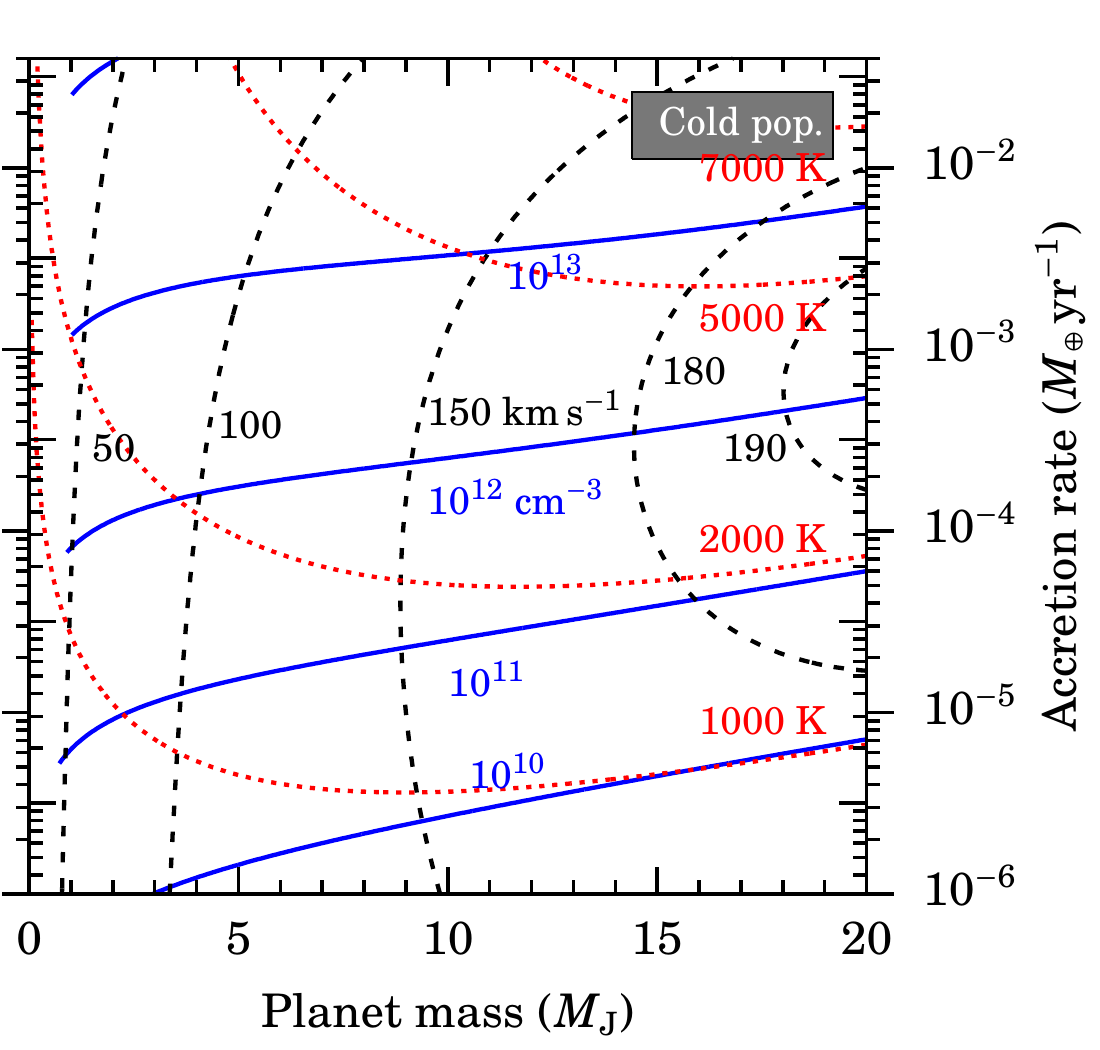}
\end{center}
\caption{
Contours of $n_0=10^{10}$--$10^{14}$~cm$^{-3}$ (blue solid lines), $v_0=50$--190~$\kms$ (black dashed), and $T_0=1000$--7000~K (red dotted)
from Equations~(\ref{eq:v0})--(\ref{eq:n0 a}) and~(\ref{eq:T0 historical...}) in the mass--accretion rate plane.
We set $\ffill=1$ and use the radius fitted
to the population synthesis results in the warm (\textit{left panel}) and cold (\textit{right panel}) cases.
The highest $v_0$, $n_0$, and $T_0$ contours are missing in the warm population due to the larger radii.
The inverse relations are shown in Figure~\ref{fig:inverse n0v0}.
Note that $1~\MdotUJ\approx3\times10^{-4}~\MdotU$.
}
\label{fig:n0v0 contours}
\end{figure*}

Figure~\ref{fig:n0v0 contours} displays contours of $n_0$ and $v_0$ in the $\Mdot$--$\MP$ plane
using the $\RP(\Mdot,\MP)$ fit (Equation~(\ref{eq:R fit})) for the warm and the cold population.
(The inverse relations are shown in Figure~\ref{fig:inverse n0v0}.)
We use $\ffill=1$.
The $v_0$ contours depend on $\Mdot$ because of the dependence of the radius on $\Mdot$.
This leads even to a non-monotonic behaviour of $v_0$ with $\Mdot$,
with, in the warm (cold) population, a maximum around $\Mdot\approx3\times10^{-5}~\MdotU$ ($\Mdot\approx10^{-3}~\MdotU$).
The maximum preshock velocity is for both populations roughly
$v_0\approx100~\kms$ ($v_0=180~\kms$) for $\MP\approx5$ ($\MP\approx15$--$20~\MJ$).
Since the radii are smaller in the cold population, the velocities are slightly
higher at a given mass but only by some tens of kilometers per second.
The curving of the $v_0$ contours at high $v_0$ (see Figure~\ref{fig:n0v0 contours})
means that a measurement of the preshock velocity $v_0$,
for instance through the Doppler broadening of the emission lines,
can be explained only by a limited range of accretion rates, assuming that a rough upper limit
on the mass exists (e.g.\ from imaging or dynamical arguments).

\subsubsection{Shock conditions}

We assume that at the shock, the preheated gas \citep{Marleau+2019b} undergoes
a hydrodynamical shock before cooling down radiatively.  %
This corresponds effectively
to the Zel'dovich spike \citep[see][]{Vaytet+2013}, and our actual computations
begin directly after the hydrodynamical shock, i.e., roughly at the tip of the Zel'dovich spike.
To obtain this immediate postshock state, we use
the classical Rankine--Hugoniot shock jump conditions, which reflect
mass, momentum, and energy conservation.
This is valid because the time needed for the gas to cross the hydrodynamical shock thickness,
of the order of a particle mean free path,
is much less than the cooling timescale of the gas.
In \citet{Marleau+2017,Marleau+2019b}, ``the shock'' referred to both the hydrodynamic jump
and the postshock cooling region; here we refer by ``shock'' \textit{only} to the hydrodynamic jump,
with the postshock cooling region the focus of this work.
The Rankine--Hugoniot relations read:
\begin{align}
 \rho_1 &= \frac{(\gamma+1)\Mach^2}{(\gamma-1)\Mach^2+2}\rho_0\\
 v_1 &= \frac{\rho_0 v_0}{\rho_1}\\
 P_1 &= \frac{2\gamma\Mach^2-(\gamma-1)}{\gamma+1}P_0\\
     &=\frac{2}{\gamma+1}\Pram-\frac{\gamma-1}{\gamma+1}P_0,
\end{align}
with the subscript ``0'' denoting the preshock and ``1'' the postshock state.
The ram pressure is $\Pram\equiv\rho_0{v_0}^2$.
The upstream Mach number is $\Mach=v_0/\cs$,
with $\cs=\sqrt{\Gamma_{1,0} \kB T_0/(\mu_0\mH)}$,
where $\Gamma_1=\left(\partial\ln P/\partial\ln\rho\right)_s$ is the first adiabatic index
and $s$ the entropy.
Across the hydrodynamic jump (but not below, in the main part of our computations),
we assume that %
the abundances and thus $\mu$ and $\Gamma_1$
remain constant. This is justified if the gas undergoes the jump (a few mean free paths thick)
on a timescale shorter than the chemical reaction time.
For simplicity,
we take $\Gamma_1=\gamma$, where $\gamma$ is the ratio of specific heats,
and we specify its value below (see Equation~(\ref{eq:gamma})).
The high-Mach number limits for $\rho_1$ and $P_1$ are
\begin{align}
 \rho_1 &= \frac{\gamma+1}{\gamma-1}\rho_0 \label{eq:rho1 high Mach}\\
 P_1 &= \frac{2\gamma\Mach^2}{\gamma+1}P_0\\
     &=\frac{2}{\gamma+1}\Pram,\label{eq:P1 high Mach}
\end{align}
which implies, still in the limit $\Mach\gg1$,
\begin{align}
 \label{eq:T1}
T_1 &= \frac{\mu_0\mH}{\kB}\frac{2(\gamma-1)}{\left(\gamma+1\right)^2} {v_0}^2\\
   &\approx 4\times10^5 \left(\frac{\MP}{10~\MJ}\right) \left(\frac{2~\RJ}{\RP}\right)~\mathrm{K},
\end{align}
taking $\mu=1.23$ and $\gamma=1.43$ (generally appropriate for the incoming gas; see Figure~\ref{fig:n0v0}) for the second line.
Expressions for the case of different $\gamma$ and $\mu$ values left and right of the shock
can be found in Equation~(4.17ff) of \citet{drake06}.
This is the same physics as for stars (see Equation~(4) of \citealt{Hartmann+2016}).
Note that $T_1$ is a \textit{non-equilibrium} temperature,
which holds over only a very small temporal and spatial scale relative to any other relevant scale.
(In the example in Figure~\ref{fig:postshock}, this is $\sim10^{-2}$~s and $\sim~10^{-6}~\RJ$.)
It is thus in no way an effective temperature $\Teff$ nor an equilibrium gas temperature. This non-LTE effect usually cannot be resolved in full radiation-hydrodynamical simulations because of the vast differences in scales.

The preshock temperature $T_0$ is needed
to set the preshock pressure $P_0$ and the preshock Mach number $\Mach$,
as well as the chemical abundances before and therefore also directly below the shock.
Using gray radiation transfer, \citet{Marleau+2017,Marleau+2019b} found that the radiation and the gas ahead of the shock were able to equilibrate. This is due to the sufficiently high Planck opacity of the gas or, for lower shock temperatures, of the dust. Therefore, we calculate the preshock temperature from $(\rho_0,v_0)$ by
\begin{equation}
 \label{eq:T0 historical...}
 T_0 = \sqrt[4]{\frac{\rho_0 v_0^3}{2\sigma}}. %
\end{equation}
Because it has $\sigma$ and not $ac=4\sigma$ on the denominator,
where $a$ is the radiation constant and $c$ the speed of light,
this expression
is higher by a factor of $4^{1/4}\approx1.4$  %
than the equilibrium shock temperature obtained analytically and numerically
by \citet[][see their Equation~(6)]{Marleau+2019b}\footnote{We
	noticed this difference only at a later stage of this work.
	Since it barely changes the results, a correction of this factor
	will be deferred to the next iteration of our models.}.
It also ignores the negligible contribution of the internal luminosity in setting
the planet's surface temperature; see e.g.\ Equation~(32) of \citet{Marleau+2019b}.
However, since the shock is strong (i.e., the Mach number is large; \citealp{Marleau+2019b}),
both $\Mach$ and $P_0$ barely affect the postshock quantities,
as Equations~(\ref{eq:rho1 high Mach}) and~(\ref{eq:P1 high Mach}) show.
Thus it is inconsequential that
\citet{Aoyama+2018} assumed a constant $T_0=200$~K.
Also, the initial postshock composition
does not affect the radiative fluxes by more than $\sim 1$\,\%,
as we have verified (not shown) by varying $T_0$ even by a factor of ten.

Finally, the adiabatic index for the mixture is given by
\begin{equation}
 \label{eq:gamma}
  \frac{1}{\gamma-1} = \sum{\frac{y_i/\yt}{\gamma_i-1}},
\end{equation}
with $\gamma=5/3$ for H, He, C, and O, and $\gamma=7/5$ for $\mathrm{H}_2$.
It enters into the Rankine--Hugoniot equations and in
the time-dependent energy equation through $P=(\gamma-1)E$, where $P$ is the pressure and $E$ the internal energy (see Equation~(7) of \citet{Aoyama+2018}).

\subsubsection{Postshock flow}
 \label{PostShock}

We assume a time-independent  %
plane-parallel one-dimensional flow after the shock. Then, the gas flows with conserved mass flux and momentum flux, implying
\begin{align}
\rho v &= \rho_1 v_1 \\  %
\rho v^2 + P &= \rho_1 v_1^2 + P_1.
\end{align}
Chemical reactions including electron level transitions are the external energy source.
Therefore, the internal-energy volume density of the gas $E$ is not conserved but evolves according to
\begin{equation}
 \label{eq:dEdt}
    \frac{dE}{dt} = \left(\Gamma - \Lambda \right) + \left[ \frac{P+E}{\rho} \frac{d\rho}{dt} \right],
\end{equation}
where $\Gamma$ and $\Lambda$ are the heating and cooling rates per unit volume, respectively.
Note that, in a 1-D flow, temporal differentiation is easily converted into spatial differentiation with flow velocity $v$.

In this study, the coolants are the dissociation of molecular hydrogen, the collisional excitation and ionization of atomic hydrogen, and the emission of radiation by CO, OH, and $\mathrm{H_2O}$. The heat sources are the formation of molecular hydrogen as well as the collisional de-excitation and collisional recombination of atomic hydrogen. For detailed expressions, see \citet{Aoyama+2018}.

\subsubsection{Radiative transfer}
 \label{RadiativeTransfer}

We consider electron level transitions between ten levels of neutral hydrogen and the ionized state. We numerically calculate the radiative transfer of 45 lines and ten recombination continua with de-exciting transitions. To integrate the flux, we use the two-stream approximation, assuming a plane-parallel 1-D flow.
We iterated the hydrodynamic simulation and the radiative transfer until the \Ha\ flux converges. Note that since the \Lya\ still changes when the other lines converge and the iteration stops, the \Lya\ intensity is less reliable, in this model.
Detailed expression and equations are given in \citet{Aoyama+2018}.

Finally, given the assumed geometry described in Section~\ref{sec:parameters}, the luminosity of hydrogen lines and recombination continua emitted from the shock-heated gas is
\begin{equation}
 \label{eq:L from F}
    L=4\pi \RP^2 \ff F,
\end{equation}
where $F$ is the photon energy flux at the shock and is the result of the radiative transfer in the postshock gas flow.

\section{Relation between our models and Case~B (Baker \&\ Menzel 1938; Storey \&\ Hummer 1995)}
 \label{sec:cfSH95}

The work of
\citet{Hummer+Storey1992} and \citet[][hereafter \citetalias{Storey+Hummer1995}]{Storey+Hummer1995}, based on the Case~B model for radiative recombination and ionization \citep{Baker+Menzel1938},
was developed for regions illuminated by a photoionising source such as planetary nebulae and \HII regions.
The emissivity tables of \citetalias{Storey+Hummer1995}
have been used to analyze accretion line intensities or their ratios in the context of CTTS \citep[e.g.,][]{k11}.
Recently,
\citet{szul20} used \citetalias{Storey+Hummer1995}
to calculate hydrogen line emissivities from accreting planets.
\srevise{However, for shock emission, \citetalias{Storey+Hummer1995} is  not applicable.}

The main reason why the approach of \citetalias{Storey+Hummer1995} cannot be used for the planetary- or CPD-surface shock is
\srevise{their treatment of the collisional excitation from the ground state. 
\citetalias{Storey+Hummer1995} %
\textit{fixed} the 
ground state population to a low level, so that 
collisional excitations from the ground are negligible relative to the recombination from the ionized state.
However, in our results, the ground state population is comparable to the abundance of ionized hydrogen %
in the region where the line is mainly emitted
(see Section~\ref{sec:Postshock structure} and Figure~\ref{fig:postshock}). 
In some cases, collisional excitations are not negligible but can be the main source of excited hydrogen (see also Section~4 in \citealp{Hummer+Storey1992} for the critical ionization fraction).
Indeed, as $v_0$ decreases, the ionization fraction becomes smaller and ground-state hydrogen more abundant than ionized hydrogen \citep{Aoyama+2018}.%
}

There is a further, secondary, issue with \citetalias{Storey+Hummer1995} 
\srevise{for some cases of emission linked to accretion onto planets or CTTSs,}
concerning the optical depth of the emission lines.
\srevise{A fundamental assumption of \citetalias{Storey+Hummer1995} is} that the gas is optically thick to photons from the Lyman series but that all other transitions are optically thin\srevise{, which is appropriate for their original objects of interests, e.g., \HII regions.}
As mentioned in Section~\ref{sec:distinguish} and shown in Figure~\ref{fig:SED_mass}, in several cases the \Ha\ line is not optically thin in the postshock region, especially towards high preshock number densities $n_0$.
\srevise{This problem of optical thickness is also reported from some recent detailed observation of hydrogen lines from CTTSs. In some cases, hydrogen lines indicate they are optically thick}
(\citealp{k11};
see the review and detailed comparisons in \citealp{Edwards+2013,antoniucci17}; see Section~4.3 of \citealt{Rigliaco+2015}).
\srevise{The} fundamental assumption of \citetalias{Storey+Hummer1995} is fulfilled in some case but in several it is not.
\srevise{For the shock emission, there is a third issue that no equilibrium is reached.} %
In the cooling region below the hydrodynamic shock,%
the cooling timescale $\tcool$ is comparable to---and not much longer than---the timescale over which
the electron level populations change, $\tpop$.
\srevise{Since} $\tpop$ is set by collisions, this is a trivial statement: collisions set the populations and at the same time are the mechanism by which the gas cools.
Namely, $\tcool\approx\Eint/\Lambda$, where \Eint is the internal energy of the gas and $\Lambda$ is the cooling rate. In turn, $\Lambda$ is given by the energy difference between the levels\footnote{%
  In our case, the \Lya\ transition is responsible for most of the cooling, such that $\Lambda\approx\LambdaLya$, where \LambdaLya is given by Equation~(B5) of \citealt{Iida+2001}. Then, $\Delta E=10.2$~eV. See Figure~\ref{fig:postshock}d for an example of $\Lambda$.}
$\Delta E$ divided by the timescale for the transition $\tpop$. Thus $\tcool\approx\Eint/\Lambda\approx\Eint/(\Delta E/\tpop)$, such that $\tcool/\tpop\approx\Eint/\Delta E$, which is around unity for $\Delta E\approx10$~eV and the %
$T\approx10^5$~K. 
\srevise{Note that the emission comes mainly from $T < 10^5$~K (see Figure~\ref{fig:postshock}).} %
Thus, the gas cools faster than an equilibrium distribution of electrons could be reached; equilibrium would require $\tpop\ll\tcool$ to let the populations adapt to the changing ambient conditions. 
\srevise{Therefore, any model using time-independent level population, including \citetalias{Storey+Hummer1995}, is inapplicable to this kind of shock emission.}

In summary, there are several reasons why the tables of \citetalias{Storey+Hummer1995} do not apply to the shock emission of accreting planets. This highlights the need for a ``zero-dimensional'' time-dependent NLTE Lagrangian radiation-hydrodynamics model (equivalent to a steady-state 1D Eulerian approach) as we have developed and applied to the CPD-surface shock \citep{Aoyama+2018} and the planet-surface shock (this work).

\section{Inverse relation between the shock-microphysical and planet-formation (macrophysical) parameters}
 \label{sec:inverse n0v0}

\begin{figure*}
\includegraphics[width=0.49\textwidth]{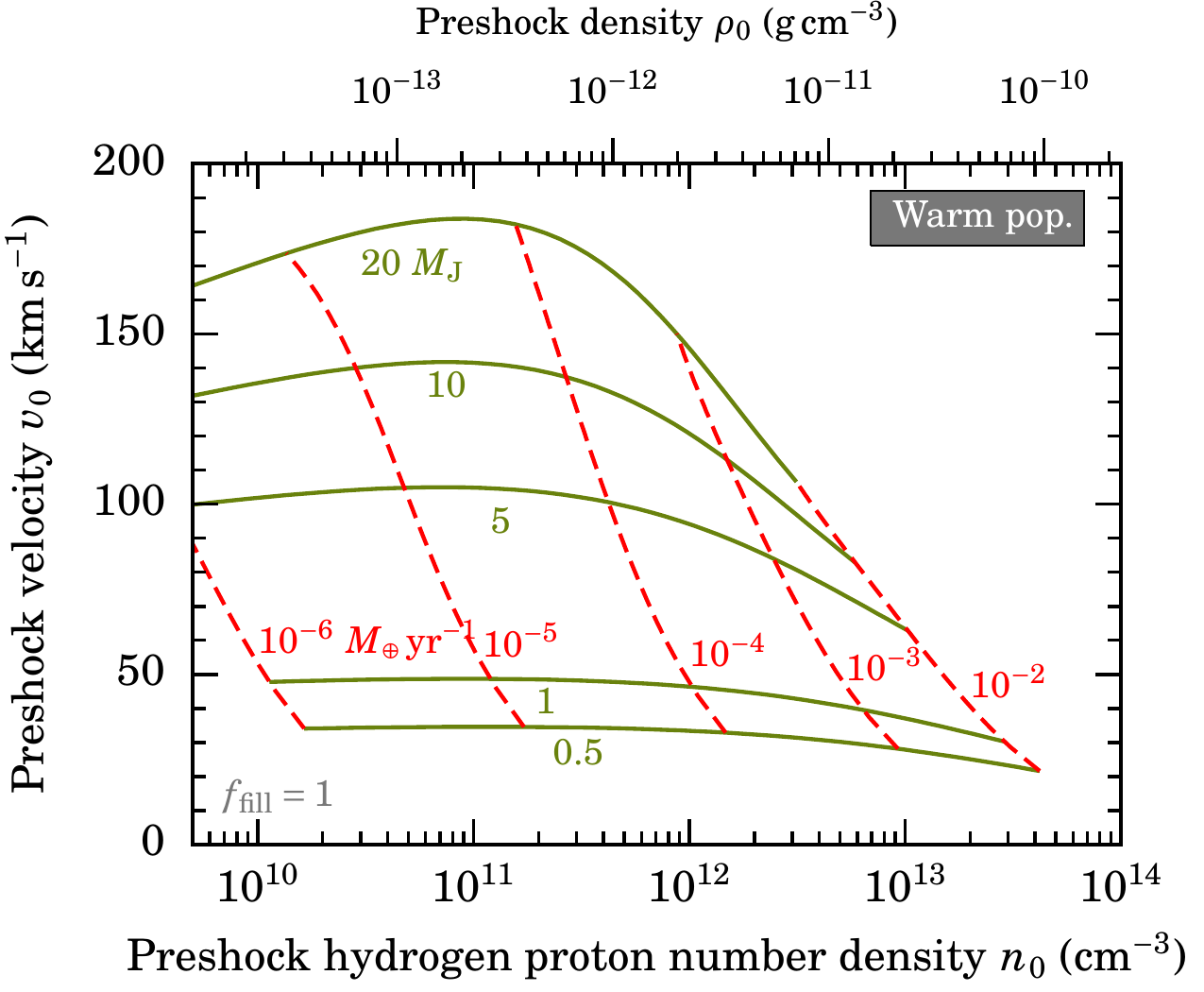}
\includegraphics[width=0.49\textwidth]{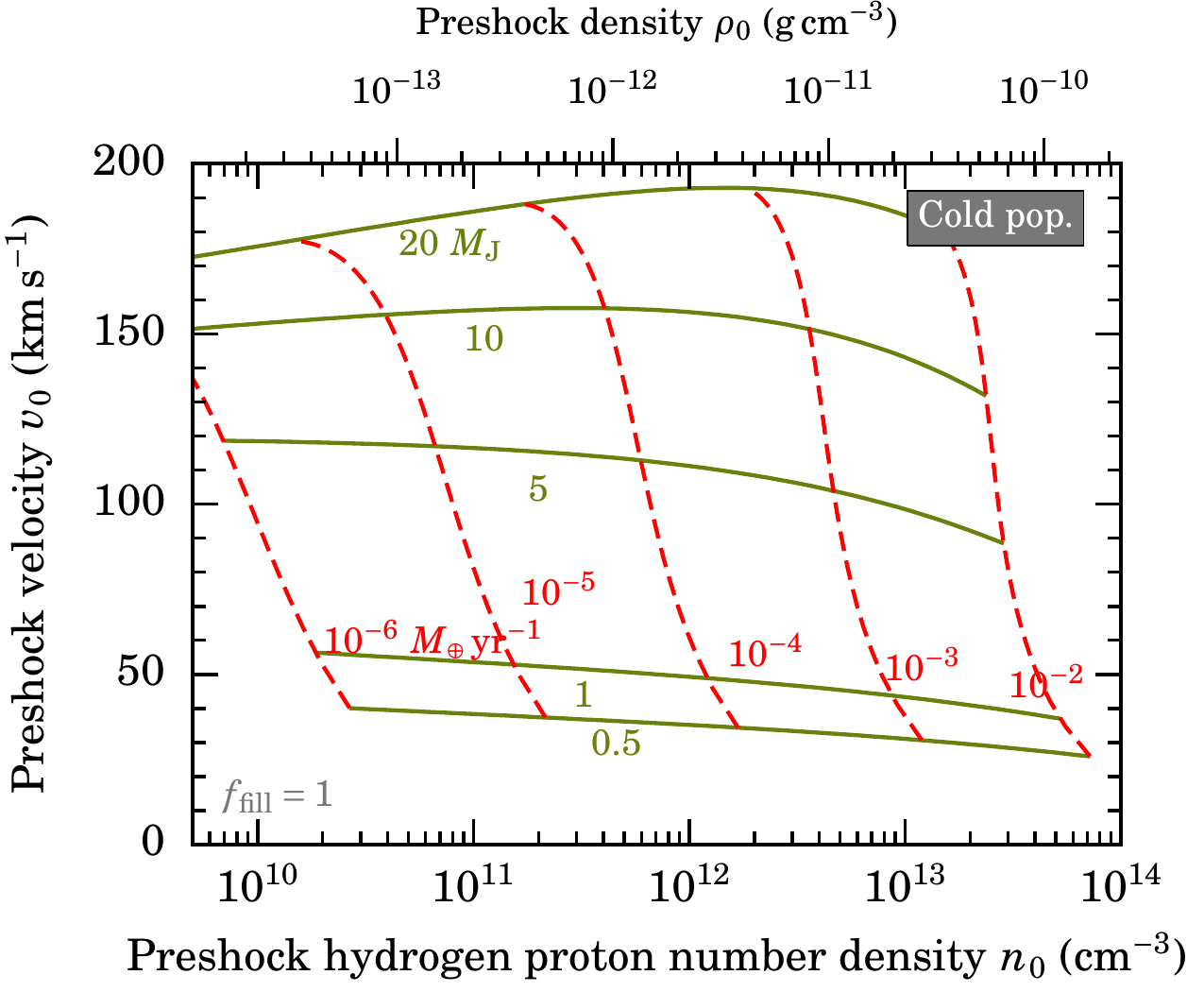}\\
\caption{
Shock parameter space $(n_0,v_0)$ or $(\rho_0,v_0)$ covered by our grid (Equation~(\ref{eq:parspac})), using Equations~(\ref{eq:v0})--(\ref{eq:n0 b})
and the radius fit of Equation~\ref{eq:R fit} for the warm (\textit{left panel})
 and the cold (\textit{right}) population respectively.
Lines of constant $\Mdot$ (dashed) and $\MP$ (solid) are labeled.
We fix $\ffill=1$.}
\label{fig:inverse n0v0}
\end{figure*}

For completeness, we show in Figure~\ref{fig:inverse n0v0} lines
of constant $\Mdot$ and $\MP$ in the $(n_0,v_0)$ or $(n_0,\rho_0)$ plane.
In the warm case, because of the larger radii, the upper right corner
(high preshock density and velocity) is not reached, contrary to the cold case.
Except for this, however, in both cases the same part of parameter space is covered.

\section{H alpha luminosity as a function of accretion rate and mass: cold-start fit}
\label{sec:HaContourboth}

\begin{figure}
    \centering
    \includegraphics[width=0.49\textwidth]{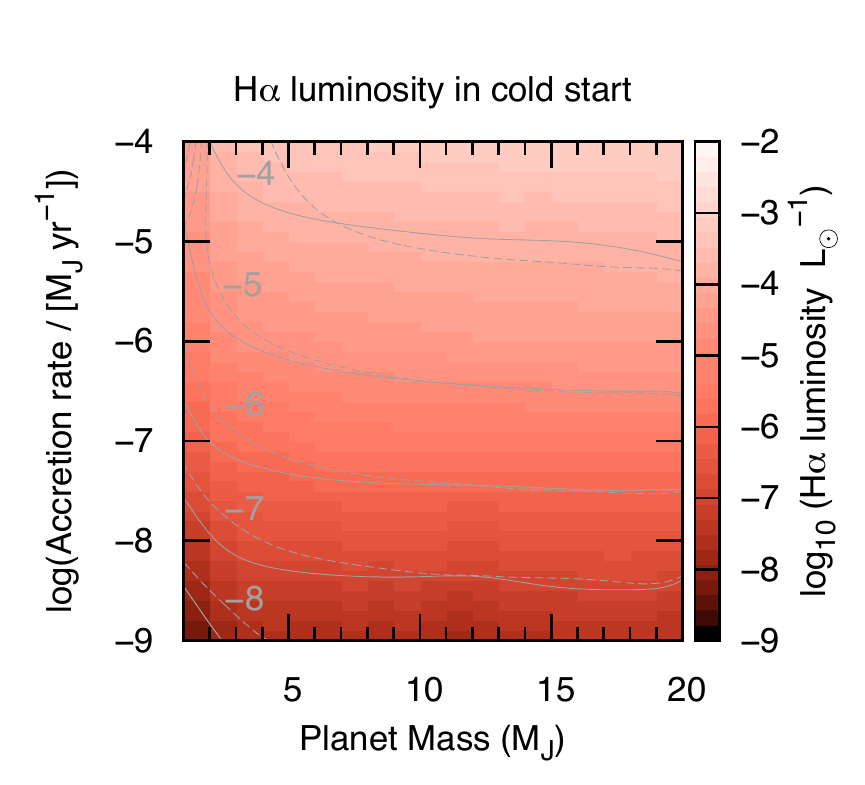}
    \caption{Same as Figure~\ref{fig:HaContour} but for the cold-start radius fit.
    The dashed gray contours show the hot-start results of Figure~\ref{fig:HaContour}.}
    \label{fig:HaContourboth}
\end{figure}

In Figure~\ref{fig:HaContourboth} we show contours as in Figure~\ref{fig:HaContour}
but for the cold-start radii (colour and solid contours).
They are very similar to the contours for the hot-start case.

\section[H alpha luminosity derived from the data of Wagner et al. (2018)]{\Ha\ luminosity derived from the data of Wagner et al. (2018)}
\label{sec:Wagner}

\citet{Wagner+2018} do not report the \Ha\ luminosity of \PDSb\ explicitly but they write that they followed the approach of \citet{Close+2014}.
Thus their luminosity, assuming isotropic emission, is given by
\begin{equation}
 \LHa = 4\pi D^2 \times 10^{\AR} \times C_\mathrm{H\,\alpha} 10^{-R_\mathrm{A}/2.5} V_0 W_\mathrm{f},
\label{eq:PDS70L}
\end{equation}
where $C_\mathrm{H\,\alpha}=(1.14\pm0.47)\times10^{-3}$ is the contrast of the \Ha\ signal at the planet's position to the signal from the primary star in the adjacent continuum (the R band; \citealp{Wagner+2018}), $\AR$ is the extinction in the R band,
$D=113.43\pm 0.52$~pc is the distance of the system \citep{GaiaBrown2018}, $V_0=2.339\times10^{-5}~\mathrm{erg\,s^{-1}\,cm^{-2}\,\upmu m^{-1}}$ is the zero-point of the Vega magnitude system in the MagAO \Ha\ filter and $W_\mathrm{f}=0.006~\mathrm{\upmu m}$ is the filter width \citep{Close+2014},
and $R_\mathrm{A}=11.7\pm0.4$~mag is the R band magnitude of \PDS \citep{Henden+2015,Wagner+2018}, respectively. Taking the case of no extinction ($\AR=0$~mag; see below), we obtain for the companion
\begin{equation}
 \label{eq:LHaClose}
    \LHa=(1.4\pm 0.6)\times 10^{-6}~\LSun. %
\end{equation}
The relative error on this \LHa is dominated by the relative uncertainty on the contrast  ($C_\mathrm{H\,\alpha}$). For the contrast itself we used value from the ``combined image''.
This value agrees with the value of \citet{Thanathibodee+2019}, $\LHa=(1.3\pm0.7)\times10^{-6}~\LSun$ assuming the same distance.

Equation~(\ref{eq:LHaClose}) is confirmed by Equation~(7) of \citet{Close2020}, which appeared while this work was in preparation. However, \citet{Close2020} mentions an upper limit on $\AR$ of 0.2~mag and uses this as the value of $\AR$, leading to $\LHa=2.0\times10^{-6}~\LSun$. This value of $\AR$ seems too high given the determination of \citet{mueller18} based on fits to the stellar spectrum using the detailed MIST models \citep{dotter16,choi16}, which yielded $\AV=0.05^{+0.05}_{-0.03}$~mag and thus
$\AR=0.06^{+0.06}_{-0.04}$~mag using the \citet{cardelli89} law with $R=3.1$ for typical ISM dust. This agrees with the results of the \texttt{Stilism} statistical tool under \url{https://stilism.obspm.fr} \citep{Lallement+2019},  %
$E(B-V)=0.01^{+0.02}_{-0.01}$~mag in the direction and at the distance of \PDS, which similarly implies $\AR=0.04^{+0.07}_{-0.04}$~mag.

\bibliographystyle{yahapj}
\bibliography{reference}
\end{document}